\documentclass[12pt,a4paper,twoside]{scrartcl}

\usepackage[utf8]{inputenc}

\usepackage[T1]{fontenc}
\usepackage{lmodern}

\usepackage[german,english]{babel}

\usepackage[tmargin=22mm,bmargin=22mm,lmargin=20mm,rmargin=20mm]{geometry}

\usepackage{latexsym,amsmath,amssymb,mathtools,textcomp}
\usepackage[figuresright]{rotating}

\usepackage{amsthm}

\newcommand{\la}{$\leftarrow$ }
\newcommand{\ipso}{IPS${}^4$o}

\newcommand{\red}[1]{\textcolor{red}{#1}}

\numberwithin{equation}{section}

\usepackage{bibgerm}

\usepackage{graphicx}
\usepackage{csquotes}
\usepackage{caption}
\usepackage{pdfpages}
\usepackage{listings}

\usepackage{array,multirow}

\usepackage{enumitem}

\setlist[enumerate]{topsep=0pt}
\setlist[itemize]{topsep=0pt}
\setlist[description]{font=\normalfont,topsep=0pt}

\setlist[enumerate,1]{label=(\roman*)}

\usepackage{tikz}
\usetikzlibrary{calc}

\usepackage{fancyhdr}
\usepackage{fancyvrb}

\fancypagestyle{plain}{
  \fancyhead{}
  \fancyfoot{}
  \fancyfoot[LE,RO]{\normalsize\thepage}

}

\fancypagestyle{normal}{
  \setlength{\headheight}{20pt}
  \setlength\footskip{32pt}
  \fancyhead{}
  \fancyhead[LE]{\normalsize\textsc{\nouppercase{\leftmark}}}
  \fancyhead[RO]{\normalsize\textsc{\nouppercase{\rightmark}}}
  \fancyfoot{}
  \fancyfoot[LE,RO]{\normalsize\thepage}

}

\usepackage{xcolor,hyperref}

\hypersetup{
  pdftitle={Enigneering faster sorters for small sets of items},
  pdfauthor={Jasper Marianczuk},
  pdfsubject={Insertion Sort, Sorting networks, Branching, Conditional Moves, Inline Assembly},
  colorlinks=true,
  pdfborder={0 0 0},
  bookmarksopen=true,
  bookmarksopenlevel=1,
  bookmarksnumbered=true,
  linkcolor=blue!60!black,
  citecolor=blue!60!black,
  urlcolor=blue!60!black,
  filecolor=green!60!black,
  pdfpagemode=UseNone,
  unicode=true,
}

\usepackage[ruled,vlined,linesnumbered,norelsize]{algorithm2e}
\DontPrintSemicolon
\def\NlSty#1{\textnormal{\fontsize{8}{10}\selectfont{}#1}}
\SetKwSty{texttt}
\SetCommentSty{emph}
\def\listalgorithmcfname{Algorithms}
\def\algorithmautorefname{Algorithm}
\let\chapter=\section 

\SetKw{KwForIn}{in}
\SetKw{KwOfSize}{of size}

\begin{document}


\pagestyle{empty} 

\begin{titlepage}

  \begin{center}\large

    \quad\includegraphics[height=17mm]{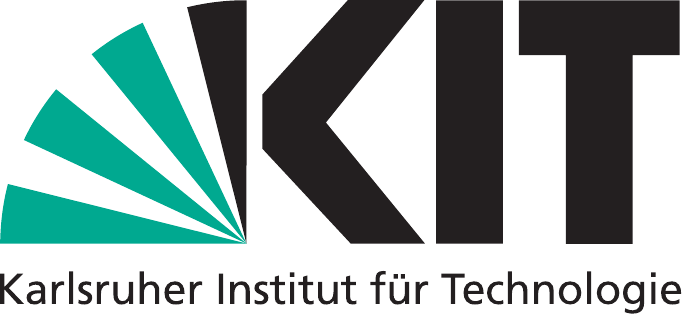} \hfill
    \includegraphics[height=20mm]{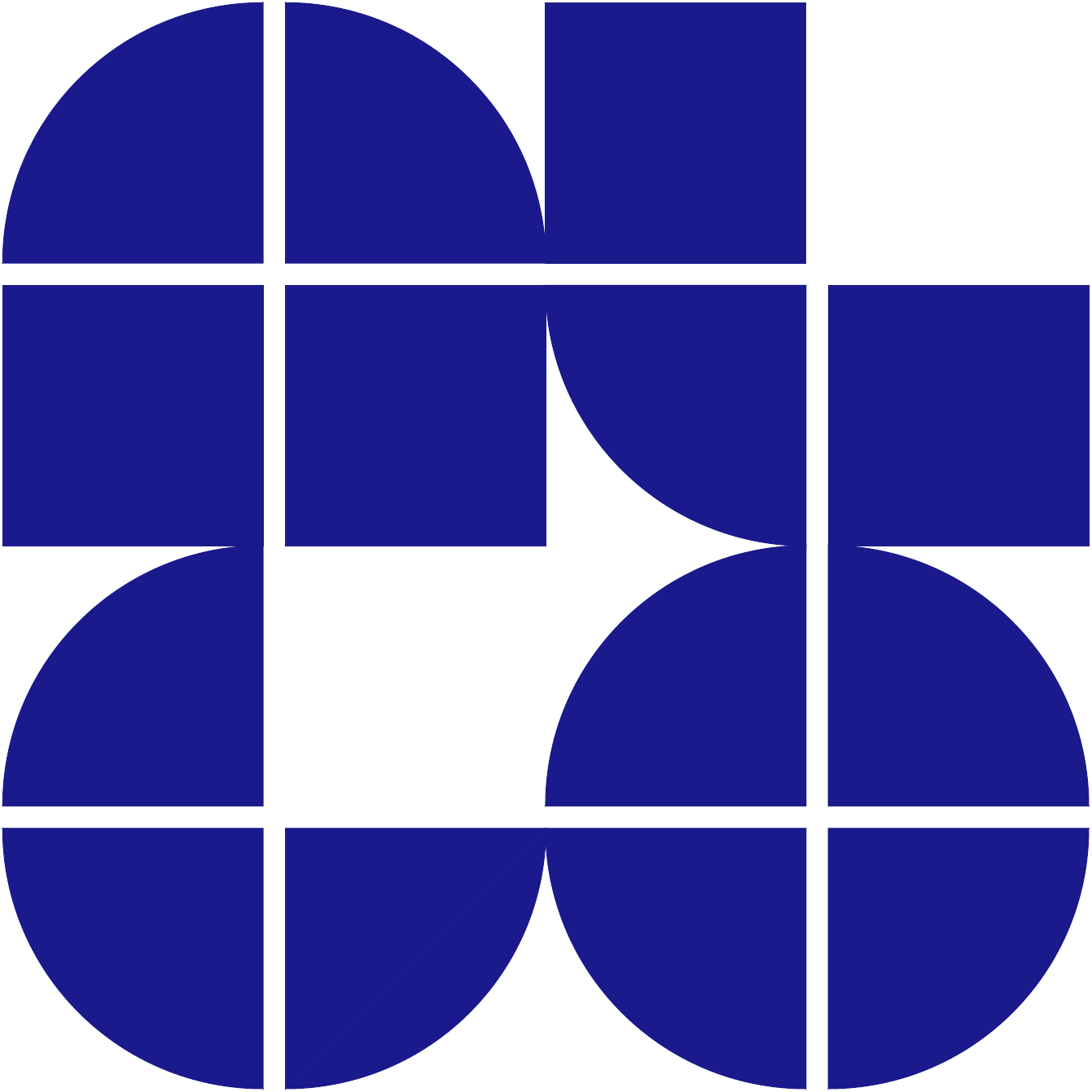}\quad\null

    \vfill

    Bachelor Thesis
    \vspace*{2cm}

    \textbf{\huge Engineering Faster Sorters} \\ \vspace{0.4cm}
    \textbf{\huge for Small Sets of Items}

    \vfill

    Jasper Anton Marianczuk

    \vspace*{15mm}

    Date: May 09, 2019

    \vspace*{45mm}

    \begin{tabular}{rl}
      Supervisors: & Prof. Dr. Peter Sanders \\
      & Dr. Timo Bingmann \\
    \end{tabular}
    
    \vspace*{10mm}


    Institute of Theoretical Informatics, Algorithmics \\
    Department of Informatics \\
    Karlsruhe Institute of Technology

    \vspace*{12mm}
  \end{center}

\end{titlepage}


\vspace*{0pt}\vfill

\hrule\medskip

Hiermit versichere ich, dass ich diese Arbeit selbständig verfasst und keine anderen, als die angegebenen Quellen und Hilfsmittel benutzt, die wörtlich oder inhaltlich übernommenen Stellen als solche kenntlich gemacht und die Satzung des Karlsruher Instituts für Technologie zur Sicherung guter wissenschaftlicher Praxis in der jeweils gültigen Fassung beachtet habe.

\bigskip

\noindent
Karlsruhe, den 09.05.2019


\vspace*{5cm}

\clearpage


\vspace*{0pt}\vfill

\selectlanguage{english}
\begin{abstract}
{\centering \textbf{Abstract}

}
\medskip
Sorting a set of items is a task that can be useful by itself or as a building block for more complex operations. That is why a lot of effort has been put into finding sorting algorithms that sort large sets as efficiently as possible. But the more sophisticated and fast the algorithms become asymptotically, the less efficient they are for small sets of items due to large constant factors.

A relatively simple sorting algorithm that is often used as a base case sorter is insertion sort, because it has small code size and small constant factors influencing its execution time.

This thesis aims to determine if there is a faster way to sort these small sets of items to provide an efficient base case sorter. We looked at sorting networks, at how they can improve the speed of sorting few elements, and how to implement them in an efficient manner by using conditional moves. Since sorting networks need to be implemented explicitly for each set size, providing networks for larger sizes becomes less efficient due to increased code sizes. To also enable the sorting of slightly larger base cases, we modified Super Scalar Sample Sort and created Register Sample Sort, to break down those larger sets into sizes that can in turn be sorted by sorting networks. 

From our experiments we found that when sorting only small sets, the sorting networks outperform insertion sort by at least 25\% for any array size between 2 and 16. When integrating sorting networks as a base case sorter into quicksort, we achieved far less performance improvements over using insertion sort, which is due to the networks having a larger code size and cluttering the L1 instruction cache. The same effect occurs when including Register Sample Sort as a base case sorter for \ipso. But for computers that have a larger L1 instruction cache of 64 KiB or more, we obtained speed-ups of 6.4\% when using sorting networks as a base case sorter in quicksort, and of 9.2\% when integrating Register Sample Sort as a base case sorter into \ipso, each in comparison to using insertion sort as the base case sorter.

In conclusion, the desired improvement in speed could only be achieved under special circumstances, but the results clearly show the potential of using conditional moves in the field of sorting algorithms.

\end{abstract}

\vfill

\selectlanguage{german}
\begin{abstract}
{\centering \textbf{Zusammenfassung}

}
\medskip
Das Sortieren einer Menge von Elementen ist ein Prozess der f\"ur sich alleine n\"utzlich sein kann oder als Baustein f\"ur komplexere Operationen dient. Deswegen wurde in den Entwurf von Sortieralgorithmen, die eine gro\ss{}e Menge an Elementen effizient sortieren, bereits gro\ss{}er Aufwand investiert. Doch je ausgefeilter und schneller die Algorithmen asymptotisch sind, desto ineffizienter werden sie beim Sortieren kleinerer Mengen aufgrund hoher konstanter Faktoren.

Ein relativ einfacher Sortieralgorithmus, der oft als Basisfall Sortierer genutzt wird, ist Insertion Sort, weil dessen Code kurz ist und er kleine konstante Faktoren hat.

Diese Bachelorarbeit hat das Ziel herauszufinden ob es einen schnelleren Algorithmus gibt um solche wenigen Elemente zu sortieren, damit dieser als effizienter Basisfall Sortierer genutzt werden kann. Wir haben uns dazu Sortiernetzwerke angeschaut, wie man durch sie das Sortieren kleiner Listen beschleunigen kann und wie man sie effizient implementiert: Durch das Ausnutzen von konditionellen \texttt{move}-Befehlen. Weil Sortiernetzwerke f\"ur jede Listengr\"o\ss{}e explizit implementiert werden m\"ussen, nimmt die Effizienz des Sortierens mittels Sortiernetwerken wegen erh\"ohter Codegr\"o\ss{}e ab je gr\"o\ss{}er die Listen sind, die sortiert werden sollen. Um auch das Sortieren etwas gr\"o\ss{}erer Basisf\"alle zu erm\"oglichen haben wir Super Scalar Sample Sort modifiziert und Register Sample Sort entworfen, welcher eine gr\"o\ss{}ere Liste in mehrere kleine Listen zerteilt, die dann von den Sortiernetzwerke sortiert werden k\"onnen.

In unseren Experimenten sind wir zu dem Ergebnis gekommen, dass, wenn nur kleine Mengen sortiert werden, die Sortiernetzwerke um mindestens 25\% schneller sind als  Insertion Sort, f\"ur alle Listen, die zwischen 2 und 16 Elementen enthalten. Beim Integrieren der Sortiernetzwerke als Basisfall Sortierer in Quicksort haben wir weit weniger Geschwindigkeitszuwachs gegen\"uber der Benutzung von Insertion Sort erhalten, was daran liegt, dass der Code der Netzwerke mehr Platz ben\"otigt und den Code f\"ur Quicksort aus dem L1 Instruktionscache verdr\"angt. Derselbe Effekt tritt auch beim Benutzen von Register Sample Sort as Basisfall Sortierer f\"ur \ipso\, auf. Allerdings konnten wir uns bei Rechnern, die \"uber einen gr\"o\ss{}eren L1 Instruktionscache von 64 KiB oder mehr verf\"ugen, mit Sortiernetzwerken bei Quicksort um 6,4\% und mit Register Sample Sort bei \ipso\, um 9,2\% gegen\"uber Insertion Sort als Basisfall Sortierer verbessern.

Zusammenfassend haben wir die angestrebte Verbesserung nur unter besonderen Bedingungen erreicht, aber die Ergebnisse weisen deutlich darauf hin, dass die konditionellen \texttt{move}-Befehle Potential im Anwendungsbereich von Sortieralgorithmen haben.

\end{abstract}
\selectlanguage{english}

\vfill\vfill\vfill
\clearpage






\pagestyle{normal}
\renewcommand\sectionmark[1]{\markboth{\thesection\quad\MakeUppercase{#1}}{\thesection\quad\MakeUppercase{#1}}}
\renewcommand\subsectionmark[1]{\markright{\thesubsection\quad\MakeUppercase{#1}}}

\tableofcontents

\clearpage


\listoffigures
\listoftables
\listofalgorithms

\clearpage


\section{Introduction} 
	\subsection{Motivation}
		Sorting, that is rearranging the elements in a set to be in a specific order, is one of the basic algorithmic problems. In school and university, basic sorting algorithms like bubble sort, insertion sort, and merge sort, as well as a simple variant of quicksort are taught at first. These algorithms are rated by the number of comparisons they require to sort a set of items. This amount of comparisons is put into relation to the input size and looked at on an asymptotic level. Only later one realizes that what looks good on paper does not have to work well in practice, so factors like average cases, cache effects, hardware setups, and constant factors need to be taken into consideration, too. A sophisticated choice on which sorting algorithm to use (for a particular use case) should be influenced by all of these factors. \\
		Complex sorting algorithms aim to sort a large number of items quickly, and a lot of them follow the divide-and-conquer idea of designing an algorithm. However, sorting small sets of items, e.g. with 16 elements or less, is usually fast enough that investing a lot of effort into optimizing sorting algorithms for those cases results in very small gains, looking at the absolute amount of time saved. \\
		The complex sorters do not perform as well when sorting small sets of items, having good asymptotic properties but larger constant factors that become more important for the small sizes. Because of that the \emph{base case} of sorting small enough subsets is performed using a simpler algorithm, which is often insertion sort. It has a worst-case run-time of $\mathcal{O}(n^2)$, but small constant factors that make it suitable to use for small $n$. If this sorter is executed many times as base case of a larger sorter though, the times do sum up to contribute to a substantial part of the sorting time. \\
		The guiding question of this thesis is:
		\vspace{-0.01cm}
		\begin{center}
			Is there a faster way to sort sets of up to 16 elements than insertion sort?
		\end{center}
		\vspace{-0.01cm}
		When sorting a set of uniformly distributed random numbers, the chance of any number being greater than another is on average 50\%. Therefore, whenever a conditional branch is influenced by one element's relation to another, one in two of those branches will be mispredicted, which leads to an overall performance penalty.\\
		This is a problem that has already been looked at by Michael Codish, Lu\'is Cruz-Filipe, Markus Nebel and Peter Schneider-Kamp in \enquote{Optimizing sorting algorithms by using sorting networks} \cite{DBLP:journals/fac/CodishCNS17} in 2017, and this thesis has taken a great deal of inspiration from it.
	\subsection{Overview of the Thesis}
		We will first look at sorting networks in section \ref{section:networks}. Section \ref{section:preliminaries:networks} gives a basis of sorting networks and assembly code. After that, we look at different ways of implementing sorting networks efficiently in C++ in section \ref{section:implementation-networks}. For that we focused on elements that consist of a key and an additional reference value. This enables the sorting of complex items, not being limited to integers. \\
		In section \ref{section:samplesort} we will take a small detour to look at Super Scalar Sample Sort and develop an efficient modified version for sets with 256 elements or less by holding the splitters in general purpose registers instead of an array. After that section \ref{section:results} discusses the results and improvements of using sorting networks we achieved in our experiments, measuring the performance of the sorting networks and sample sort individually, and also including them as base cases into quicksort and \ipso~ \cite{DBLP:conf/esa/AxtmannWF017}. After that we conclude the results of this thesis in section \ref{section:conclusion}.
	

\section{Sorting Networks} \label{section:networks}
	\subsection{Preliminaries} \label{section:preliminaries:networks}
		Sorting algorithms can generally be classified into two groups: Those of which the behaviour depends on the input, e.g. quicksort where the sorting speed depends on how well the chosen pivot partitions the set into equally-sized halves, and those of which the behaviour is not influenced by the configuration of the input. The latter are also called \emph{data-oblivious}.
		
		One example of a data-oblivious sorting algorithm is the sorting network. A sorting network of size $n$ consists of a number of $n$ so-called channels numbered $1$ to $n$, each representing one of the inputs, and connections between the channels, called comparators. Where two channels are connected by a comparator it means that the values are to be compared, and if the channel with the lower number currently holds a value that is greater than the value of the channel with the higher number, the values are to be exchanged between the channels. The comparators are given in a fixed order that determines the sequence of executing these \emph{conditional swaps}, so that in the end
		\begin{enumerate}
			\item the channels contain a permutation of the original input, and
			\item the values held by the channels are in nondecreasing order.
		\end{enumerate}
		Sorting networks are data-oblivious because all the comparisons are always performed, and in the same order, no matter which permutation of an input is given.
		
		For any sorting network, two metrics can be used to quantify it: the \emph{length} and the \emph{depth}. A network's length refers to the number of comparators it contains, and a network's depth describes the minimal amount of levels a network can be divided into. \\
		Where two comparators are ordered one after the other, and no channel is used by both comparators, they can be combined into a level. In other words: the result of the second comparator does not depend upon the result of the first. Inductively, any comparator can be merged into a level that executes right before or after it, if its channels are not already used by any comparator in the level. Since now all the comparators in a level are independent from one another, they can be executed in parallel.
		\subsubsection{Networks in Practice}
			\begin{itemize} 
				\item \textbf{Best known networks:} For networks of up to size 16 there exist proven optimal lengths and proven optimal depths. For example, the network for 10 elements with optimal length 29 has depth 9, the one with optimal depth 7 has length 31 \cite{DBLP:books/lib/Knuth98a, DBLP:conf/ictai/CodishCFS14}. For networks of greater size there only exist currently known lowest numbers of length or depth. Those best networks are acquired through optimizations that were initially done by hand and nowadays are realized e.g.~with the help of computers and evolutionary algorithms \cite{SorterHunter}.
				\item \textbf{Recursive networks:} For creating sorting networks there also exist algorithms that work in a recursive divide-and-conquer way: split the input into two parts, sort each part recursively, and merge the two parts together in the end. Representatives for this kind of approach are the constructions of R.J. Nelson and B.C. Bose \cite{DBLP:journals/jacm/BoseN62} and the algorithm by K.E. Batcher \cite{DBLP:conf/afips/Batcher68}. Bose and Nelson split the input sequence into first and second half, while Batcher partitions into elements with an even index and elements with an odd index. The advantage of those recursive networks over the specially optimized ones is that they can easily be created even for large network sizes. While the generated networks may have more comparators than the best known networks, the number of comparators in a network acquired from either Bose-Nelson or Batcher of size $n$ has an upper bound of $\mathcal{O}(n\, (\log n)^2)$ \cite{DBLP:books/lib/Knuth98a}.
			\end{itemize}
			\begin{figure}[!htb]
				\begin{center}
					\includegraphics[width=0.7\textwidth]{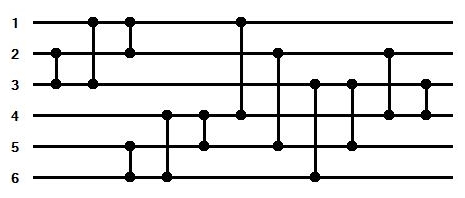}
				\end{center}
				\caption{Sorting network by Bose and Nelson for 6 elements} \label{network:bosenelson:6}
			\end{figure}
			Sorting networks are usually depicted by using horizontal lines for the channels, and vertical connections between these lines for the comparators. A network by Bose and Nelson for 6 elements displayed like that can be seen in figure \ref{network:bosenelson:6}. \\
			
		\subsubsection{Improving the Speed of Sorting through Sorting Networks}
			An important question to ask is how sorting networks can improve the sorting speed on a set of elements (on average), if they can not take any shortcuts for \enquote{good} inputs, like an insertion sort that would leverage an already sorted input and do one comparison per element. The answer to this question is \emph{branching}. Because the compiler knows in advance which comparisons are going to be executed in which order, the control flow does not contain conditional branches, in particular getting rid of expensive branch mispredictions. On uniformly distributed random inputs, the chances that any number is smaller than another is 50\% on average, making branches unpredictable. In the case of insertion sort that means not knowing in advance with how many elements the next one has to be compared until it is inserted into the right place. \\
			Even though with sorting networks the compiler knows in advance when to execute which comparator, implementing the compare-and-swap operation in a naive way (as seen in \ref{section:preliminaries:compare-and-swap}) the compiler might still generate branches. In that case, the sorting networks are no faster than insertion sort, or even slower.
		\subsubsection{Compare-and-Swap} \label{section:preliminaries:compare-and-swap}
			For sorting networks, the basic operation used is to compare two values against each other. If they are in the wrong order (the \enquote{smaller} element occurs after the \enquote{bigger} one in the sequence), they are swapped. Intuitively, one might implement the operation in C++ like this:
			
			\begin{verbatim}
				void ConditionalSwap(TValueType& left, TValueType& right)
				{
				    if (left > right) { std::swap(left, right); }
				}
			\end{verbatim} 
			Here \verb|TValueType| is a template typename and can be instantiated with any type that implements the \verb|>| operator. \\
			As suggested in \cite{DBLP:journals/fac/CodishCNS17}, the same piece of code can be rewritten like this:
			\begin{verbatim}
				void ConditionalSwap2(TValueType& left, TValueType& right)
				{
				    TValueType temp = left;
				    if (temp > right) { left = right; }
				    if (temp > right) { right = temp; }
				}
			\end{verbatim}
			At first glance it looks like we now have two branches that can be taken. But the code executed if the condition is true now only consists of a single assignment each, which can be expressed in x86-Architecture through a \emph{conditional move} instruction. In AT\&T syntax (see section \ref{section:preliminaries:asm}), a conditional move (\verb|cmov a,b|) will write a value in register \verb|a| into register \verb|b|, if a condition is met. If the condition is not met, no operation takes place (still taking the same number of CPU cycles as the move operation would have). Since the address of the next instruction no longer depends upon the previously evaluated condition, the control flow now does not contain branches. The only downside of the conditional move is that it can take longer than a normal move instruction on certain architectures, and can only be executed when the comparison has performed and its result is available. \\
			When the elements to be sorted are only integers, some compilers do generate code with conditional moves for those operations. When the elements are more general (in this thesis we will look at pairs of an unsigned 64 bit integer key and an unsigned 64 bit reference value, which could be a pointer or an address in an array), gcc 7.3.0, the compiler used for the experiments, does not generate conditional moves. To force the usage of conditional moves, a feature of gcc was used that allows the programmer to specify small amounts of assembly code to be inserted into the regular machine code generated by gcc, called inline assembly \cite{GccInlineAssembly}. This mechanic and the notation is further explained in section \ref{section:preliminaries:asm}.
	
		\subsubsection{Assembly Code} \label{section:preliminaries:asm}
		Assembly code represents the machine instructions executed by the CPU. It can be given as the actual opt-codes or as human-readable text. There are two different conventions for the textual representation, the Intel syntax or MASM syntax and the AT\&T syntax. The main differences are: \medskip \\
		\begin{tabular}{c | p{6.2cm} | p{6.2cm}}
							& \multicolumn{1}{c |}{Intel} & \multicolumn{1}{c}{AT\&T} \\ \hline
			Operand size 	& The size of the operand does not have to be specified & The size of the operand is appended to the instruction: \verb|b| (byte = 8 bit), \verb|l| (long = 32 bit), \verb|q| (quad-word = 64 bit) \\ \hline
			Parameter order & The destination is written first, then the source of the value: \verb|mov dest,src| & The source is written first, then the destination: \verb|movq src,dest| \\
		\end{tabular} \bigskip \\
		In this thesis only the AT\&T syntax will be used. \\
		The gcc C++ compiler has a feature that allows the programmer to write assembly instructions in between regular C++ code, called \enquote{inline assembly} (\verb|asm|) \cite{GccInlineAssembly}. A set of assembly instructions to be executed must be given, followed by a definition for input and output variables and a list of clobbered registers. This extra information is there to communicate to the compiler what is happening inside the \verb|asm| block. Gcc itself does not parse or optimize the given assembly statements, they are only after compilation added into the generated assembly code by the GNU Assembler. 
		A variable being in the output list means that the value will be modified, a clobbered register is one where gcc cannot assume that the value it held before the \verb|asm| block will be the same as after the block. In this thesis, the clobbered registers will almost always be the conditional-codes registers (\verb|cc|), which include the carry-flag, zero-flag and the signed-flag, which are modified during a compare-instruction. This way of specifying the input, output and clobbered registers is also called \emph{extended asm}. \\
		Taking the code from \ref{section:preliminaries:compare-and-swap}, and assuming \verb|TValueType = uint64_t|, the statement
		\begin{verbatim}
			if (temp > right) { left = right; }
		\end{verbatim}
		can now be written as
		\begin{verbatim}
			__asm__(
			  "cmpq %[temp],%[right]\n\t"        //performs right - temp internally
			  "cmovbq %[right],%[left]\n\t"      //left = right, if right < temp
			  : [left] "=&r"(left)               //output
			  : "0"(left), [right] "r"(right), [temp] "r"(temp) //input
			  : "cc"                             //clobber
			);
		\end{verbatim}
		In extended \verb|asm|, one can define C++ variables as input or output operands, and gcc will assign a register for that value (if it has the \verb|"r"| modifier), and also write the value in an output register back to the given variable after the \verb|asm| block. Note that the names in square brackets are symbolic names only valid in the context of the assembly instructions and independent from the names in the C++ code before. The link between the C++ names and the symbolic names happens in the input and output declarations. \\
		With the conditional moves it is important to properly declare the input and output variables, because they perform a task that is a bit unusual: an output variable may be overwritten, and also may not. For the output register for \verb|left|, two things must apply:
		\medskip
		\begin{enumerate}
			\item if the condition is false, it must hold the value of \verb|left|, and \label{cmov:condition:1}
			\item if the condition is true, it must hold the value of \verb|right|.
		\end{enumerate}
		\medskip
		For optimizations purposes, the compiler might reduce the number of registers used by placing the output of one operation into a register that previously held the input for some other operation. To prevent this, the declaration for the output \verb|[left] "=&r"(left)| has the \verb|"&"| modifier added to it, meaning it is an \enquote{early clobber} register and that no other input can be placed in that register. In combination with \verb|"0"(left)| in the input line, it is also tied to an input, so that the previous value of \verb|left| is loaded into the register beforehand, to comply with constraint \ref{cmov:condition:1}. Because we already declared it as output, instead of giving it a new symbolic name we tie it to the output by referencing its index in the output list, which since it is the first output variable is \verb|"0"|. The \verb|"="| in the output declaration solely means that this register will be written to. Any output needs to have the \verb|"="| modifier.  \\
		We see that each assembly instruction is postfixed with \verb|\n\t|. That is because the instruction strings are appended into a single instruction string during compilation and \verb|\n\t| tells the GNU assembler where one instruction ends and the next begins. \\
		
		The \verb|cmov| instruction is postfixed with a \verb|b| in this example, which stands for \enquote{below}. So the \verb|cmov| will be executed if \verb|right| is below \verb|temp| (unsigned comparison \verb|right < temp|). Apart from below we will also see not equal (\verb|ne|) and  carry (\verb|c|) as a postfix. \\
		In addition to that, both the \verb|cmp| and the \verb|cmovb| are postfixed with a \verb|q| (quad-word) to indicate that the operands are 64-bit values. \\
		When a subtraction $\mathtt{minuend} - \mathtt{subtrahend}$ is performed and \texttt{subtrahend} is larger than \texttt{minuend} (interpreted as unsigned numbers), the operation causes an underflow which results in the carry flag being set to 1. The check for that carry flag being 1 can be used as a condition by itself, and the carry flag influences other condition checks like \emph{below}. This property of the comparison setting the carry flag will be used in section \ref{section:samplesort:impl}. 


	\subsection{Implementation of Sorting Networks} \label{section:implementation-networks}
		\subsubsection{Providing the Network Frame}
			\begin{figure}[!p]
				\includegraphics[width=1.0\textwidth]{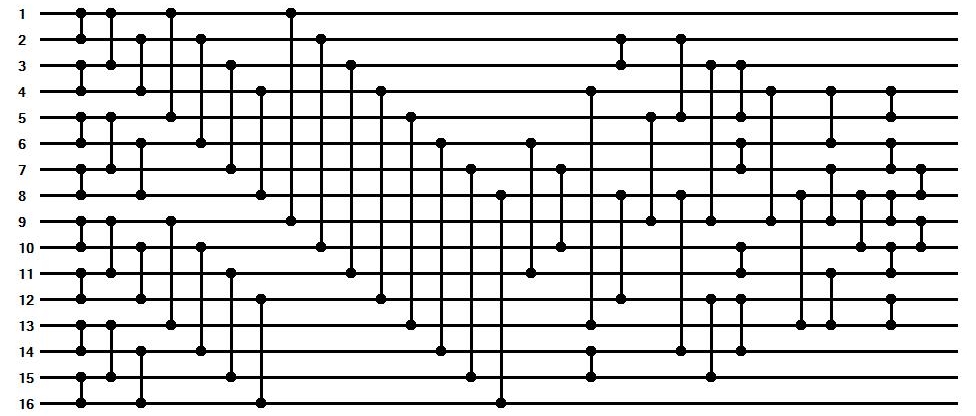}
				\caption{Best network with optimal length for 16 elements} \label{network:best:16}
			\end{figure}
			\begin{figure}[!p]
				\includegraphics[width=1.0\textwidth]{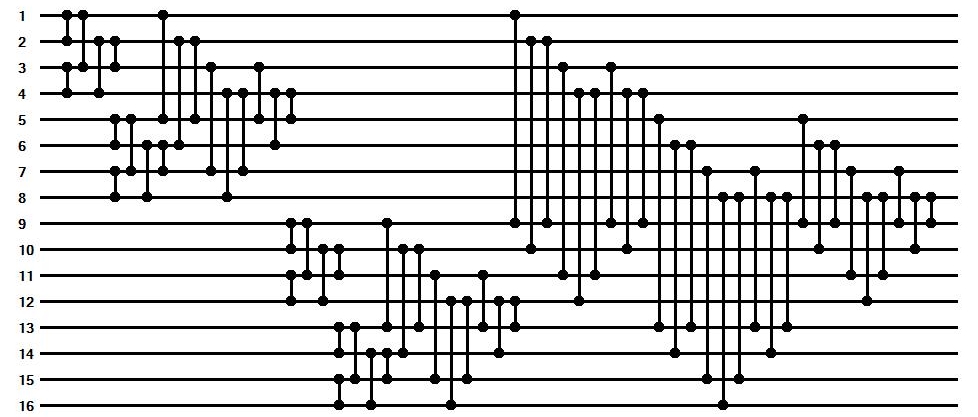}
				\caption{Bose Nelson network for 16 elements optimizing locality} \label{network:bosenelson:16}
			\end{figure}
			\begin{figure}[!p]
				\includegraphics[width=1.0\textwidth]{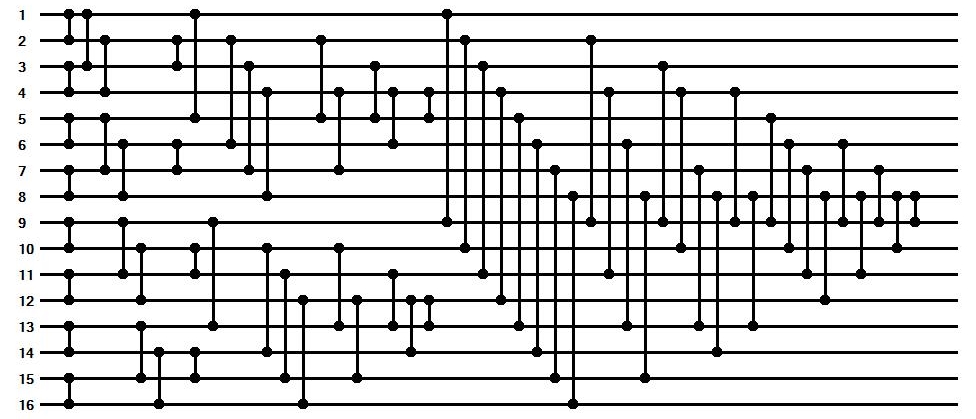}
				\caption{Bose Nelson network for 16 elements optimizing parallelism} \label{network:bosenelson:parl:16}
			\end{figure}
			The best networks for sizes of up to 16 elements were taken from John Gamble's Website \cite{JgambleNetworks} and are length-optimal. \\
			The Bose Nelson networks have been generated using the instructions from their paper \cite{DBLP:journals/jacm/BoseN62}. \\
			For sizes of 8 and below the best and generated networks have the same amount of comparators and levels. For sizes larger than 8 the generated networks are at a disadvantage because they have more comparators and/or levels. As a trade-off their recursive structure makes it possible to leverage a different trait: locality. Instead of optimizing them to sort as parallel as possible, we can first sort the first half of the set, then the second half, and then apply the merger. This way, chances are higher that all $\frac{n}{2}$ elements of the first half might fit into the processor's general purpose registers. During this part of the sorting routine, no accesses to memory or cache are required. To determine if there is a visible speed-up, the networks were generated optimizing (a) locality and (b) parallelism. \\
			As an extra idea, the Bose Nelson networks were generated in a way that one can pass the elements as separate parameters instead of as an array. That way one can sort elements that are not contiguously placed in memory. Because the networks were implemented as method calls to the smaller sorters and merge methods, there would be a large overhead in placing many elements onto the call stack for each method call. While we hoped this would make a difference by reducing code size, the overhead for the method call was too large. That is why all the methods are declared \verb|inline| which results in the same flat sequence of swaps for each size the networks optimizing locality have. \\
			Examples of networks for 16 elements can be seen in figures \ref{network:best:16}, \ref{network:bosenelson:16} and \ref{network:bosenelson:parl:16}.
			
			All networks are implemented so that they have an entry method that takes a pointer to an array \verb|A| and an array size \verb|n| as input and delegates the call to the specific method for that number of elements, which in turn executes all the comparators. To measure different implementations for the conditional swaps, the network methods and the swap are templated, so that when calling the network with an array of a specific type the respective specialized conditional-swap implementation will be used. \\
			\clearpage
			Our approach differs from the work in \cite{DBLP:journals/fac/CodishCNS17} in the type of elements that were sorted. While they measured the sorting of \verb|int|s, which are usually 32-bit sized integers, we made the decision to sort elements that consist of a 64-bit integer key and a 64-bit integer reference value, enabling not only the sorting of numbers but also the sorting of complex elements, when giving a pointer or an array index as the reference value to the original set. This was implemented by creating \verb|struct|s that contain a key and reference value each, having the following structure:
			\begin{verbatim}
				struct SortableRef
				{
				    uint64_t key, reference;
				}
			\end{verbatim}
			They also define the operators~~\verb|>, >=, ==, <, <=|~~and~~\verb|!=|~~for reasons of usability, and templated methods \verb|uint64_t GetKey(TSortable)| and \verb|uint64_T GetReference(TSortable)| are available.
		\subsubsection{Implementing the Conditional Swap} \label{section:implementation-conditionalswap}
			The \texttt{ConditionalSwap} is implemented as a templated method like this:
			\begin{verbatim}
				template <typename TValueType>
				inline
				void ConditionalSwap(TValueType& left, TValueType& right)
				{
				    //body
				}
			\end{verbatim}
			The following variants will represent the body of one specialization of the template function for a specific struct. Each of them was given a three letter abbreviation to name them in the results. We implemented the following approaches:
			\newcommand{\Def}{using std::swap (\texttt{Def})}
			\newcommand{\QMa}{using inline if statements (\texttt{QMa})}
			\newcommand{\Tie}{using std::tie and std::tuple (\texttt{Tie})}
			\newcommand{\JXc}{using jmp and xchg (\texttt{JXc})}
			\newcommand{\FCm}{using four cmovs and temp variables (\texttt{4Cm})}
			\newcommand{\FCS}{using four cmovs split from one another and temp variables (\texttt{4CS})}
			\newcommand{\SCm}{using six cmovs and temp variables (\texttt{6Cm})}
			\newcommand{\Cla}{moving pointers with cmov instead of values (\texttt{Cla})}
			\newcommand{\CPr}{moving pointers and supporting a predicate (\texttt{CPr})}
			\begin{itemize}
				\item \Def
				\item \QMa
				\item \Tie
				\item \JXc
				\item \FCm
				\item \FCS
				\item \SCm
				\item \Cla
				\item \CPr
			\end{itemize}
			The details of implementation can be seen in the following paragraphs.
			\paragraph*{\Def} The default implementation for the template makes use of the defined \verb|<| operator:
				\begin{verbatim}
					if (right < left)
					    std::swap(left, right);
				\end{verbatim}
				This is the intuitive way of writing the conditional swap we already saw in section \ref{section:preliminaries:compare-and-swap}, without any inline assembly.
			\paragraph*{\QMa}
				\begin{verbatim}
					bool r = (left > right);
					auto temp = left;
					left = r ? right : left;
					right = r ? temp : right;
				\end{verbatim}
				Here it was attempted to convince the compiler to generate conditional moves by using the \emph{inline if}-statements with trivial values in the else part.
			\paragraph*{\Tie}
				\begin{verbatim}
					std::tie(left, right) =
					    (right < left) ? std::make_tuple(right, left) : std::make_tuple(left, right);
				\end{verbatim}
				This approach uses assignable tuples (tie).
			\paragraph*{\JXc}
				\begin{verbatim}
					__asm__(
					    "cmpq %[left_key],%[right_key]\n\t" 
					    "jae %=f\n\t"
					    "xchg %[left_key],%[right_key]\n\t" 
					    "xchg %[left_reference],%[right_reference]\n\t"
					    "%=:\n\t" 
					    : [left_key] "=&r"(left.key), [right_key] "=&r"(right.key), 
					      [left_reference] "=&r"(left.reference), 
					      [right_reference] "=&r"(right.reference)
					    : "0"(left.key), "1"(right.key), "2"(left.reference), "3"(right.reference)
					    : "cc" 
					);
				\end{verbatim}
				The \verb|%=| generates a unique label for each instance of the \verb|asm| statement, so that the jumps go where they belong.
			\paragraph*{\FCm}
				\begin{verbatim}
					uint64_t tmp = left.key;
					uint64_t tmpRef = left.reference;
					__asm__( 
					    "cmpq %[left_key],%[right_key]\n\t" 
					    "cmovbq %[right_key],%[left_key]\n\t" 
					    "cmovbq %[right_reference],%[left_reference]\n\t"
					    "cmovbq %[tmp],%[right_key]\n\t"
					    "cmovbq %[tmp_ref],%[right_reference]\n\t"
					    : [left_key] "=&r"(left.key), [right_key] "=&r"(right.key), 
					      [left_reference] "=&r"(left.reference), 
					      [right_reference] "=&r"(right.reference)
					    : "0"(left.key), "1"(right.key), "2"(left.reference), "3"(right.reference), 
					      [tmp] "r"(tmp), [tmp_ref] "r"(tmpRef)
					    : "cc" 
					);
				\end{verbatim}
				
			\paragraph*{\FCS}
				\begin{verbatim}
					uint64_t tmp = left.key;
					uint64_t tmpRef = left.reference;
					__asm__ volatile ( 
					    "cmpq %[left_key],%[right_key]\n\t"
					    : 
					    : [left_key] "r"(left.key), [right_key] "r"(right.key)
					    : "cc" 
					);
					__asm__ volatile (
					    "cmovbq %[right_key],%[left_key]\n\t"
					    : [left_key] "=&r"(left.key)
					    : "0"(left.key), [right_key] "r"(right.key)
					    : 
					);
					__asm__ volatile (
					    "cmovbq %[right_reference],%[left_reference]\n\t"
					    : [left_reference] "=&r"(left.reference)
					    : "0"(left.reference), [right_reference] "r"(right.reference)
					    :
					);
					__asm__ volatile (
					    "cmovbq %[tmp],%[right_key]\n\t"
					    : [right_key] "=&r"(right.key)
					    : "0"(right.key), [tmp] "r"(tmp)
					    : 
					);
					__asm__ volatile (
					    "cmovbq %[tmp_ref],%[right_reference]\n\t"
					    : [right_reference] "=&r"(right.reference)
					    : "0"(right.reference), [tmp_ref] "r"(tmpRef)
					    : 
					);
				\end{verbatim}
				Because we split the \verb|asm| blocks, they have to be declared \verb|volatile| so that the optimizer does not move them around or out of order. Without declaring them \verb|volatile|, some of the networks were not sorting correctly. The blocks were split because we hoped the compiler would be able to insert operations that do not affect the conditional codes and are unrelated to the current conditional swap between the \verb|cmp|-instruction and the conditional moves, to reduce the amount of wait cycles that have to be performed. This was successful as can be seen in the experimental results in section \ref{section:experiments:normal}. \\
				\newpage
			\paragraph*{\SCm}
				\begin{verbatim}
					uint64_t tmp;
					uint64_t tmpRef;
					__asm__ ( 
					    "cmpq %[left_key],%[right_key]\n\t" 
					    "cmovbq %[left_key],%[tmp]\n\t"
					    "cmovbq %[left_reference],%[tmp_ref]\n\t"
					    "cmovbq %[right_key],%[left_key]\n\t" 
					    "cmovbq %[right_reference],%[left_reference]\n\t"
					    "cmovbq %[tmp],%[right_key]\n\t"
					    "cmovbq %[tmp_ref],%[right_reference]\n\t"
					    : [left_key] "=&r"(left.key), [right_key] "=&r"(right.key), 
					      [left_reference] "=&r"(left.reference), 
					      [right_reference] "=&r"(right.reference), 
					      [tmp] "=&r"(tmp), [tmp_ref] "=&r"(tmpRef)
					    : "0"(left.key), "1"(right.key), "2"(left.reference), "3"(right.reference), 
					      "4"(tmp), "5"(tmpRef)
					    : "cc" 
					); 
				\end{verbatim}
				
			\paragraph*{\Cla} This idea came from a result created by the clang compiler from the special code as seen in the ConditionalSwap2 method in \ref{section:preliminaries:compare-and-swap}. For the transformation to gcc, we took only the minimal necessary instructions concerning the conditional move into the \verb|asm| block: 
				\begin{verbatim}
					SortableRef_ClangVersion* leftPointer = &left;
					SortableRef_ClangVersion* rightPointer = &right;
					uint64_t rightKey = right.key;
					SortableRef_ClangVersion tmp = left;
					__asm__ volatile(
					    "cmpq %[tmp_key],%[right_key]\n\t"
					    "cmovbq %[right_pointer],%[left_pointer]\n\t"
					    : [left_pointer] "=&r"(leftPointer)
					    : "0"(leftPointer), [right_pointer] "r"(rightPointer), 
					      [tmp_key] "m"(tmp.key), [right_key] "r"(rightKey)
					    : "cc"
					);
					left = *leftPointer;
					leftPointer = &tmp;
					__asm__ volatile(
					    "cmovbq %[left_pointer],%[right_pointer]\n\t"
					    : [right_pointer] "=&r"(rightPointer)
					    : "0"(rightPointer), [left_pointer] "r"(leftPointer)
					    :
					);
					right = *rightPointer;
				\end{verbatim}
			\newpage
			\paragraph*{\CPr} Instead of performing the comparison inside the \texttt{asm} block, which requires knowledge of the datatype of the key, it can also be done over a predicate, using the result of that comparison inside the inline assembly:
				\begin{verbatim}
					SortableRef_ClangPredicate* leftPointer = &left;
					SortableRef_ClangPredicate* rightPointer = &right;
					SortableRef_ClangPredicate temp = left;
					int predicateResult = (int) (right < temp);
					__asm__ volatile(
					    "cmp $0,%[predResult]\n\t"
					    "cmovneq %[right_pointer],%[left_pointer]\n\t"
					    : [left_pointer] "=&r"(leftPointer)
					    : "0"(leftPointer), [right_pointer] "r"(rightPointer), 
					      [predResult] "r"(predicateResult)
					    : "cc"
					);
					left = *leftPointer;
					leftPointer = &temp;
					__asm__ volatile(
					    "cmovneq %[left_pointer],%[right_pointer]\n\t"
					    : [right_pointer] "=&r"(rightPointer)
					    : "0"(rightPointer), [left_pointer] "r"(leftPointer)
					    :
					);
					right = *rightPointer;
				\end{verbatim}
				For the \texttt{Cla} implementation the \verb|b| in \verb|cmovb| was used to execute the conditional move if \verb|right_key| was smaller than \verb|temp_key|. If that is the case, the predicate will return true, or as an int a value not equal to zero. When comparing this result to 0, the \verb|cmov| is to be executed if the result was any value other than zero, so the postfix here is \verb|ne| (not equal).	\\
				Note that while the knowledge of how to compare the elements is still present by doing the comparison directly (\verb|right < temp|), the compiler now needs to take the result from the comparison, and put it into an integer that is then used in the \verb|asm| block. The only addition to make it completely independent from the sorted elements would be to pass a predicate to do the comparison, which would also involve modifying the network frame to take and pass the predicate. To measure on the same network frame we took this shortcut of doing the comparison using the \verb|<| operator. 
\newpage
\section{Register Sample Sort} \label{section:samplesort}
	\subsection{Preliminaries} \label{section:samplesort:preliminaries}
		Sample sort is a sorting algorithm that follows the divide-and-conquer principle. The input is separated into $k$ subsets, that each contain elements within an interval of the total ordering, with the intervals being distinct from one another. That is done by first choosing a subset $S$ of $a \cdot k$ elements and sorting $S$. Afterwards the splitters $\{s_0, s_1, \ldots, s_{k-1}, s_k\} = \{-\infty, S_a, S_{2a}, \ldots, S_{(k-1)a}, \infty\}$ are taken from $S$. The parameter $a$ denotes the oversampling factor. Oversampling is used to get a better sample of splitters to achieve more evenly-sized partitions, trading for the time that is required to sort the larger sample. \\
		With the splitters the elements $e_i$ are then \emph{classified}, placing them into buckets $b_j$, where $j \in \{1, \ldots, k\}$ and $s_{j-1} < e_i \leq s_j$. For $k$ being a power of 2, this placement can be achieved by viewing the splitters as a binary tree, with $s_{k/2}$ being the root, all $s_l$ with $l < k/2$ representing the left subtree and those with $l > k/2$ the right one. To place an element, one must only traverse this binary tree, resulting in a binary search instead of a linear one \cite{DBLP:conf/esa/SandersW04}.\\
		Quicksort is therefore a specialization of sample sort with fixed parameter $k = 2$, having only one splitter, the pivot, and splitting the input into two partitions.
	\subsection{Implementing Sample Sort for Medium-Sized Sets} \label{section:samplesort:impl}
		The motivation to look at sample sort was that we wanted to see how well the sorting networks perform when using them as a base case for the \textit{In-Place Parallel Super Scalar Samplesort} (\ipso) by Michael Axtmann, Sascha Witt, Daniel Ferizovic and Peter Sanders \cite{DBLP:conf/esa/AxtmannWF017}. The problem that occured is that \ipso~can go into the base case with sizes larger than 16, while the networks we looked at only sort sets of up to 16 elements. \\
		To close that gap, we created a sequential version of Super Scalar Sample Sort \cite{DBLP:conf/esa/SandersW04} that can reduce base case sizes of up to 256 down to blocks of 16 or less in an efficient manner.
		
		Since the total size was expected to not be much greater than 256, not much effort was made to keep the algorithm in-place. The central idea was to place the splitters not into an array, as described in \cite{DBLP:conf/esa/SandersW04}, but to hold them in general purpose registers for the whole duration of the element classification.
		
		The question now arose as to which splitter an element needs to be compared to after the first comparison with the middle splitter. When the splitters are organized in a binary heap in an array, that can be done by using array indices, the children of splitter $j$ being at positions $2j$ and $2j + 1$. If an element is smaller than $s_j$, it would afterwards be compared to $s_{2j}$, otherwise to $s_{2j+1}$. But this way of accessing the splitters does not work when they are placed in registers. The solution was to create a copy of the left subtree, and to conditionally overwrite that with the right subtree, should the element be greater than the root node. The next comparison is then made against the root of the temporary tree that now contains the correct splitters to compare that element against. For 3 splitters that requires 1 conditional move, and for 7 splitters would require 3 conditional moves after the first comparison and 1 more after the second comparison, per element.
		
		After finding the correct splitters to compare to, we are left with one more problem: How to know in which bucket the element is to be placed into at the end. In \cite{DBLP:conf/esa/SandersW04} this was done by making use of the calculated index determining the next splitter to compare to. We chose an approach similar to creating this index, using the correlation between binary numbers and the tree-like structure of the splitters. We will be viewing the splitters not as a binary heap but just as a list where the middle of the list represents the root node of the tree, its children being the middle element of the left and the middle element of the right list. \\
		If an element $e_i$ is larger than the first splitter $s_{k/2}$ (with $k-1$ being the number of splitters), it must be placed in a bucket $b_j$ with $j \geq \frac{k}{2}$ (assuming 0-based indexing for $b$). That also means that the index of that bucket, represented as a binary number, must have its bit at position $l := \log \frac{k}{2}$ set to 1. That way, the result of the comparison ($e_i > s_{k/2}$) can be interpreted as an integer (1 for \verb|true|, 0 for \verb|false|) and added to j. If that was not the last comparison, $j$ is then multiplied by 2 (meaning its bits are shifted left by one position). This means the bit from the first comparison makes its way \enquote{left} in the binary representation while the comparison traverses down the tree, and so forth with the other comparisons. After traversing the splitter tree to the end, $e_i$ will have been compared to the correct splitters and $j$ will hold the index of the bucket that $e_i$ belongs into. These operations can be implemented without branches by making use of the way comparisons are done: \\
		At the end of section \ref{section:preliminaries:asm} we explained that when comparing (unsigned) numbers (which is nothing but a subtraction), and the \texttt{subtrahend} being greater than the \texttt{minuend}, the operation causes an underflow and the carry flag is set. We also notice that when converting the result of the predicate ($e_i > s_{k/2}$) to an integer value, the integer will be 1 for \verb|true| and 0 for \verb|false|. So in assembly code, we can compare the result from evaluating the predicate to the value 0: \verb|cmp %[predResult],%[zero]| where \verb|zero| is just a register that holds the value 0. This trick is needed because the \verb|cmp| instruction needs the second operand to be a register. This will execute $0 - \mathtt{predResult}$, which underflows for the predicate returning true. This way we can postfix the \verb|cmov| needed for moving the next splitters with a \verb|c| checking for a set carry flag. The second instruction we make use of is the \emph{rotate carry left} (\verb|rcl|) instruction, which performs a \emph{rotate left} instruction on $j$, but includes the carry flag as an additional bit after the least significant bit of the integer. This exactly takes the predicate result and puts it at the bottom of $j$, with the previous content being shifted one to the left beforehand. That means it performs two necessary operations at once. \\
		As an addition to the efficient classification, while looping over the elements we allow to place multiple elements into buckets per loop, allowing for all the registers in the machine to be used. This additional parameter is called \emph{blockSize}.
		
		There is one downside to this approach: The keys of the splitters (since we only need a splitter's key for classifying an element) must be small enough to fit into a general purpose register. Needing more than one register per key would mean either running out of registers or spending extra time to conditionally move the splitter keys around. For three splitters the needed number of registers for block sizes 1 to 5 are as seen in table \ref{table:samplesort:registerusage}. We can see that the trade-off for classifying multiple elements at the same time is the amount of registers needed. \\
		If we were to use 7 splitters instead of three, the number of registers required for classifying just 1 element at a time would go up to 15. Also, with 8 buckets, if we get recursive subproblems with sizes just over 16, classifying into 8 buckets again would be greatly inefficient, resulting in many buckets containing very few. This is why we decided to only use three splitters for this particular sorter. \\
		Pseudocode to implement the classification can be seen as an example for an array of integers and \texttt{blockSize = 1} in algorithm \ref{algo:samplesort:cversion}. $j$ is here called \texttt{state}, and the temporary subtree consists of one splitter which we gave the name \texttt{splitterx}. For the branchless implementation we used the \verb|cmovc| for line \ref{algo:samplesort:cmov} and the \verb|rcl| instruction for line \ref{algo:samplesort:rcl}. At the last level of classification no more moving of splitters is required, so instead of doing another comparison against the predicate result and using \verb|rcl|, we can just shift \texttt{state} left by one position and add the predicate's result to it (line \ref{algo:samplesort:laststate}). Alternatively we could use a bitwise \verb|OR| or \verb|XOR| after the shift, which would have the same result. But we decided that adding the predicate result was more readable. \\
		For sorting the splitter sample, the same sorting method can be used as for the base case. \\
		\begin{table}[!h]
			\begin{center}
				\begin{tabular}{ r | c c c c c | c c c c c}
					& \multicolumn{5}{c |}{3 splitters} & \multicolumn{5}{c}{7 splitters} \\ \hline
					& \multicolumn{5}{c |}{block size} & \multicolumn{5}{c}{block size} \\
					& 1 & 2 & 3 & 4 & 5 & 1 & 2 & 3 & 4 & 5 \\ \hline
					splitters 				& 3 & 3 & 3 & 3 & 3 & 7 & 7 & 7 & 7 & 7 \\
					buckets pointer 		& 1 & 1 & 1 & 1 & 1 & 1 & 1 & 1 & 1 & 1 \\
					current element index 	& 1 & 1 & 1 & 1 & 1 & 1 & 1 & 1 & 1 & 1 \\
					element count			& 1 & 1 & 1 & 1 & 1 & 1 & 1 & 1 & 1 & 1 \\ \hline
					state 					& 1 & 2 & 3 & 4 & 5 & 1 & 2 & 3 & 4 & 5 \\
					predicate result				& 1 & 2 & 3 & 4 & 5 & 1 & 2 & 3 & 4 & 5 \\
					splitterx				& 1 & 2 & 3 & 4 & 5 & 3 & 6 & 9 & 12 & 15 \\ \hline
					sum						& 9 & 12 & 15 & 18 & 21 & 15 & 20 & 25 & 30 & 35
				\end{tabular}
			\end{center}
			\caption{Registers required by Register Sample Sort with three or seven splitters} \label{table:samplesort:registerusage}
		\end{table}
		\begin{algorithm}[!h]
			\caption{Register Sample Sort Classification(\texttt{array}, \texttt{elementCount}, \texttt{predicate})} \label{algo:samplesort:cversion}
			
			int splitter0, splitter1, splitter2 \la determineSplitters() \\
			int state, predicateResult, splitterx \\
			int* b${}_0$, b${}_1$, b${}_2$, b${}_3$ \la allocateBuckets(elementCount) \\
			\For{$1 \leq i \leq \mathtt{elementCount}$}{
				state \la 0\\
				predicateResult \la (int) predicate(splitter1 < array[i]) \\
				splitterx \la splitter0 \\
				\If{predicateResult > 0}{
					splitterx \la splitter2 \label{algo:samplesort:cmov} \\
					state \la (state \texttt{<<} 1) + 1 \label{algo:samplesort:rcl}
				}
				predicateResult \la (int) predicate(splitterx < array[i]) \\
				state \la (state \texttt{<<} 1) + predicateResult \label{algo:samplesort:laststate}\\
				place array[i] in buckets $b_{state}$
			}
		\end{algorithm}

\clearpage
\section{Experimental Results} \label{section:results}
	In the tests we ran, different sorting algorithms and conditional-swap implementations were compared. For the details about the different sorters and swaps refer to section \ref{section:implementation-networks}. \\
	The names of the sorters are built in an abbrevatory way that matches the following format:
	{ \setlength\parskip{\medskipamount}
	\begin{enumerate}
		\item It starts with an \verb|I| or an \verb|N|, indicating if the used algorithm is insertion sort or a sorting network. \label{enumeration:experimentnaming:algtype}
		\begin{itemize}
			\item In case of sorting networks, if it is a \verb|Best| network or a Bose Nelson network (\verb|BoNe|).
			\begin{itemize}
				\item For a Bose Nelson network whether it was optimized for Locality (\verb|L|), Parallelism (\verb|P|) or generated to take the items as single parameters \verb|M| (see section \ref{section:implementation-networks})
			\end{itemize}
		\end{itemize}
		\item Then follows the type of benchmark, \verb|-N| for sorting one set of items (\enquote{normal sort}, section \ref{section:experiments:normal}), \verb|-I| for sorting many continuous sets of items (\enquote{inrow sort}, section \ref{section:experiments:inrow}), \verb|-S| for sorting with Sample Sort (section \ref{section:experiments:samplesort}), \verb|-Q| for sorting with quicksort (section \ref{section:experiments:quicksort}) and \verb|-4| for sorting with \ipso~ (section \ref{section:experiments:ipso}).
		\begin{itemize}
			\item In case of Sample Sort, the parameters \texttt{numberOfSplitters}, \texttt{oversamplingFactor} and \texttt{blockSize} are appended as numbers
		\end{itemize}
		\item Lastly, the name of the \verb|struct| used for the template specialization is appended (see section \ref{section:preliminaries:compare-and-swap} for the abbreviations for conditional swaps) as well as a single \verb|K| for elements that have only a key and \verb|KR| for those that have a key and a reference value.
	\end{enumerate}
	} \medskip
	Where for comparison \verb|std::sort| was run, the name in step \ref{enumeration:experimentnaming:algtype} is \verb|StdSort|. 
	
	For example, when measuring sample sort with parameters 332 and a Bose Nelson network optimizing parallelism as the base case with conditional swap \verb|4CS|, the sorter name would be \verb|N BoNeP -S332 KR 4CS| .

	\subsection{Environment}
		\begin{table}[!h]
			\begin{tabular}{r | r | r | r}
				Machine Name & \multicolumn{1}{c|}{A} & \multicolumn{1}{c|}{B} & \multicolumn{1}{c}{C} \\ \hline 
				\multirow{2}{*}{CPU} & 2 x Intel Xeon 8-core & 2 x Intel Xeon 12-core & AMD Ryzen 8-core \\
				& E5-2650 v2 2.6 GHz & E5-2670 v3 2.3 GHz & 1800X 3.6 GHz \\
				RAM & 128 GiB DDR3 & 128 GiB DDR4 & 32GB DDR4 \\
				L1 Cache (per Core) & 32 KiB I + 32 KiB D & 32 KiB I + 32 KiB D & 64 KiB I + 32 KiB D \\
				L2 Cache (per Core) & 256 KiB & 256 KiB & 512 KiB \\
				L3 Cache (total) & 20 MiB & 30 MiB & 16 MiB [8 MiB]
			\end{tabular}
			\caption{Hardware properties of the machines used} \label{table:machines}
		\end{table}
		As compiler the gcc C++ compiler in version 7.3.0 was used with the \verb|-O3| flag. \\
		The measurements were done with only essential processes running on the machine apart from the measurement. To prevent the process from being swapped to another core during execution it was run with \verb|taskset 0x1|. \\
		In total, three different machines were used to do the measurements. Their hardware properties can be seen in table \ref{table:machines}. \enquote{I} and \enquote{D} refer to dedicated Instruction and Data caches. Also note that while the AMD Ryzen's L3 cache has a total size of 16 MiB, it is divided into two 8 MiB caches that are exclusive to 4 cores each. Since all measurements were done on a single core, the L3 cache size in brackets is the one available to the program. The operating system on all machine was Ubuntu 18.04.
	\newpage
	\subsection{Generating Plots}
		Due to the high number of dimensions in the measurements (machine the measurement is run on, type of network, conditional swap implementation, array size) the results could not always be plotted two-dimensionally. We used box-plots where applicable to show more than just an average value for a measurement. The box incloses all values between the first quartile (\texttt{1Q}) and third quartile (\texttt{3Q}). The line in the middle shows the median. Further the inter-quartile-range (\texttt{IQR}) is calculated as the distance between first and third quartile. The lines (called whiskers) left and right of the boxes go until the smallest value greater than $\mathtt{1Q} - 1.5 \cdot \mathtt{IQR}$ and the greatest value smaller than $\mathtt{3Q} + 1.5 \cdot \mathtt{IQR}$ respectively. Values below these ranges are called outliers and shown as individual dots.
		
	\subsection{Conducting the Measurements} \label{section:measurements}
		\paragraph*{Random Numbers} In order to measure the time needed to sort some data, one has to have data first. For these measurements, the data consisted of pairs of a 64-bit unsigned integer key and a 64-bit unsigned integer reference value. Those were generated as uniformly distributed random numbers by a lightweight implementation of the std::minstd\_rand generator from the C++ <random> library that works as follows: \\
		First a \texttt{seed} is set, taken e.g. from the current time. When a new random number is requested, the generator calculates $\mathtt{seed} = \mathtt{seed}~ \cdot~ 48271~ \%~ 2147483647$ and returns the current \texttt{seed}. \\
		The numbers generated like that do not use all 64 bits available, which is only for practicality with the permutation check as will be seen below. \\
		For each measurement $i$, a new $\mathtt{seed}_i$ is taken from the current time. The same $\mathtt{seed}_i$ is then set before the execution of each sorter, to provide all sorters with the same random inputs.
		\paragraph*{Measuring} The actual measuring was done via linux's PERF\_EVENT interface that allows to do fine-grained measurements. Here, the number of cpu cycles spent on sorting was the unit of measurement. That also means that the results do not depend on clock speeds (e.g. when overclocking), but only on the CPU's architecture.
		\paragraph*{Compilation} When we started this project, it was only a single source file (.cpp) with an increasing amount of headers that were all included in that single file. That is also due to the fact that templated methods cannot be placed in source files because they need to be visible to all including files at compile time. The increasing amount of code and the many different templates brought the compiler to a point where it took over a minute to compile the project. The problem we encountered was that the compiler only gives itself a limited amount of time for compiling a single source file. In order to stay within the time boundaries for a single file, the optimization became poor. We saw measurements being slower for no apparent reason. To solve that problem, we used code generation to create source files that contain an acceptable amount of methods that initiate part of a measurement in a wrapper method. This way, from the main source file we only need to call the correct wrapper methods to perform the measurements, and this way we achieved results that were more stable and reproducible. \\
		For compilation, the flag \verb|-O3| was used to achieve high optimization and speed. That also means that, without using the sorted data in some way, the compiler would deem the result unimportant and skip the sorting altogether. That is why after each sort, to generate a side-effect, the set is checked for two properties: That it is sorted, and that it is a permutation of the previously generated set. The first can easily be done by checking for each value that it is not greater than the value before it.
		\paragraph*{Permutation Check} The permutation check is done probabilistically: At design time, a (preferably large) prime number $p$ is chosen. \\
		Before sorting, $v = \prod_{i = 1}^{n} (z - a_i) \mod p$ is calculated for a number $z$ and values $a = \{a_1, \ldots, a_n\}$. \\
		To check the permutation after sorting and obtaining $a' = \{a'_1, \ldots, a'_n\}$, $w = \prod_{i = 1}^{n} (z - a'_i) \mod p$ is calculated. If $v \neq w$, $a'$ cannot be a permutation of $a$. If $v = w$, we claim that $a'$ is a permutation of $a$. \\
		To minimize the chances of $a'$ not being a permutation of $a$, but $v$ being equal to $w$, $v = 0$ was disallowed in the first step. If $v$ is zero, $z$ is incremented by one and the product calculated again, until $v \neq 0$.
		\paragraph*{Benchmarks} The benchmark seen in algorithm \ref{algo:normal} was used for most of the measurements. \\
		To reduce the chance of cache misses at the beginning of the measurement, one warmup run of random generation, sorting and sorted checking is done beforehand (lines \ref{algo:normal:warmup:start} to \ref{algo:normal:warmup:end}). The array is then sorted \texttt{numberOfIterations} times and checked for the sorted and permutation properties. After that only the generation of the random numbers and the sorted and permutation checking is measured, to later subtract the time from the previously measured one, resulting in the time needed for the sorting alone. Since this is not deterministic in time, and both measurements are subjects to their own deviation, it can occasionally happen that the second measurement takes longer than the first, even though less work has been done. We get those negative times more often for the sorters with small array sizes, where the sorting itself takes relatively little time compared to the random generation and sorted checking. The negative times show up as outliers in the results. \\
		The function \texttt{simulateCheckSorted} checks the permutation like \texttt{checkSorted}, but since randomly generated arrays are rarely ordered, instead of checking for each element if it is smaller than its predecessor, it checks for equality. That should never happen with the random number generator used, and thus run for the same amount of cycles. \\
		The function MeasureSorting is called a total of \texttt{numberOfMeasures} times for each \texttt{arraySize} that is sorted. \\
		For the measurements shown in section \ref{section:experiments:inrow} the benchmark was slightly modified as can be seen in algorithm \ref{algo:inrow}. Here the goal was to look at cache- and memory-effects by creating an array that does not fit into the CPU's L3-cache, and then filling the cache with something else, in this case the reference array. We then split the original array into many blocks of size \texttt{arraySize} and sort each independently. Because we have to create the whole array at the beginning, we can generate the numbers before and check for correct sorting after measuring, so there is no need to do a second measurement like in the first benchmark (lines \ref{algo:normal:random:start} to \ref{algo:normal:random:end} in algorithm \ref{algo:normal}). \\
		Here, instead of giving a \texttt{numberOfIterations} parameter to indicate how often the sorting is to be executed, we provide a \texttt{numberOfArrays} value that says how many arrays of size \texttt{arraySize} are to be created contiguously. This parameter is chosen for each \texttt{arraySize} in a way that $\mathtt{numberOfArrays} ~\times~ \mathtt{arraySize}$ does not fit into the L3 cache of the machine the measurement is performed on.
		\begin{algorithm}[!p]
			\caption{MeasureSorting(\texttt{arraySize}, \texttt{numberOfIterations}, \texttt{seed})} \label{algo:normal}
			
			\ForEach{sorter}{
				setSeed(seed) \\
				arr \la makeArray(arraySize) \\
				numberOfBadSorts \la 0 \\
				arr \la generateRandomArray() \label{algo:normal:warmup:start} \\
				sorter(arr) \\
				checkSorted(arr) // create side-effect \label{algo:normal:warmup:end} 
				
				\medskip
				startMeasuring() \\
				\For{i \la 0 \KwTo numberOfIterations}{
					arr \la generateRandomArray() \\
					sorter(arr) \\
					checkSorted(arr) // create side-effect
				}
				stopMeasuring() \\
				outputResult() \\
				
				\medskip
				setSeed(seed) \label{algo:normal:random:start} \\
				startMeasuring() \\
				\For{i \la 0 \KwTo numberOfIterations}{
					arr \la generateRandomArray() \\
					simulateCheckSorted(arr) // create side-effect
				}
				stopMeasuring() \\
				outputResult() \label{algo:normal:random:end} \\
			}
		\end{algorithm}
		\begin{algorithm}[!p]
			\caption{MeasureSortingInRow(\texttt{arraySize}, \texttt{numberOfArrays}, \texttt{seed})} \label{algo:inrow}
			
			\ForEach{sorter}{
				SetSeed(seed) \\
				arr \la makeArray(arraySize $\times$ numberOfArrays) \\
				arr \la GenerateRandomArray() \\
				compareArr \la makeArray(arraySize $\times$ numberOfArrays) \\
				compareArr \la CopyArray(arr) \\
				\ForEach{currentArr \KwForIn compareArr \KwOfSize arraySize}{
					sort(currentArray, arraySize) //sort reference array
				}
				//warmup on single array of size \textit{arraySize} like in algorithm \ref{algo:normal}, lines \ref{algo:normal:warmup:start} to \ref{algo:normal:warmup:end}
				
				\medskip
				StartMeasuring() \\
				\ForEach{currentArr \KwForIn arr \KwOfSize arraySize}{
					sorter(currentArray, arraySize)	
				}
				StopMeasuring() \\
				
				\medskip
				CheckArraysForEquality(arr, compareArr) //check correct sorting, create side-effect
				OutputResult() \\
			}
		\end{algorithm}
	\clearpage
	\subsection{Sorting One Set of 2-16 items} \label{section:experiments:normal}
\begin{figure}[!tbp]
			\includegraphics[width=1.0\textwidth]{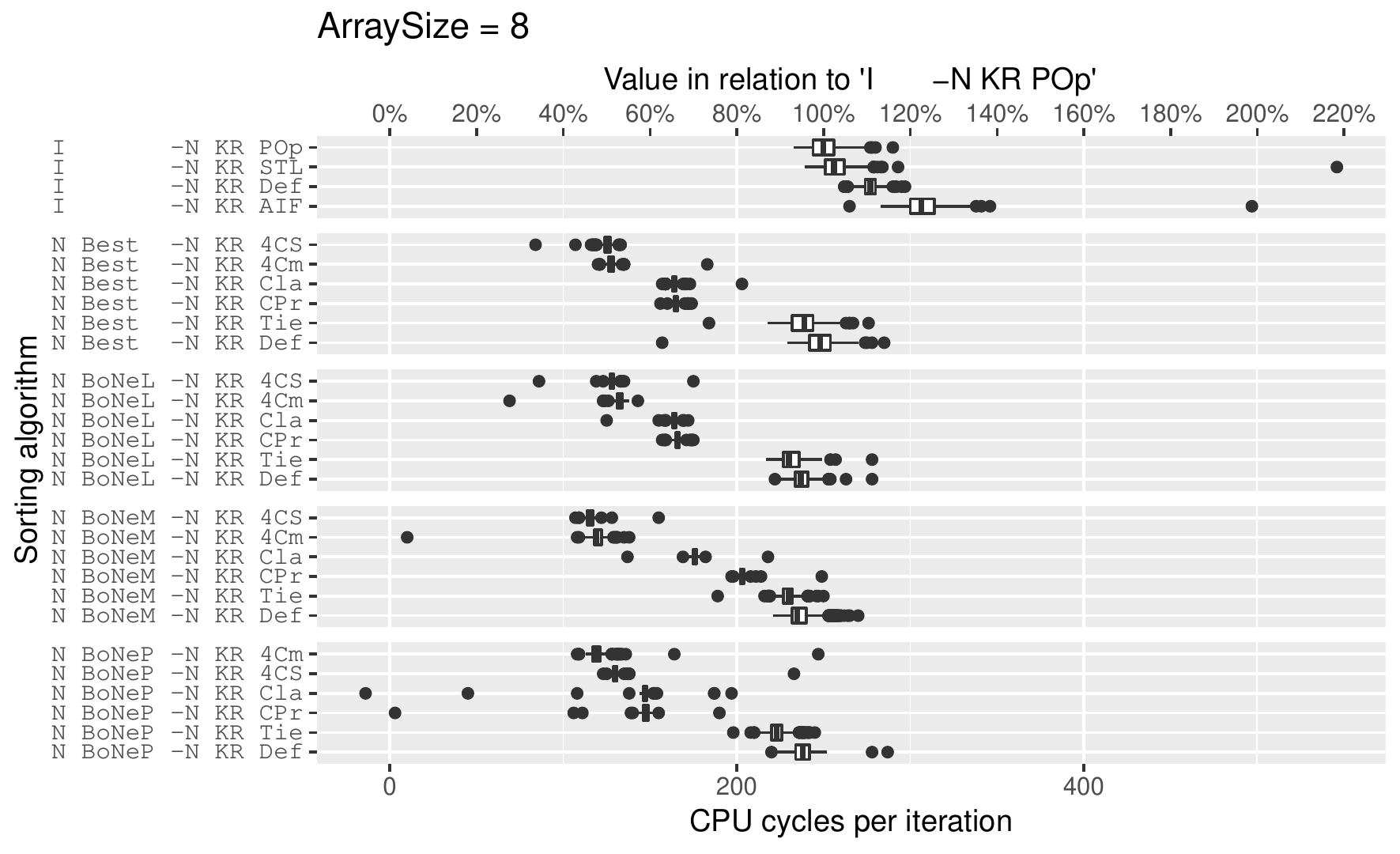}
			\caption{Single sort for array size = 8 on machine A} \label{plot:normal:8:A}
		\end{figure}

\begin{figure}[!tbp]
			\includegraphics[width=1.0\textwidth]{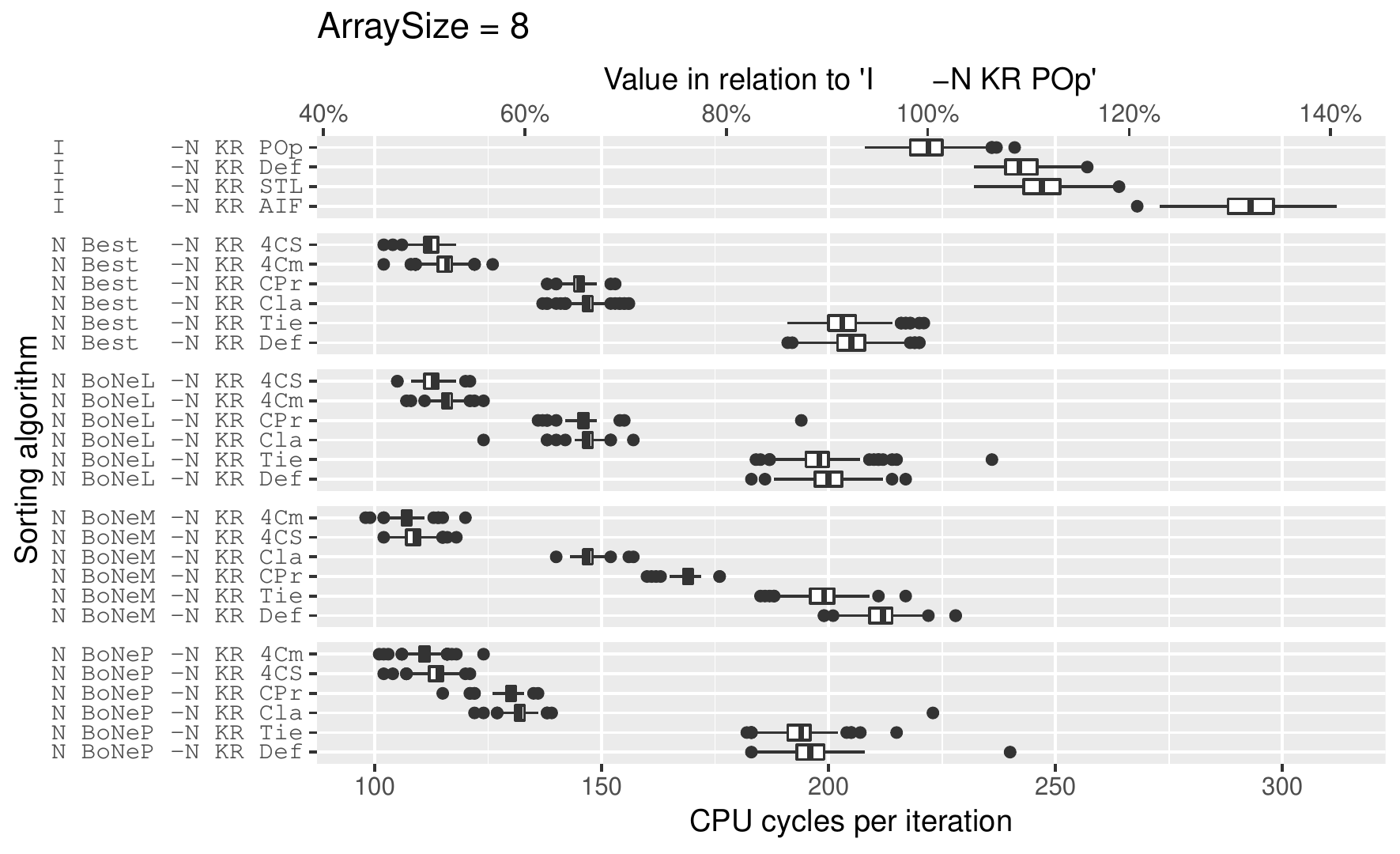}
			\caption{Single sort for array size = 8 on machine B} \label{plot:normal:8:B}
		\end{figure}

\begin{figure}[!tbp]
			\includegraphics[width=1.0\textwidth]{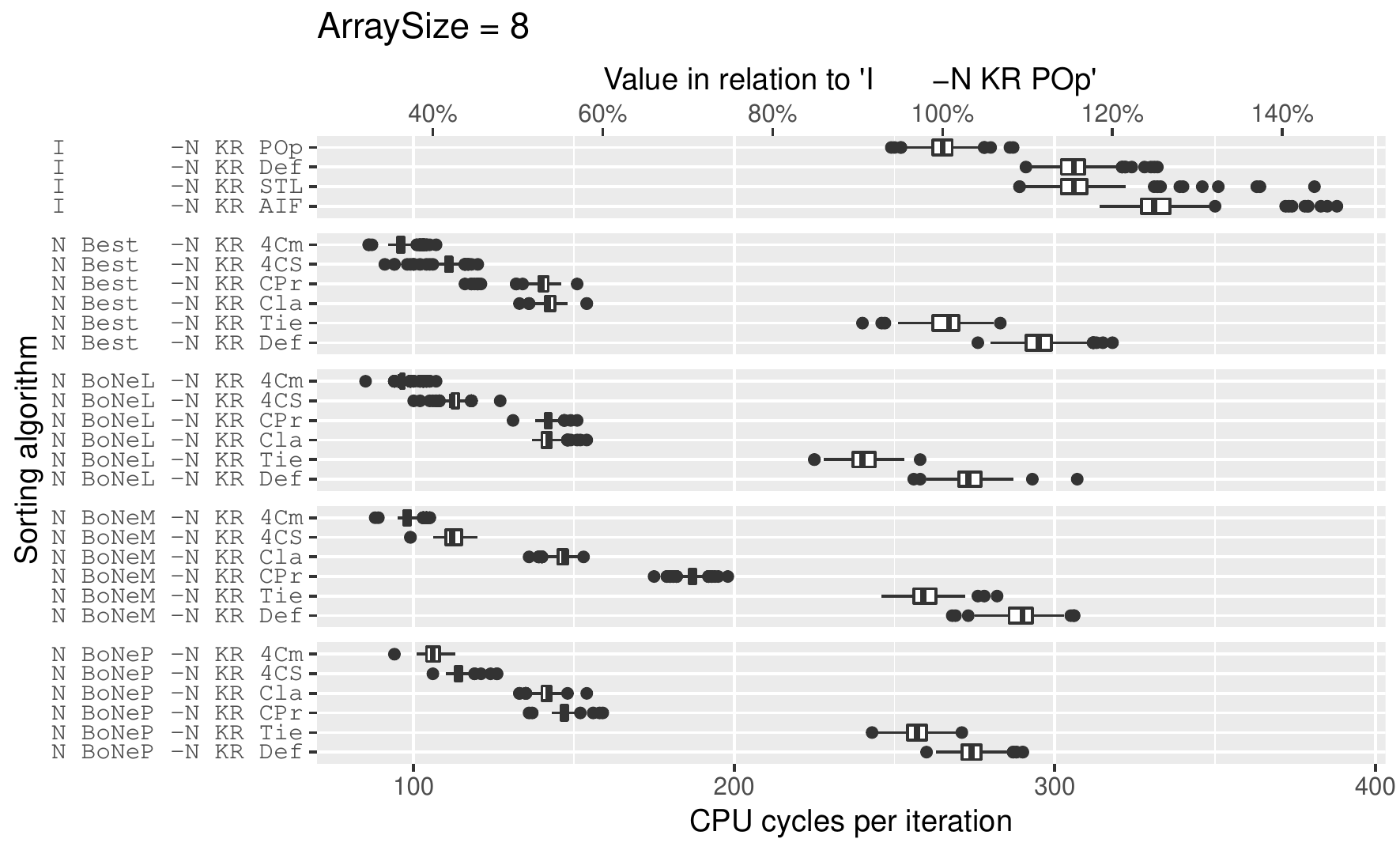}
			\caption{Single sort for array size = 8 on machine C} \label{plot:normal:8:C}
		\end{figure}

\begin{figure}[!tbp]
			\includegraphics[width=1.0\textwidth]{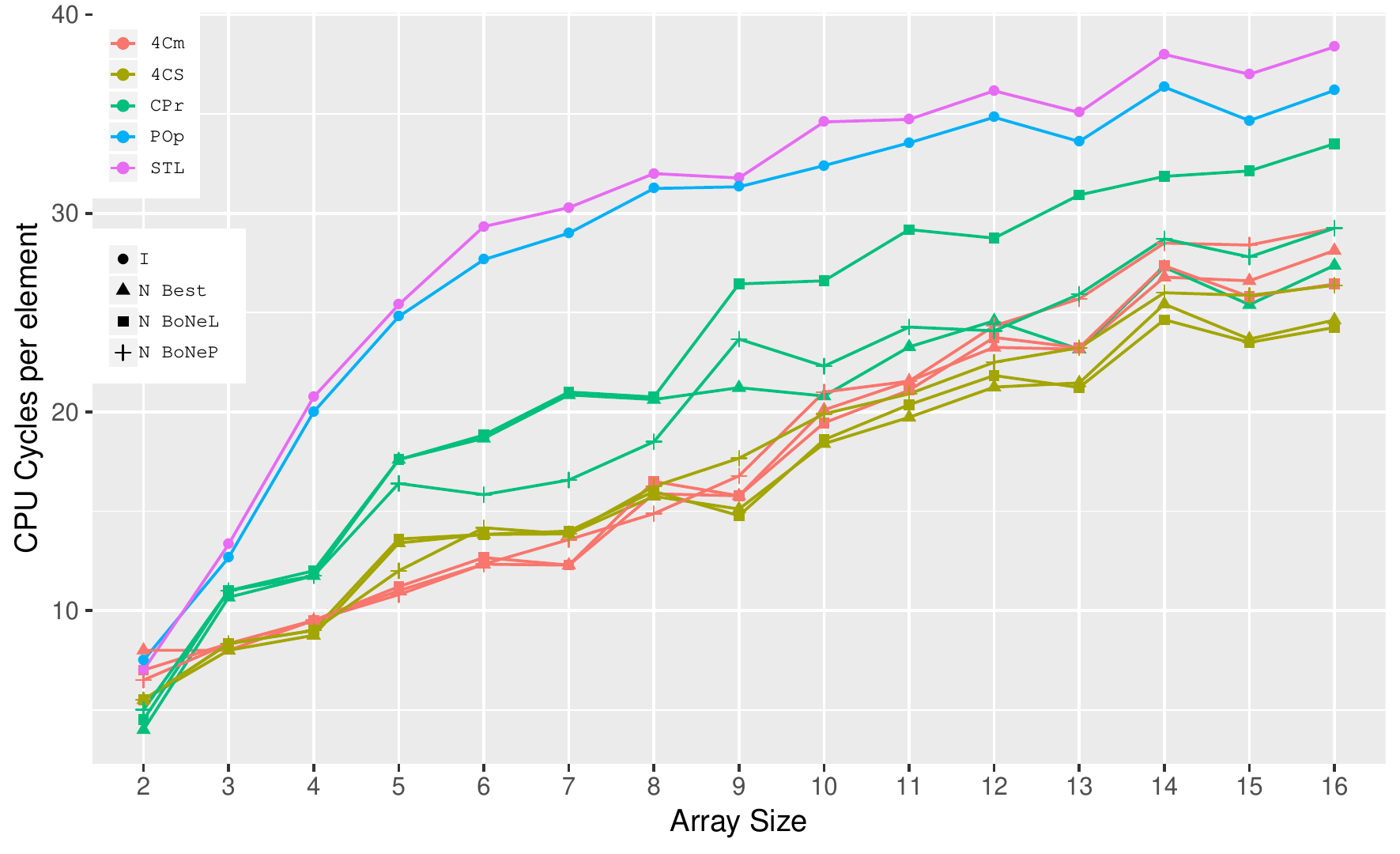}
			\caption{Single sort of array sizes 2 to 16 on machine A} \label{plot:normal:lineplot:A}
		\end{figure}

\begin{figure}[!tbp]
			\includegraphics[width=1.0\textwidth]{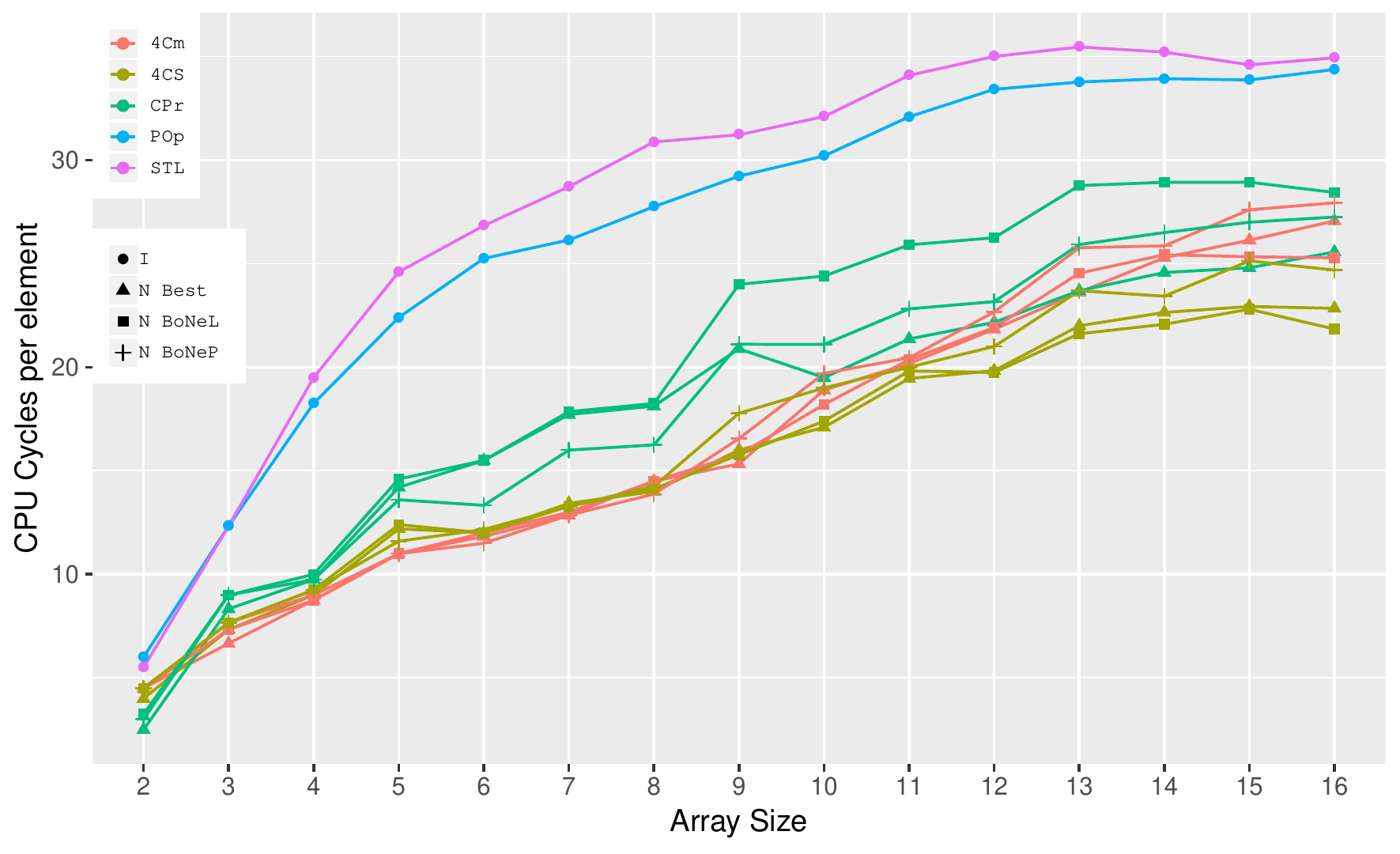}
			\caption{Single sort of array sizes 2 to 16 on machine B} \label{plot:normal:lineplot:B}
		\end{figure}

\begin{figure}[!tbp]
			\includegraphics[width=1.0\textwidth]{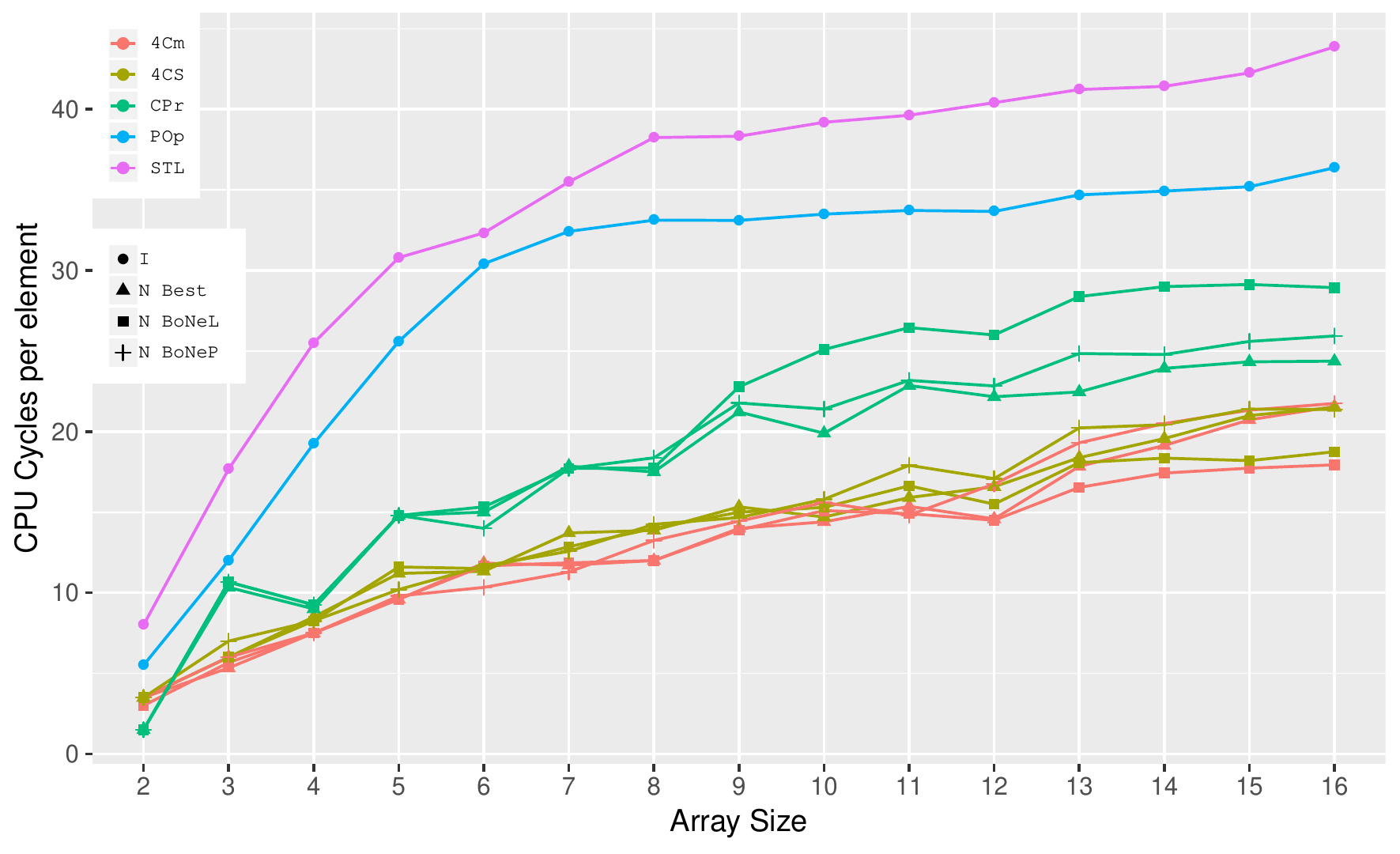}
			\caption{Single sort of array sizes 2 to 16 on machine C} \label{plot:normal:lineplot:C}
		\end{figure}

		The benchmark from algorithm \ref{algo:normal} was used with parameters
		\begin{itemize}
			\item $\mathtt{numberOfIterations} = 100$
			\item $\mathtt{numberOfMeasures} = 500$
			\item $\mathtt{arraySize} \in \{2, \ldots, 16\}$.
		\end{itemize}
		The results seen in tables \ref{table:normalsort:avg:A}, \ref{table:normalsort:avg:B} and \ref{table:normalsort:avg:C} contain the name of the sorter and the average number of cycles per iteration, over the total of all measurements, for machines A, B and C. The algorithm that performed best in a column is marked in bold font, and for each column the value relative to the best in that column was calculated. For each row the geometric mean is calculated over the relative values and from that the rank is determined. \\
		Table \ref{table:normalsort:avg:all} contains the geometric mean and rank taking the results from all three machines into consideration. \\
		Here it becomes visible that the implementations that have conditional branches and those that do not are clearly separated by rank, the former occupy the lower share of the ranks, while the latter get all the higher ranks. We see that the claim from section \ref{section:implementation-conditionalswap} for the \verb|4CS| conditional swap is true for machines A and B, but not for machine C. We also see in table \ref{table:normalsort:avg:all} that the first three ranks have the same geometric mean, so the Bose Nelson networks can compete with the optimized networks that have fewer comparators due to their locality. \\
		The boxplots for array size 8 are given for each machine in figures \ref{plot:normal:8:A}, \ref{plot:normal:8:B} and \ref{plot:normal:8:C}, showing that these higher-ranked implementations are not only faster on average, but that their distribution is almost entirely faster than any of the insertion sort implementations, together with a lower variance. To improve readability, the variants \verb|JXc|, \verb|6Cm| and \verb|QMa| are omitted. Also one outlier was removed from dataset of machine B for the \texttt{'N BoNeL -N KR Cla'} sorter with value $-42.6$ so that the plot has a scale similar to those of the other two machines, to improve comparability. The result set for machine A contains a lot of outliers that we did not want to exclude. To be able to compare it easily with the other two plots we added an additional axis at the top that shows the CPU cycles per iteration as percentages where the average of the best insertion sort is 100\%.  \\
		To see a trend in increasing array size, we chose a few Conditional Swap implementations that do best for more than one network and array size on all machines. Their average sorting times can be seen in figures \ref{plot:normal:lineplot:A}, \ref{plot:normal:lineplot:B} and \ref{plot:normal:lineplot:C}. For visibility reasons, we omitted the Bose Nelson Parameter networks in these plot. What we already saw from the tables is here visible as well, the \verb|4Cm| and \verb|4CS| implementations have good performance and are almost always faster on average than insertion sort (apart from \verb|arraySize = 2| on machine A). \\
		These results indicate that there is potential in using sorting networks, showing an improvement of 32\% of the best network over the best insertion sort, on average, for any array size. Problems with this way of measurement are that the same space in memory is sorted over and over again, which is rarely a use case when sorting a base case. Because of this, the measurements probably reflect unrealistic conditions regarding cache accesses and cache misses. To get a bit closer to actual base case sorting, the next section has a different approach to not sort the same space in memory twice.
		\begin{sidewaystable}[!p]
			\begin{scriptsize}
\begin{tabular}{l | r @{~~} r | r@{~~}r@{~~}r@{~~}r@{~~}r@{~~}r@{~~}r@{~~}r@{~~}r@{~~}r@{~~}r@{~~}r@{~~}r@{~~}r@{~~}r@{~~}r|}
 & \multicolumn{2}{c|}{Overall} & \multicolumn{15}{c}{Array Size} \\
 & Rank & GeoM & 2&3&4&5&6&7&8&9&10&11&12&13&14&15&16\\ \hline
\verb+I       -N KR POp+ & 22 & 1.85 & 15.21&37.82&80.52&124.17&166.14&204.37&250.08&282.97&323.87&369.31&417.57&437.93&509.37&520.20&579.44\\
\verb+I       -N KR STL+ & 26 & 1.92 & 13.82&39.99&83.75&128.44&178.50&213.03&257.62&287.15&346.35&382.08&434.47&455.29&532.36&554.80&614.56\\
\verb+I       -N KR Def+ & 33 & 2.07 & 17.23&40.53&84.80&132.12&178.09&220.69&277.29&311.27&378.66&412.51&475.48&499.27&595.68&614.74&693.90\\
\verb+I       -N KR AIF+ & 36 & 2.21 & 16.55&50.76&90.56&148.37&202.86&252.22&307.29&342.66&400.98&442.48&485.63&518.62&593.07&609.29&672.98\smallskip \\
\verb+N Best  -N KR 4CS+ & 1 & 1.07 & 11.59&\textbf{24.11}&\textbf{34.35}&66.58&82.54&96.95&125.56&134.92&\textbf{183.73}&\textbf{218.14}&\textbf{254.95}&278.53&356.75&353.24&395.58\\
\verb+N Best  -N KR 4Cm+ & 5 & 1.12 & 16.33&24.21&38.05&54.80&74.74&\textbf{85.23}&127.40&141.99&201.85&238.39&279.15&301.47&375.39&399.06&450.54\\
\verb+N Best  -N KR Cla+ & 8 & 1.23 & 8.91&31.22&40.41&86.73&110.04&144.05&163.77&188.67&220.54&240.00&298.76&285.42&347.36&\textbf{349.35}&400.98\\
\verb+N Best  -N KR CPr+ & 10 & 1.27 & \textbf{8.13}&32.20&46.88&87.99&112.46&146.10&164.91&190.61&208.00&256.03&296.58&301.14&382.17&381.71&438.18\\
\verb+N Best  -N KR 6Cm+ & 13 & 1.37 & 17.20&25.57&46.86&64.68&96.36&107.56&146.34&176.34&259.56&289.24&339.02&392.64&490.05&502.11&585.95\\
\verb+N Best  -N KR Def+ & 21 & 1.84 & 20.09&37.08&73.94&113.14&144.92&179.09&248.33&268.15&302.44&338.76&417.69&429.27&555.59&552.84&712.22\\
\verb+N Best  -N KR Tie+ & 25 & 1.90 & 20.47&38.06&63.58&98.92&139.21&182.82&238.59&271.77&316.34&369.96&477.94&519.14&597.69&639.07&753.42\\
\verb+N Best  -N KR JXc+ & 32 & 2.04 & 18.44&36.50&68.67&113.06&167.99&207.12&264.82&293.56&347.16&409.67&506.11&522.21&680.63&711.20&791.41\\
\verb+N Best  -N KR QMa+ & 37 & 2.60 & 17.72&44.69&96.02&149.03&207.73&252.95&341.19&397.81&438.98&573.64&681.13&700.09&832.27&910.77&1057.45\smallskip \\
\verb+N BoNeL -N KR 4CS+ & 2 & 1.08 & 11.45&24.99&35.82&67.94&82.41&98.05&128.16&\textbf{132.68}&186.46&224.66&262.69&\textbf{275.88}&\textbf{344.76}&352.46&\textbf{387.59}\\
\verb+N BoNeL -N KR 4Cm+ & 3 & 1.11 & 13.53&25.18&38.33&55.62&76.06&86.06&132.02&142.06&193.99&232.70&284.86&302.12&383.21&386.96&423.71\\
\verb+N BoNeL -N KR 6Cm+ & 15 & 1.42 & 15.96&28.18&45.90&73.18&90.18&115.24&148.93&214.27&278.32&298.28&365.05&414.27&493.26&508.85&560.16\\
\verb+N BoNeL -N KR Cla+ & 16 & 1.42 & 8.72&31.04&40.71&82.99&112.46&143.75&163.76&239.80&270.36&325.30&354.56&403.83&452.67&493.90&550.58\\
\verb+N BoNeL -N KR CPr+ & 17 & 1.44 & 9.03&33.01&47.21&88.37&113.62&147.21&166.12&238.67&265.87&321.12&347.10&401.81&446.13&482.28&536.51\\
\verb+N BoNeL -N KR Tie+ & 27 & 1.93 & 20.87&40.57&64.35&99.65&137.68&173.18&231.53&265.85&343.47&383.88&472.81&513.56&636.85&676.79&782.22\\
\verb+N BoNeL -N KR Def+ & 28 & 1.94 & 20.11&40.52&78.60&102.58&139.42&170.73&237.18&265.27&366.93&372.58&478.09&481.87&637.88&636.35&764.29\\
\verb+N BoNeL -N KR JXc+ & 31 & 2.04 & 18.95&36.18&68.86&108.92&160.18&196.61&256.54&314.83&368.23&427.12&504.82&570.41&642.64&662.73&789.99\\
\verb+N BoNeL -N KR QMa+ & 38 & 2.67 & 18.16&45.95&93.38&143.03&196.39&241.82&326.07&406.05&514.55&578.58&685.08&776.92&912.15&998.34&1163.99\smallskip \\
\verb+N BoNeM -N KR 4Cm+ & 7 & 1.22 & 16.08&27.11&38.84&54.86&82.51&94.36&119.90&214.28&252.78&251.85&284.48&318.74&415.51&394.31&500.68\\
\verb+N BoNeM -N KR 4CS+ & 11 & 1.28 & 11.79&24.70&43.96&73.23&83.76&130.14&\textbf{115.48}&204.19&273.27&286.72&281.43&314.56&487.85&464.53&518.94\\
\verb+N BoNeM -N KR 6Cm+ & 18 & 1.51 & 15.65&27.98&52.71&93.07&85.76&113.21&134.35&223.11&340.69&340.96&393.85&434.99&575.98&499.30&630.78\\
\verb+N BoNeM -N KR Cla+ & 19 & 1.63 & 13.05&32.46&54.07&90.32&116.27&149.86&175.80&278.17&314.13&348.62&395.85&448.91&559.24&566.87&639.09\\
\verb+N BoNeM -N KR CPr+ & 20 & 1.67 & 15.10&33.91&46.42&103.89&120.13&153.62&203.55&287.35&322.22&347.58&374.67&423.75&521.46&565.28&726.59\\
\verb+N BoNeM -N KR Def+ & 29 & 1.94 & 18.38&39.34&75.91&113.68&157.87&199.64&237.50&259.60&352.19&369.31&455.58&479.48&596.34&633.54&748.67\\
\verb+N BoNeM -N KR Tie+ & 30 & 2.01 & 21.34&38.64&62.66&96.05&135.19&172.29&229.37&265.31&368.96&452.05&554.50&544.95&769.84&767.68&788.78\\
\verb+N BoNeM -N KR JXc+ & 35 & 2.18 & 19.47&38.29&70.01&108.73&147.29&184.76&252.61&358.55&454.97&478.92&594.31&589.14&769.05&751.62&924.53\\
\verb+N BoNeM -N KR QMa+ & 39 & 2.71 & 24.28&54.14&100.79&136.42&204.57&251.56&325.24&403.05&480.82&547.65&651.81&739.13&864.35&966.34&1092.30\smallskip \\
\verb+N BoNeP -N KR 4CS+ & 4 & 1.11 & 11.41&24.83&35.67&61.12&84.51&96.59&130.30&159.81&199.35&230.24&271.39&302.50&363.86&388.34&422.56\\
\verb+N BoNeP -N KR 4Cm+ & 6 & 1.14 & 13.00&25.08&38.14&\textbf{53.95}&\textbf{74.04}&94.47&119.49&151.09&209.73&237.70&291.55&334.29&398.82&426.06&468.23\\
\verb+N BoNeP -N KR Cla+ & 9 & 1.25 & 8.81&31.30&41.47&80.30&100.54&130.65&147.00&211.21&233.02&265.58&286.03&320.45&363.05&385.07&438.35\\
\verb+N BoNeP -N KR CPr+ & 12 & 1.28 & 9.68&33.04&47.46&82.36&94.76&116.33&147.20&212.97&223.40&267.75&289.20&337.12&401.57&417.11&468.35\\
\verb+N BoNeP -N KR 6Cm+ & 14 & 1.41 & 15.17&27.86&45.69&66.43&86.38&102.07&151.31&207.81&275.71&317.69&375.96&418.47&505.54&545.92&617.54\\
\verb+N BoNeP -N KR Tie+ & 23 & 1.88 & 20.56&37.00&64.49&97.05&131.03&171.37&223.33&270.21&324.68&395.20&462.10&534.33&590.53&642.31&744.88\\
\verb+N BoNeP -N KR Def+ & 24 & 1.88 & 19.80&41.86&74.79&111.41&144.15&174.45&237.91&265.89&325.01&350.58&420.83&473.65&563.51&599.15&727.96\\
\verb+N BoNeP -N KR JXc+ & 34 & 2.08 & 20.22&36.90&69.50&109.37&150.28&188.40&251.74&293.71&379.19&439.07&525.51&585.25&719.42&744.81&844.24\\
\verb+N BoNeP -N KR QMa+ & 40 & 2.79 & 24.23&52.09&99.01&148.85&192.75&263.79&338.27&395.25&517.71&584.97&677.76&794.20&938.78&1006.31&1166.80\\
\end{tabular}

			\end{scriptsize}
			\caption{Average number of CPU cycles per iteration of single array sorting on machine A} \label{table:normalsort:avg:A}
		\end{sidewaystable}
		\begin{sidewaystable}[!p]
			\begin{scriptsize}
\begin{tabular}{l | r @{~~} r | r@{~~}r@{~~}r@{~~}r@{~~}r@{~~}r@{~~}r@{~~}r@{~~}r@{~~}r@{~~}r@{~~}r@{~~}r@{~~}r@{~~}r@{~~}r|}
 & \multicolumn{2}{c|}{Overall} & \multicolumn{15}{c}{Array Size} \\
 & Rank & GeoM & 2&3&4&5&6&7&8&9&10&11&12&13&14&15&16\\ \hline
\verb+I       -N KR POp+ & 25 & 1.84 & 12.56&36.78&73.37&111.91&151.52&183.05&221.86&263.07&302.35&353.36&399.83&439.54&475.60&508.11&550.21\\
\verb+I       -N KR STL+ & 29 & 1.93 & 10.97&37.26&78.04&122.75&161.05&201.27&247.30&280.78&321.94&376.15&420.50&461.52&493.07&519.96&559.69\\
\verb+I       -N KR Def+ & 32 & 1.98 & 14.03&40.14&76.93&117.68&157.59&198.50&242.60&280.73&322.72&375.54&422.67&465.00&511.51&558.47&599.95\\
\verb+I       -N KR AIF+ & 36 & 2.32 & 14.98&54.78&92.20&135.59&195.79&245.14&292.73&327.82&380.85&438.38&489.94&538.38&573.31&612.57&656.83\smallskip \\
\verb+N Best  -N KR 4CS+ & 1 & 1.06 & 8.00&22.10&35.70&60.36&71.61&93.40&112.30&144.17&\textbf{171.34}&\textbf{214.03}&238.42&285.88&316.71&344.10&364.40\\
\verb+N Best  -N KR 4Cm+ & 3 & 1.10 & 8.75&\textbf{20.34}&\textbf{34.47}&\textbf{54.70}&70.88&90.27&115.48&\textbf{137.92}&189.67&222.14&262.03&307.09&353.77&391.72&432.70\\
\verb+N Best  -N KR CPr+ & 8 & 1.19 & \textbf{5.23}&25.15&38.71&71.13&92.24&124.12&145.15&188.31&194.53&234.72&266.15&307.63&343.99&372.53&409.52\\
\verb+N Best  -N KR Cla+ & 9 & 1.20 & 6.86&28.30&38.70&75.69&92.84&129.24&146.96&190.31&196.30&233.05&261.56&\textbf{278.43}&\textbf{306.94}&\textbf{335.57}&363.08\\
\verb+N Best  -N KR 6Cm+ & 14 & 1.36 & 9.52&24.23&39.97&69.73&88.59&111.52&136.40&174.35&228.98&283.96&321.57&402.13&458.39&499.02&553.25\\
\verb+N Best  -N KR Def+ & 21 & 1.78 & 16.38&33.75&64.57&95.65&125.05&164.26&204.86&246.37&265.34&336.21&399.28&428.57&508.38&550.37&631.51\\
\verb+N Best  -N KR Tie+ & 22 & 1.82 & 16.04&33.74&57.37&88.86&114.24&165.85&203.09&250.32&280.61&352.47&435.29&508.75&539.74&598.76&685.05\\
\verb+N Best  -N KR JXc+ & 33 & 2.00 & 16.52&33.44&63.44&98.88&138.36&179.19&218.44&282.01&309.11&385.21&480.67&541.52&626.83&665.20&766.23\\
\verb+N Best  -N KR QMa+ & 37 & 2.53 & 14.98&42.43&86.31&140.41&184.73&235.35&301.62&364.06&383.06&539.44&623.15&659.71&744.50&846.43&932.15\smallskip \\
\verb+N BoNeL -N KR 4CS+ & 2 & 1.08 & 8.79&23.49&37.04&62.21&72.40&93.37&112.70&142.31&173.74&218.67&\textbf{237.75}&281.41&309.13&342.61&\textbf{349.19}\\
\verb+N BoNeL -N KR 4Cm+ & 4 & 1.11 & 9.06&21.77&35.45&55.14&72.10&90.98&116.37&142.51&181.85&223.36&263.63&319.19&355.23&380.37&404.78\\
\verb+N BoNeL -N KR CPr+ & 13 & 1.34 & 6.54&27.47&39.79&72.94&93.51&125.24&145.99&215.92&243.90&285.57&315.19&374.44&404.65&434.14&454.70\\
\verb+N BoNeL -N KR Cla+ & 15 & 1.39 & 7.30&29.84&39.35&76.71&92.91&130.63&147.23&226.02&247.07&293.28&324.55&393.59&415.47&460.03&477.18\\
\verb+N BoNeL -N KR 6Cm+ & 16 & 1.43 & 10.65&25.90&40.56&70.90&87.99&113.19&137.92&217.54&260.44&292.33&353.98&427.97&475.74&503.83&529.54\\
\verb+N BoNeL -N KR Tie+ & 26 & 1.87 & 17.59&33.75&57.62&87.79&115.16&157.70&198.00&245.89&307.31&367.01&444.25&506.00&579.13&658.81&722.07\\
\verb+N BoNeL -N KR Def+ & 27 & 1.89 & 16.64&36.37&67.23&90.58&120.61&158.18&200.27&250.83&320.55&355.67&453.90&487.58&575.63&625.45&688.18\\
\verb+N BoNeL -N KR JXc+ & 31 & 1.97 & 16.28&33.15&62.43&91.86&133.58&171.44&214.22&279.84&338.71&398.70&456.86&536.44&615.27&660.74&748.77\\
\verb+N BoNeL -N KR QMa+ & 38 & 2.63 & 14.51&43.23&82.70&134.53&178.88&225.65&290.45&391.15&462.34&547.55&646.34&742.97&835.85&940.49&1057.89\smallskip \\
\verb+N BoNeM -N KR 4Cm+ & 7 & 1.17 & 9.86&21.49&34.89&55.43&72.68&\textbf{89.56}&\textbf{106.83}&212.18&224.25&233.20&260.33&325.45&378.21&405.97&438.89\\
\verb+N BoNeM -N KR 4CS+ & 12 & 1.28 & 8.90&23.66&41.74&72.84&72.88&129.91&108.65&206.26&248.00&262.31&262.53&321.93&431.00&445.25&452.85\\
\verb+N BoNeM -N KR 6Cm+ & 18 & 1.56 & 11.72&25.59&48.66&95.36&84.11&110.15&133.02&234.05&318.24&343.21&382.43&447.82&527.23&528.53&599.74\\
\verb+N BoNeM -N KR CPr+ & 19 & 1.58 & 11.92&29.69&39.66&93.65&94.49&144.63&168.80&258.49&301.06&331.26&341.15&432.34&444.36&531.04&566.47\\
\verb+N BoNeM -N KR Cla+ & 20 & 1.60 & 12.11&33.08&47.86&88.18&94.52&149.64&147.27&249.49&280.22&330.82&350.09&443.00&490.15&546.82&537.15\\
\verb+N BoNeM -N KR Def+ & 28 & 1.92 & 15.74&35.40&67.25&105.51&137.72&179.21&211.66&250.63&320.06&350.38&443.83&481.71&549.25&621.84&689.81\\
\verb+N BoNeM -N KR Tie+ & 30 & 1.96 & 16.95&33.72&56.67&86.72&115.31&154.85&198.75&245.12&339.09&443.65&526.52&553.74&695.92&741.39&723.53\\
\verb+N BoNeM -N KR JXc+ & 35 & 2.14 & 16.46&35.71&60.93&95.72&123.50&167.92&218.13&359.78&433.33&449.88&539.73&577.80&689.96&729.62&862.97\\
\verb+N BoNeM -N KR QMa+ & 39 & 2.66 & 20.64&48.64&90.61&126.40&185.62&229.86&300.87&373.85&431.10&518.01&615.57&735.18&780.40&895.37&951.65\smallskip \\
\verb+N BoNeP -N KR 4CS+ & 5 & 1.12 & 8.46&23.34&37.09&57.80&73.68&92.70&113.69&159.60&190.06&219.94&252.55&307.79&328.41&377.27&392.42\\
\verb+N BoNeP -N KR 4Cm+ & 6 & 1.14 & 8.95&22.69&35.57&55.35&\textbf{69.60}&90.41&111.13&149.56&197.46&225.24&271.98&335.76&362.40&414.60&447.03\\
\verb+N BoNeP -N KR CPr+ & 10 & 1.22 & 5.97&26.99&39.44&67.44&80.12&111.68&129.78&190.36&211.08&251.35&278.01&336.65&370.57&405.37&436.74\\
\verb+N BoNeP -N KR Cla+ & 11 & 1.23 & 7.10&29.97&39.96&73.81&84.93&116.40&131.89&200.48&213.50&243.83&265.25&317.23&337.10&377.03&402.82\\
\verb+N BoNeP -N KR 6Cm+ & 17 & 1.44 & 10.32&25.62&40.47&69.19&84.75&110.24&134.78&210.52&252.52&313.76&368.79&434.89&481.18&543.99&580.31\\
\verb+N BoNeP -N KR Def+ & 23 & 1.83 & 16.91&38.39&65.18&97.07&120.63&156.82&195.89&242.43&288.94&336.14&403.45&458.41&522.17&589.12&656.89\\
\verb+N BoNeP -N KR Tie+ & 24 & 1.84 & 17.37&32.93&57.10&86.35&114.55&156.12&193.53&248.25&299.72&369.58&429.28&509.05&566.42&628.05&716.46\\
\verb+N BoNeP -N KR JXc+ & 34 & 2.04 & 16.51&32.47&60.17&95.97&137.22&171.44&221.18&283.43&339.45&422.44&492.87&576.73&655.46&723.94&784.93\\
\verb+N BoNeP -N KR QMa+ & 40 & 2.75 & 20.44&48.92&87.15&142.84&174.98&248.01&298.49&385.09&462.66&544.00&626.85&763.33&847.00&940.79&1027.04\\
\end{tabular}

			\end{scriptsize}
			\caption{Average number of CPU cycles per iteration of single array sorting on machine B} \label{table:normalsort:avg:B}
		\end{sidewaystable}
		\begin{sidewaystable}[!p]
			\begin{scriptsize}
\begin{tabular}{l | r @{~~} r | r@{~~}r@{~~}r@{~~}r@{~~}r@{~~}r@{~~}r@{~~}r@{~~}r@{~~}r@{~~}r@{~~}r@{~~}r@{~~}r@{~~}r@{~~}r|}
 & \multicolumn{2}{c|}{Overall} & \multicolumn{15}{c}{Array Size} \\
 & Rank & GeoM & 2&3&4&5&6&7&8&9&10&11&12&13&14&15&16\\ \hline
\verb+I       -N KR POp+ & 21 & 2.45 & 11.39&36.05&77.28&128.96&181.15&227.59&265.00&298.11&335.56&370.76&404.10&450.91&488.88&529.10&582.69\\
\verb+I       -N KR Def+ & 33 & 2.90 & 15.33&48.24&95.20&148.27&195.18&249.58&305.80&347.41&395.45&434.98&479.17&524.59&575.01&628.20&688.63\\
\verb+I       -N KR STL+ & 35 & 2.97 & 16.64&52.56&102.10&154.79&194.80&248.41&306.96&346.42&393.33&436.37&486.46&536.80&580.03&634.94&702.84\\
\verb+I       -N KR AIF+ & 36 & 3.23 & 16.77&56.45&109.56&160.66&215.97&268.93&332.20&379.01&432.62&479.89&540.53&601.36&643.93&708.10&776.50\smallskip \\
\verb+N Best  -N KR 4Cm+ & 2 & 1.11 & 7.31&\textbf{16.27}&29.45&47.03&70.03&82.22&\textbf{96.48}&126.39&\textbf{143.46}&169.87&176.58&233.61&268.71&310.79&345.07\\
\verb+N Best  -N KR 4CS+ & 5 & 1.19 & 7.07&18.10&34.24&55.89&67.86&93.54&110.54&137.94&146.50&175.16&195.19&239.20&274.19&315.00&343.90\\
\verb+N Best  -N KR 6Cm+ & 8 & 1.25 & 6.85&17.84&33.82&62.44&79.27&97.94&116.81&142.67&165.77&184.17&206.94&256.17&295.75&316.26&352.90\\
\verb+N Best  -N KR CPr+ & 11 & 1.43 & \textbf{3.09}&30.86&36.06&73.57&89.71&124.63&140.12&190.47&199.37&251.14&265.94&292.43&335.35&364.68&389.62\\
\verb+N Best  -N KR Cla+ & 12 & 1.45 & 4.73&26.60&37.62&75.06&87.90&116.55&142.33&194.22&196.00&251.06&260.73&294.26&326.31&356.21&384.08\\
\verb+N Best  -N KR Tie+ & 26 & 2.75 & 15.76&38.05&73.01&109.62&144.16&211.29&266.32&314.66&362.63&424.85&500.30&595.58&687.34&733.87&897.51\\
\verb+N Best  -N KR JXc+ & 27 & 2.79 & 15.56&42.27&75.44&114.30&147.68&205.71&254.78&309.67&375.45&433.90&491.17&605.14&703.28&774.28&901.10\\
\verb+N Best  -N KR Def+ & 30 & 2.85 & 15.40&37.48&80.60&126.48&158.41&222.60&294.91&346.24&358.13&456.70&514.14&561.78&694.25&758.85&846.88\\
\verb+N Best  -N KR QMa+ & 38 & 3.79 & 11.05&54.53&110.76&180.78&239.95&310.40&389.88&468.14&494.45&665.40&741.91&783.56&920.78&1000.01&1110.08\smallskip \\
\verb+N BoNeL -N KR 4Cm+ & 1 & 1.07 & 6.05&16.77&29.70&47.32&70.14&83.57&96.87&\textbf{125.01}&149.19&164.25&\textbf{174.84}&\textbf{216.16}&\textbf{244.09}&\textbf{266.22}&\textbf{288.18}\\
\verb+N BoNeL -N KR 4CS+ & 4 & 1.15 & 6.51&18.51&33.64&57.84&69.08&90.70&112.78&134.28&153.70&182.76&185.91&234.77&256.08&272.71&299.99\\
\verb+N BoNeL -N KR 6Cm+ & 9 & 1.31 & 8.37&20.04&35.32&67.72&77.64&99.97&120.89&156.16&177.52&199.67&217.96&260.26&282.87&316.31&348.27\\
\verb+N BoNeL -N KR CPr+ & 17 & 1.58 & 3.27&32.08&37.08&73.85&91.57&124.20&141.78&205.08&251.50&290.99&312.06&369.32&405.85&437.23&462.54\\
\verb+N BoNeL -N KR Cla+ & 18 & 1.59 & 4.48&27.21&37.60&76.08&89.28&117.83&141.78&200.79&240.48&289.48&306.41&371.91&403.27&443.11&457.93\\
\verb+N BoNeL -N KR JXc+ & 22 & 2.56 & 15.24&35.21&66.96&100.40&131.09&189.59&235.21&304.23&364.99&418.31&472.06&536.86&655.17&704.77&777.83\\
\verb+N BoNeL -N KR Tie+ & 24 & 2.69 & 15.15&37.39&64.10&112.33&139.39&191.19&240.18&314.21&375.78&435.35&500.68&610.78&682.65&765.66&882.64\\
\verb+N BoNeL -N KR Def+ & 31 & 2.89 & 14.65&39.25&78.65&115.49&145.62&206.33&273.29&347.40&409.92&460.74&560.31&611.72&725.22&827.99&932.90\\
\verb+N BoNeL -N KR QMa+ & 39 & 3.81 & 10.33&50.08&97.42&168.02&223.17&286.64&359.03&484.01&578.52&663.41&766.77&906.75&1002.35&1112.73&1223.32\smallskip \\
\verb+N BoNeM -N KR 4Cm+ & 7 & 1.23 & 7.29&17.72&\textbf{29.35}&\textbf{45.03}&\textbf{61.88}&82.52&98.43&202.36&212.83&200.65&243.93&238.92&308.91&320.33&364.00\\
\verb+N BoNeM -N KR 6Cm+ & 15 & 1.50 & 7.00&21.07&40.85&84.36&80.01&99.31&117.10&197.20&266.33&235.02&258.16&323.18&415.37&352.34&399.50\\
\verb+N BoNeM -N KR 4CS+ & 16 & 1.51 & 6.96&17.60&40.66&80.80&70.93&145.19&112.51&208.82&246.55&256.60&238.25&282.13&411.16&416.42&433.64\\
\verb+N BoNeM -N KR CPr+ & 19 & 1.95 & 8.86&33.28&38.15&94.93&98.62&147.79&186.70&251.71&298.36&323.52&341.95&417.32&448.73&548.35&592.07\\
\verb+N BoNeM -N KR Cla+ & 20 & 1.96 & 13.01&32.55&47.51&90.72&93.87&154.40&146.24&232.55&276.11&316.92&340.40&421.70&482.01&533.05&533.90\\
\verb+N BoNeM -N KR JXc+ & 28 & 2.82 & 16.39&37.79&66.83&110.69&134.03&194.48&243.91&385.89&444.32&496.26&538.42&586.56&705.07&767.01&877.98\\
\verb+N BoNeM -N KR Tie+ & 29 & 2.84 & 17.52&41.29&65.86&105.65&145.14&198.25&259.15&322.99&398.23&468.78&547.83&627.10&796.34&798.66&842.88\\
\verb+N BoNeM -N KR Def+ & 34 & 2.95 & 11.71&37.08&84.63&136.06&176.03&231.43&289.16&351.31&413.91&456.64&542.57&641.53&720.54&826.37&908.10\\
\verb+N BoNeM -N KR QMa+ & 37 & 3.77 & 17.63&58.32&107.62&142.40&237.64&264.96&384.33&435.37&500.00&574.65&726.09&825.43&913.20&980.12&1129.61\smallskip \\
\verb+N BoNeP -N KR 4Cm+ & 3 & 1.15 & 6.86&17.92&29.84&49.01&61.90&\textbf{79.90}&106.49&129.82&156.20&\textbf{163.84}&201.19&251.63&287.49&319.13&345.12\\
\verb+N BoNeP -N KR 4CS+ & 6 & 1.21 & 6.63&20.99&33.47&50.87&69.91&89.23&114.03&132.59&155.05&197.48&204.60&262.52&285.92&321.02&342.46\\
\verb+N BoNeP -N KR 6Cm+ & 10 & 1.31 & 7.80&20.58&34.62&60.07&83.27&101.94&126.02&143.48&174.07&207.35&228.25&262.75&287.39&344.92&355.76\\
\verb+N BoNeP -N KR CPr+ & 13 & 1.48 & 3.24&31.77&37.47&74.46&84.18&123.89&146.83&196.26&213.12&254.61&274.11&323.22&347.05&384.41&414.97\\
\verb+N BoNeP -N KR Cla+ & 14 & 1.48 & 4.77&27.88&38.17&69.83&75.02&116.28&141.25&191.49&221.49&250.44&274.70&325.92&350.47&394.49&424.23\\
\verb+N BoNeP -N KR JXc+ & 23 & 2.69 & 13.90&35.16&65.39&105.11&139.71&197.48&255.87&294.85&364.97&461.81&505.71&605.46&686.42&814.11&919.46\\
\verb+N BoNeP -N KR Tie+ & 25 & 2.72 & 15.69&34.72&68.55&112.03&142.03&209.70&257.04&312.52&367.93&434.85&509.64&578.63&702.64&752.57&892.05\\
\verb+N BoNeP -N KR Def+ & 32 & 2.90 & 14.81&39.78&78.61&127.92&181.05&231.43&274.24&346.11&379.57&453.44&491.92&610.98&708.39&773.08&859.50\\
\verb+N BoNeP -N KR QMa+ & 40 & 3.99 & 19.09&56.49&99.26&180.09&222.01&298.15&371.18&475.00&549.86&660.00&745.95&881.24&994.07&1082.76&1196.13\\
\end{tabular}

			\end{scriptsize}
			\caption{Average number of CPU cycles per iteration of single array sorting on machine C} \label{table:normalsort:avg:C}
		\end{sidewaystable}
		\begin{sidewaystable}[!p]
			\begin{scriptsize}
\begin{tabular}{l | r @{~~} r | r@{~~}r@{~~}r@{~~}r@{~~}r@{~~}r@{~~}r@{~~}r@{~~}r@{~~}r@{~~}r@{~~}r@{~~}r@{~~}r@{~~}r@{~~}r|}
 & \multicolumn{2}{c|}{Overall} & \multicolumn{15}{c}{Array Size} \\
 & Rank & GeoM & 2&3&4&5&6&7&8&9&10&11&12&13&14&15&16\\ \hline
\verb+I       -N KR POp+ & 21 & 1.97 & 13.31&36.84&77.52&121.73&166.39&204.93&245.35&281.53&320.20&364.03&407.29&443.03&491.19&519.29&570.38\\
\verb+I       -N KR STL+ & 29 & 2.18 & 13.99&44.13&88.72&136.05&178.19&221.54&271.65&306.46&354.48&399.11&448.19&485.94&536.15&570.51&626.51\\
\verb+I       -N KR Def+ & 34 & 2.22 & 15.49&43.53&85.92&133.04&177.05&222.94&275.77&313.38&366.00&407.47&459.20&496.26&561.57&600.20&659.81\\
\verb+I       -N KR AIF+ & 36 & 2.48 & 16.19&53.96&98.14&148.52&205.27&255.57&311.14&350.39&404.94&454.16&505.94&553.38&604.04&645.66&703.20\smallskip \\
\verb+N Best  -N KR 4CS+ & 1 & 1.08 & 9.49&21.14&34.61&61.39&74.12&93.88&116.89&138.90&\textbf{166.86}&\textbf{201.46}&\textbf{229.80}&267.67&318.35&338.59&367.90\\
\verb+N Best  -N KR 4Cm+ & 4 & 1.10 & 11.80&\textbf{20.24}&\textbf{33.80}&\textbf{51.24}&71.67&\textbf{85.75}&113.85&\textbf{135.38}&178.33&208.90&238.10&278.08&333.61&367.57&413.45\\
\verb+N Best  -N KR Cla+ & 8 & 1.26 & 7.07&28.71&38.97&80.04&96.97&129.58&150.98&191.32&204.88&242.45&273.71&286.36&326.97&347.76&382.99\\
\verb+N Best  -N KR CPr+ & 9 & 1.27 & \textbf{5.62}&29.14&40.79&78.30&98.99&133.63&150.52&189.73&200.78&247.67&278.47&300.13&353.26&373.06&412.66\\
\verb+N Best  -N KR 6Cm+ & 12 & 1.31 & 11.86&21.84&40.18&65.61&87.83&105.09&133.64&163.41&217.18&252.83&286.92&353.76&417.93&429.41&497.07\\
\verb+N Best  -N KR Tie+ & 23 & 2.07 & 17.79&36.53&64.86&99.31&132.75&186.83&236.91&278.75&320.47&382.78&471.00&542.17&608.62&658.26&779.34\\
\verb+N Best  -N KR Def+ & 24 & 2.07 & 17.67&36.25&73.23&111.95&143.00&189.50&250.12&287.76&308.47&378.63&443.90&474.71&587.42&626.30&731.54\\
\verb+N Best  -N KR JXc+ & 33 & 2.19 & 17.11&37.58&69.36&108.46&151.93&197.17&245.69&295.11&343.64&409.43&492.56&556.92&670.48&716.70&818.76\\
\verb+N Best  -N KR QMa+ & 37 & 2.86 & 14.49&47.71&97.71&157.50&211.46&266.79&344.29&410.69&439.47&592.67&682.56&714.70&833.13&919.82&1034.29\smallskip \\
\verb+N BoNeL -N KR 4Cm+ & 2 & 1.08 & 9.98&21.36&34.70&51.99&73.39&87.23&116.70&135.64&174.36&205.56&240.54&275.67&330.05&346.83&375.29\\
\verb+N BoNeL -N KR 4CS+ & 3 & 1.08 & 9.11&22.19&35.65&62.92&75.16&94.38&119.00&136.72&173.22&211.53&232.70&\textbf{265.93}&\textbf{304.89}&\textbf{326.46}&\textbf{348.16}\\
\verb+N BoNeL -N KR 6Cm+ & 14 & 1.37 & 12.10&24.61&41.19&70.46&84.83&108.23&135.38&196.49&245.98&259.64&310.46&366.85&429.46&437.78&485.76\\
\verb+N BoNeL -N KR Cla+ & 16 & 1.43 & 6.67&29.23&39.13&79.28&99.19&130.78&150.76&222.19&252.77&303.20&328.71&389.41&422.62&465.09&493.86\\
\verb+N BoNeL -N KR CPr+ & 17 & 1.43 & 6.24&30.74&41.61&79.86&100.76&134.10&151.81&219.75&254.20&299.55&327.56&382.07&421.80&452.26&485.46\\
\verb+N BoNeL -N KR Tie+ & 25 & 2.08 & 18.14&37.39&61.98&100.26&130.55&174.16&223.09&275.50&342.71&395.67&472.43&544.67&633.90&701.33&794.77\\
\verb+N BoNeL -N KR JXc+ & 27 & 2.12 & 17.03&34.89&65.91&100.48&142.81&186.18&235.30&299.63&356.66&414.47&478.47&548.58&637.53&677.42&771.92\\
\verb+N BoNeL -N KR Def+ & 28 & 2.15 & 17.48&38.62&74.44&103.11&135.11&178.95&237.15&288.96&365.87&397.72&498.02&529.04&646.10&699.94&796.40\\
\verb+N BoNeL -N KR QMa+ & 38 & 2.92 & 14.43&46.63&90.93&149.35&199.66&251.64&325.05&427.89&518.88&596.66&699.04&810.17&917.14&1017.55&1148.28\smallskip \\
\verb+N BoNeM -N KR 4Cm+ & 7 & 1.19 & 11.67&22.36&34.35&51.89&72.19&89.01&\textbf{109.41}&208.81&231.50&227.94&263.31&294.41&367.71&374.03&431.20\\
\verb+N BoNeM -N KR 4CS+ & 13 & 1.32 & 9.54&21.89&42.09&76.18&77.20&136.41&112.08&206.68&256.74&270.60&260.43&306.41&441.37&442.21&467.06\\
\verb+N BoNeM -N KR 6Cm+ & 18 & 1.49 & 11.67&24.90&47.78&90.53&83.76&107.69&128.11&219.37&311.29&297.59&345.98&401.83&503.23&457.17&541.58\\
\verb+N BoNeM -N KR Cla+ & 19 & 1.68 & 12.87&32.68&50.31&89.81&103.25&151.29&157.12&253.04&290.36&332.44&361.42&437.62&510.11&548.59&572.66\\
\verb+N BoNeM -N KR CPr+ & 20 & 1.69 & 12.25&32.26&42.04&98.50&105.12&149.09&186.85&265.49&308.33&334.27&355.31&424.85&476.11&548.73&628.61\\
\verb+N BoNeM -N KR Def+ & 30 & 2.18 & 15.33&37.26&76.07&119.07&157.49&203.52&246.69&289.45&362.59&392.89&482.09&536.33&621.77&697.33&783.48\\
\verb+N BoNeM -N KR Tie+ & 31 & 2.18 & 18.97&38.05&61.85&96.35&131.68&175.20&229.28&278.30&368.86&455.01&542.54&576.28&753.78&769.33&784.82\\
\verb+N BoNeM -N KR JXc+ & 35 & 2.30 & 17.73&37.10&65.93&104.73&135.12&182.40&237.82&368.48&444.19&474.94&558.74&584.25&722.61&749.45&889.37\\
\verb+N BoNeM -N KR QMa+ & 39 & 2.92 & 21.00&53.83&99.47&135.04&209.36&248.78&337.02&403.97&470.19&546.77&664.62&768.21&853.46&946.93&1058.06\smallskip \\
\verb+N BoNeP -N KR 4Cm+ & 5 & 1.12 & 10.10&21.80&34.53&52.94&\textbf{69.20}&88.78&112.80&143.67&188.17&210.52&252.70&303.77&355.74&392.22&419.03\\
\verb+N BoNeP -N KR 4CS+ & 6 & 1.13 & 9.31&23.11&35.51&56.29&77.11&92.99&120.16&151.19&179.63&217.68&244.75&292.50&328.03&362.46&387.20\\
\verb+N BoNeP -N KR Cla+ & 10 & 1.28 & 6.86&29.69&39.90&74.93&87.15&122.21&139.99&201.27&222.60&253.53&275.37&321.67&350.16&385.47&421.30\\
\verb+N BoNeP -N KR CPr+ & 11 & 1.30 & 6.55&30.51&42.04&75.27&86.68&117.72&140.35&199.82&216.17&258.35&280.93&331.44&373.93&402.71&440.35\\
\verb+N BoNeP -N KR 6Cm+ & 15 & 1.38 & 11.51&24.77&40.86&65.58&84.98&105.34&137.82&187.49&235.82&276.33&324.97&372.66&435.94&470.59&525.82\\
\verb+N BoNeP -N KR Tie+ & 22 & 2.06 & 18.15&34.93&63.42&98.77&129.21&179.81&224.98&277.16&331.25&399.71&466.61&540.55&620.63&675.01&785.68\\
\verb+N BoNeP -N KR Def+ & 26 & 2.11 & 17.48&39.97&72.50&112.59&149.29&187.89&236.08&284.89&332.48&380.62&439.68&515.16&600.16&656.77&749.71\\
\verb+N BoNeP -N KR JXc+ & 32 & 2.19 & 17.01&34.81&65.09&103.38&143.31&185.32&242.27&290.60&361.36&441.20&508.06&589.21&687.21&762.41&849.86\\
\verb+N BoNeP -N KR QMa+ & 40 & 3.05 & 21.39&52.56&94.71&158.02&196.77&270.19&335.88&419.22&509.74&596.41&684.62&813.48&926.26&1010.03&1129.40\\
\end{tabular}

			\end{scriptsize}
			\caption{Average number of CPU cycles per iteration of single array sorting across all machines} \label{table:normalsort:avg:all}
		\end{sidewaystable}
	
	\clearpage
	\subsection{Sorting Many Continuous Sets of 2-16 Items} \label{section:experiments:inrow}
		Here the benchmark shown in algorithm \ref{algo:inrow} was used. Instead of sorting a single array multiple times, multiple arrays are created adjacent to each other and sorted in series. \\
		The number of arrays used is chosen in a way that all of them do not fit into the CPU's L3 cache. Since the reference array is sorted before the measurement, the original array should not be present in the cache, causing a cache miss on every access.
		
\begin{figure}[!tbp]
			\includegraphics[width=1.0\textwidth]{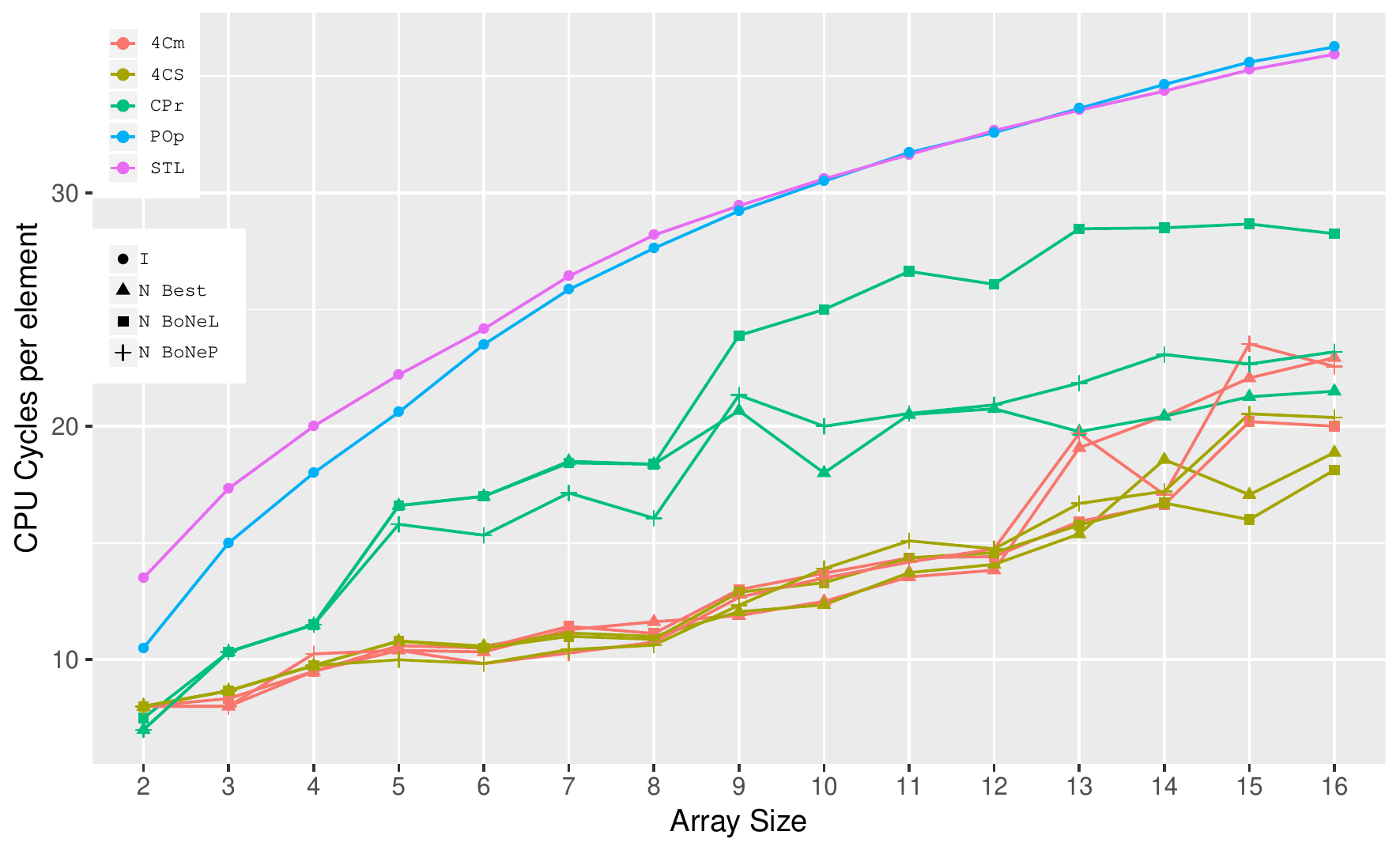}
			\caption{Continuous sorting of array sizes 2 to 16 on machine A} \label{plot:inrow:lineplot:A}
		\end{figure}

\begin{figure}[!tbp]
			\includegraphics[width=1.0\textwidth]{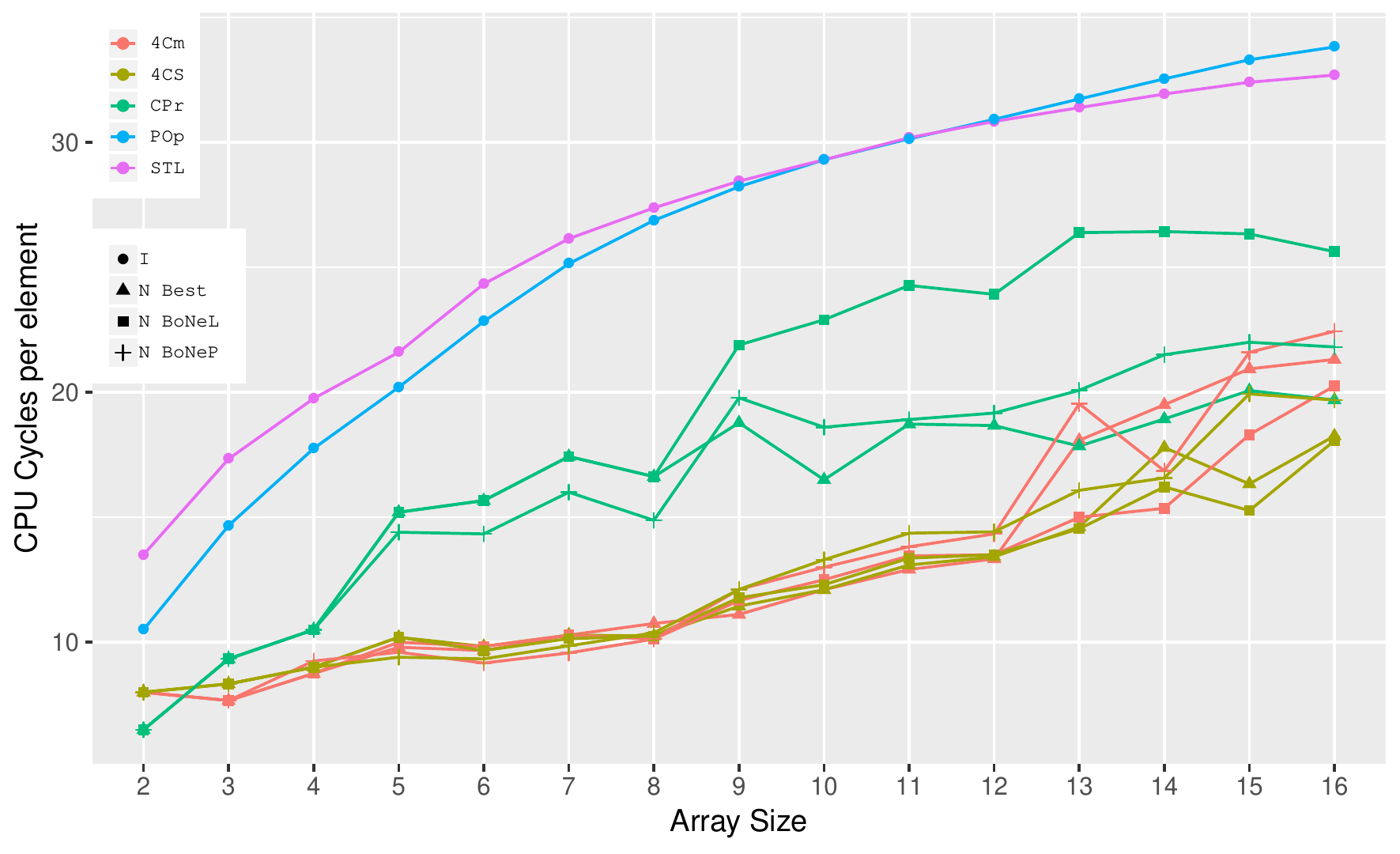}
			\caption{Continuous sorting of array sizes 2 to 16 on machine B} \label{plot:inrow:lineplot:B}
		\end{figure}

\begin{figure}[!tbp]
			\includegraphics[width=1.0\textwidth]{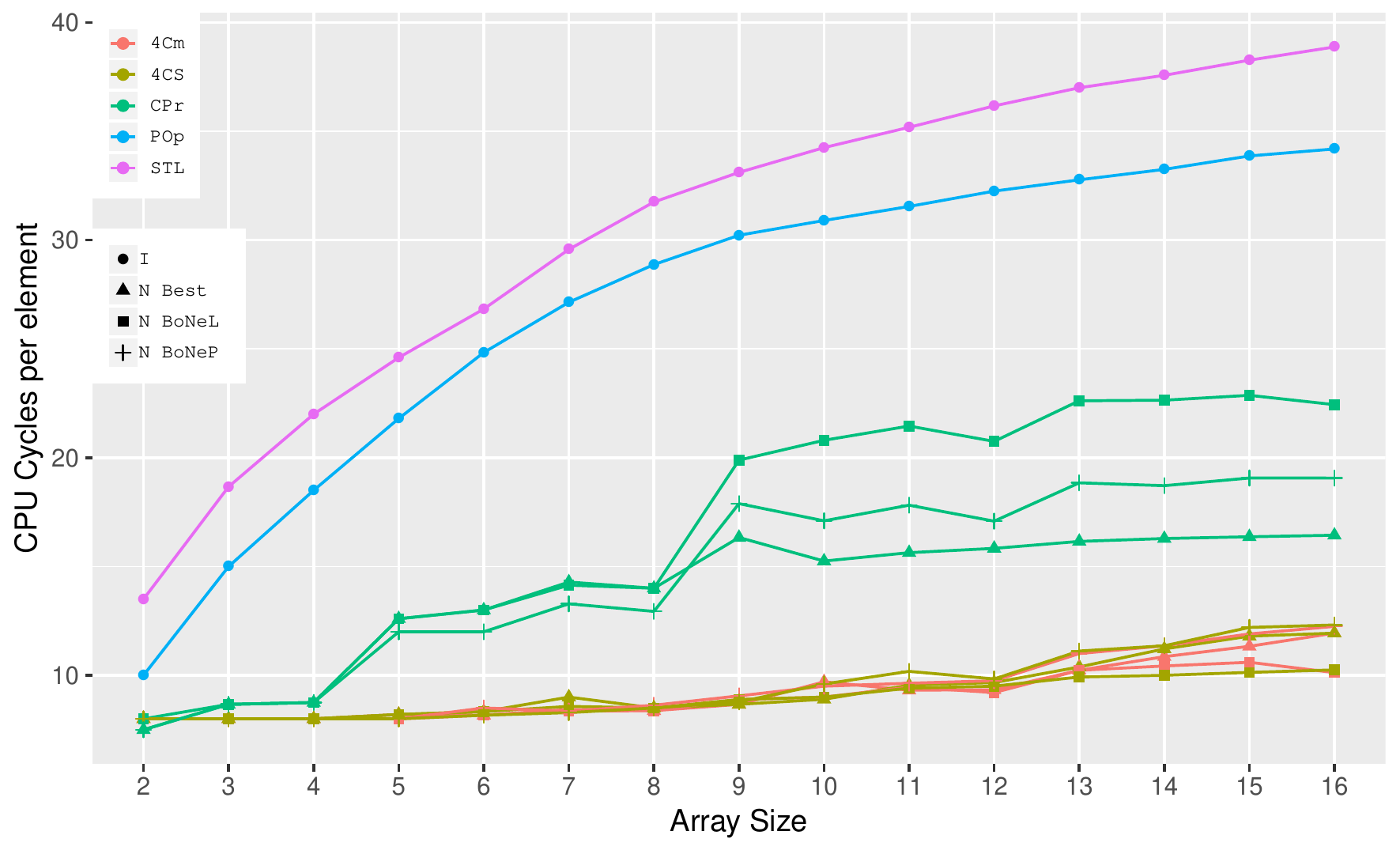}
			\caption{Continuous sorting of array sizes 2 to 16 on machine C} \label{plot:inrow:lineplot:C}
		\end{figure}

		The results are similar to the previous ones. A difference we can see when comparing figures \ref{plot:inrow:lineplot:A}, \ref{plot:inrow:lineplot:B} and \ref{plot:inrow:lineplot:C} to figures \ref{plot:normal:lineplot:A}, \ref{plot:normal:lineplot:B} and \ref{plot:normal:lineplot:C} from the single sort measurement is that the \verb|CPr| swap that operates on pointers and moves values around in memory became worse compared to the \verb|4Cm| and \verb|4CS| implementations for array sizes greater than 2. Here the values can probably get pre-loaded for the next conditional swap while the current one is finishing, while \verb|CPr| accesses the element's reference value only when the destination address is calculated, which results in less pre-loading that can be done. \\
		The complete overview over the average values of each sorter across all three machines can be seen in table \ref{table:inrowsort:avg:all}. We see speed-ups for using the sorting networks from 25\% at array size 2 all the way up to 59\% at array size 15.
		\begin{sidewaystable}[!tbp]
			\begin{scriptsize}
\begin{tabular}{l | r @{~~} r | r@{~~}r@{~~}r@{~~}r@{~~}r@{~~}r@{~~}r@{~~}r@{~~}r@{~~}r@{~~}r@{~~}r@{~~}r@{~~}r@{~~}r@{~~}r|}
 & \multicolumn{2}{c|}{Overall} & \multicolumn{15}{c}{Array Size} \\
 & Rank & GeoM & 2&3&4&5&6&7&8&9&10&11&12&13&14&15&16\\ \hline
\verb+I       -I KR POp+ & 26 & 2.39 & 20.67&44.67&72.23&104.33&142.53&182.47&222.07&262.90&302.13&342.53&382.87&425.57&468.73&513.80&556.07\\
\verb+I       -I KR Def+ & 27 & 2.42 & 22.17&47.30&74.60&106.70&143.40&183.47&223.37&263.23&302.53&342.47&383.63&425.73&467.77&511.97&554.67\\
\verb+I       -I KR STL+ & 34 & 2.57 & 27.07&53.47&82.40&114.07&150.63&191.53&232.87&273.00&313.67&355.40&398.73&441.80&484.63&529.97&573.47\\
\verb+I       -I KR AIF+ & 35 & 2.57 & 23.67&50.00&79.53&111.97&149.87&193.77&237.77&280.87&324.37&367.43&412.23&456.13&499.43&544.10&589.47\smallskip \\
\verb+N Best  -I KR 4CS+ & 2 & 1.05 & 15.97&25.00&35.67&48.73&57.60&70.90&79.30&96.50&\textbf{111.23}&133.43&148.70&175.07&222.03&226.00&261.80\\
\verb+N Best  -I KR 4Cm+ & 5 & 1.08 & 15.97&\textbf{23.67}&34.87&47.20&56.30&70.20&82.20&\textbf{95.07}&114.20&\textbf{131.30}&\textbf{146.00}&205.50&236.97&271.73&299.80\\
\verb+N Best  -I KR 6Cm+ & 8 & 1.29 & 15.97&25.70&37.03&55.13&66.80&85.50&94.37&117.63&139.07&160.90&181.33&270.23&301.70&347.80&379.53\\
\verb+N Best  -I KR Cla+ & 11 & 1.40 & 14.00&25.30&39.90&70.83&87.67&114.43&128.97&169.67&172.10&201.00&218.60&231.20&254.33&278.03&295.50\\
\verb+N Best  -I KR CPr+ & 13 & 1.43 & \textbf{13.90}&28.37&41.07&74.03&91.33&117.27&130.77&167.40&165.93&201.17&220.90&233.23&259.73&288.53&307.23\\
\verb+N Best  -I KR Def+ & 22 & 2.33 & 23.67&44.33&72.00&96.07&129.27&167.40&205.97&245.73&262.57&328.70&373.57&408.03&478.33&531.47&605.10\\
\verb+N Best  -I KR Tie+ & 23 & 2.34 & 25.97&43.77&69.00&92.00&119.37&166.30&200.83&235.50&274.13&324.33&380.43&435.33&486.90&555.63&639.40\\
\verb+N Best  -I KR JXc+ & 32 & 2.52 & 24.60&45.30&70.63&99.83&137.83&173.73&213.70&253.73&295.23&363.17&418.60&479.23&560.80&638.80&696.60\\
\verb+N Best  -I KR QMa+ & 37 & 3.10 & 24.67&46.00&83.13&130.07&176.83&223.30&279.90&333.30&357.03&500.63&575.63&580.67&684.17&773.97&845.93\smallskip \\
\verb+N BoNeL -I KR 4CS+ & 1 & 1.04 & 16.00&25.00&35.63&48.50&57.10&69.30&78.83&100.57&115.27&136.13&150.63&\textbf{174.33}&200.53&\textbf{207.23}&\textbf{247.57}\\
\verb+N BoNeL -I KR 4Cm+ & 3 & 1.06 & 15.97&24.00&34.97&47.70&57.00&70.50&79.10&100.80&117.30&136.77&148.63&180.77&\textbf{198.23}&245.17&268.67\\
\verb+N BoNeL -I KR 6Cm+ & 9 & 1.32 & 16.00&26.00&37.60&55.70&66.93&85.33&94.57&124.93&145.57&173.90&194.77&269.73&328.20&350.90&355.43\\
\verb+N BoNeL -I KR Cla+ & 17 & 1.65 & 13.93&25.13&40.00&70.97&87.63&114.33&129.00&200.47&223.80&268.67&286.87&344.30&365.33&400.97&416.60\\
\verb+N BoNeL -I KR CPr+ & 18 & 1.67 & 14.60&28.40&41.10&74.00&91.40&117.00&130.73&197.03&229.17&265.43&282.93&335.90&362.13&389.33&407.03\\
\verb+N BoNeL -I KR Tie+ & 25 & 2.36 & 25.97&43.63&68.33&92.87&120.40&166.40&200.63&237.40&282.20&331.43&384.17&444.20&494.67&562.53&639.57\\
\verb+N BoNeL -I KR Def+ & 28 & 2.42 & 24.10&45.40&74.37&95.90&128.67&166.37&205.67&248.33&306.17&330.20&414.60&438.80&525.63&575.17&630.03\\
\verb+N BoNeL -I KR JXc+ & 31 & 2.47 & 24.33&45.33&70.67&99.67&137.83&173.80&213.67&260.80&303.70&347.07&402.80&459.87&533.17&576.37&640.67\\
\verb+N BoNeL -I KR QMa+ & 39 & 3.27 & 25.00&46.50&83.20&130.23&177.13&223.37&280.17&371.97&434.07&505.10&582.43&688.13&756.13&833.53&945.70\smallskip \\
\verb+N BoNeM -I KR 4Cm+ & 7 & 1.19 & 16.00&23.93&\textbf{34.33}&47.33&55.90&66.60&\textbf{76.57}&170.50&177.00&167.90&194.00&194.70&253.57&262.33&307.97\\
\verb+N BoNeM -I KR 4CS+ & 12 & 1.41 & 16.00&24.63&41.20&63.03&55.77&111.00&77.50&165.30&203.97&204.63&200.43&294.33&321.33&349.40&382.30\\
\verb+N BoNeM -I KR 6Cm+ & 16 & 1.52 & 16.00&25.93&36.90&53.23&65.30&125.07&92.47&193.63&244.90&239.67&236.83&344.47&348.43&383.13&363.77\\
\verb+N BoNeM -I KR Cla+ & 19 & 1.86 & 16.67&29.33&41.20&83.67&89.00&135.63&128.30&220.20&259.97&297.53&311.73&402.10&427.67&484.07&477.33\\
\verb+N BoNeM -I KR CPr+ & 20 & 1.93 & 16.27&29.73&41.90&92.67&91.70&137.23&157.13&240.17&275.77&304.70&309.10&399.97&404.70&479.83&531.63\\
\verb+N BoNeM -I KR Tie+ & 29 & 2.45 & 26.00&44.33&69.03&92.70&121.00&162.47&197.57&239.17&305.27&373.97&427.80&516.20&569.87&566.87&648.47\\
\verb+N BoNeM -I KR Def+ & 30 & 2.46 & 23.33&45.40&73.73&105.90&144.73&182.37&212.07&246.03&305.07&329.70&410.60&432.07&501.03&575.37&633.07\\
\verb+N BoNeM -I KR JXc+ & 36 & 2.64 & 24.40&42.67&70.63&123.73&130.33&168.57&212.20&281.13&333.73&432.47&458.37&492.10&583.77&632.10&741.07\\
\verb+N BoNeM -I KR QMa+ & 38 & 3.20 & 29.00&52.00&89.77&126.97&177.13&225.23&278.20&346.27&400.33&467.37&533.50&611.23&702.33&774.00&878.87\smallskip \\
\verb+N BoNeP -I KR 4CS+ & 4 & 1.08 & 16.00&24.97&35.70&\textbf{45.70}&\textbf{54.77}&66.70&78.73&99.60&122.90&145.37&156.00&190.20&210.63&263.43&279.30\\
\verb+N BoNeP -I KR 4Cm+ & 6 & 1.09 & 16.00&23.80&36.73&46.93&54.97&\textbf{65.80}&78.73&101.50&119.97&138.13&155.37&218.03&211.63&285.87&305.37\\
\verb+N BoNeP -I KR 6Cm+ & 10 & 1.35 & 16.33&26.00&37.30&52.43&63.17&79.57&93.27&124.70&151.13&174.90&241.00&287.13&329.40&385.30&411.77\\
\verb+N BoNeP -I KR Cla+ & 14 & 1.45 & 14.63&25.03&40.03&66.00&81.47&108.33&116.67&178.70&188.40&215.53&227.33&269.33&285.97&316.60&333.97\\
\verb+N BoNeP -I KR CPr+ & 15 & 1.47 & 13.90&28.33&41.13&70.27&83.43&108.47&116.93&177.00&185.57&210.00&228.57&263.30&295.20&318.77&341.63\\
\verb+N BoNeP -I KR Def+ & 21 & 2.32 & 23.67&46.10&71.00&96.90&128.83&160.97&196.50&241.30&278.13&321.10&364.47&419.27&478.67&544.53&597.17\\
\verb+N BoNeP -I KR Tie+ & 24 & 2.35 & 25.50&44.67&69.63&90.33&124.73&160.10&193.67&236.70&284.33&327.73&383.80&440.30&500.37&557.47&633.60\\
\verb+N BoNeP -I KR JXc+ & 33 & 2.54 & 24.67&43.33&70.60&100.33&133.17&169.03&217.23&258.40&318.80&377.17&428.93&489.80&564.90&625.70&713.63\\
\verb+N BoNeP -I KR QMa+ & 40 & 3.29 & 26.33&49.67&87.57&130.00&165.33&224.03&274.07&367.53&428.20&509.07&570.77&684.67&776.17&849.17&927.27\\
\end{tabular}

			\end{scriptsize}
			\caption{Average number of CPU cycles per array of continuous sorting across all machines} \label{table:inrowsort:avg:all}
		\end{sidewaystable}
	\clearpage
	\subsection{Sorting a Large Set of Items with Quicksort} \label{section:experiments:quicksort}
\begin{figure}[!tbp]
			\includegraphics[width=1.0\textwidth]{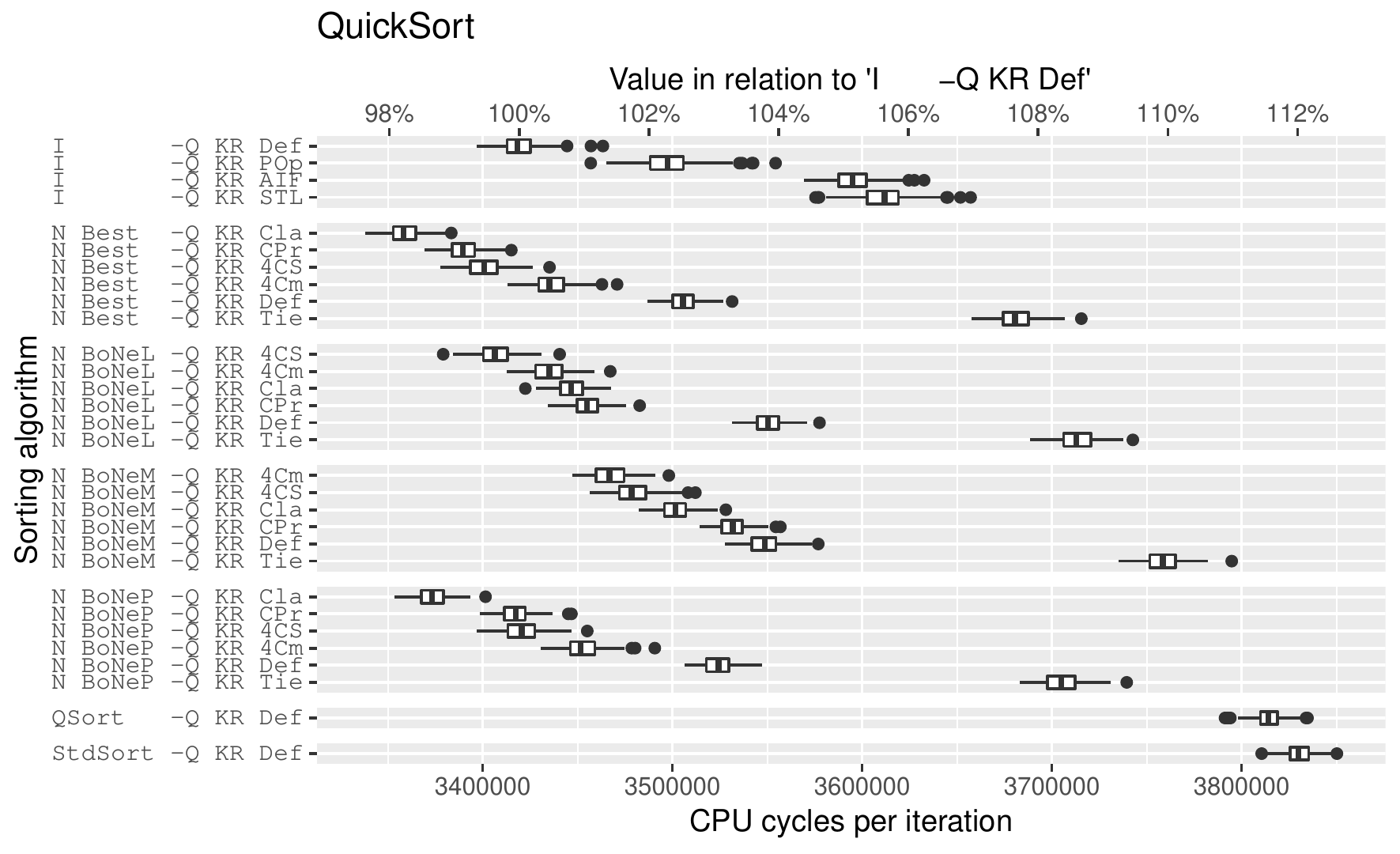}
			\caption{Sorting times of quicksort with different base cases on machine A} \label{plot:quicksort:A} 
		\end{figure}

\begin{figure}[!tbp]
			\includegraphics[width=1.0\textwidth]{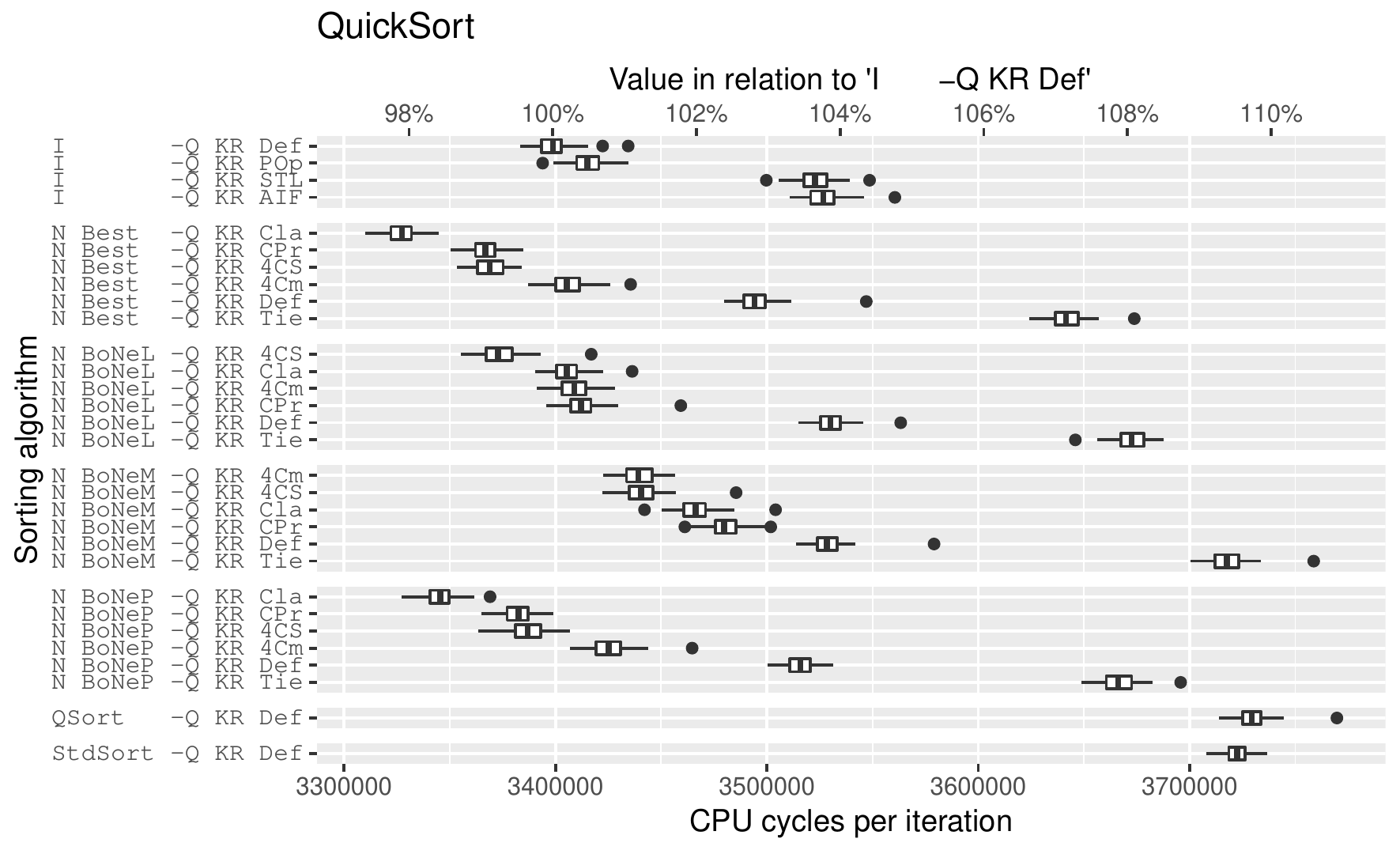}
			\caption{Sorting times of quicksort with different base cases on machine B} \label{plot:quicksort:B}
		\end{figure}

\begin{figure}[!tbp]
			\includegraphics[width=1.0\textwidth]{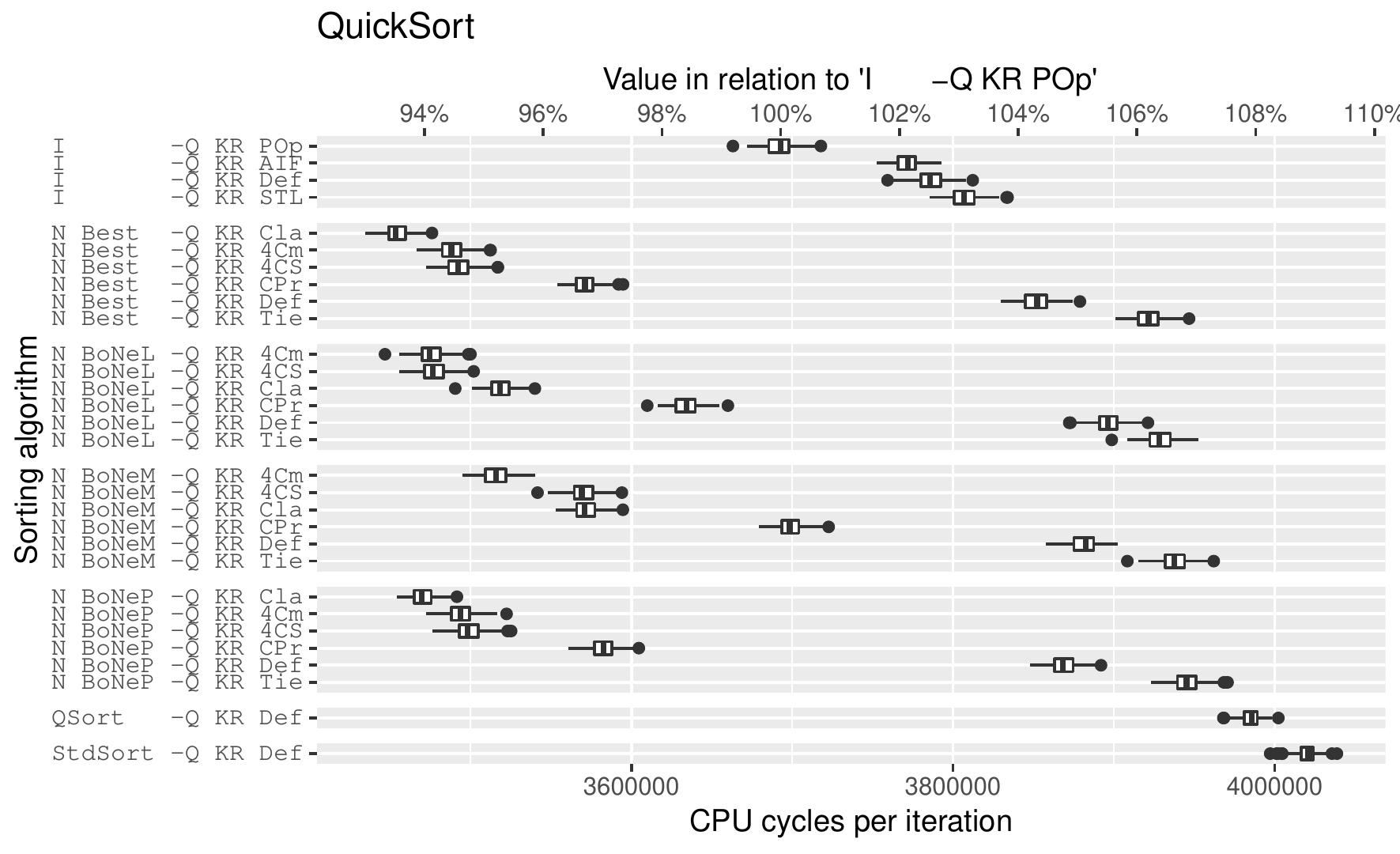}
			\caption{Sorting times of quicksort with different base cases on machine C} \label{plot:quicksort:C}
		\end{figure}

		After seeing the first two results, we wanted to know how the base case sorters perform when used inside a scalable sorting algorithm. For that we modified introsort, a quicksort implementation from the STL library, as follows: Introsort calls insertion sort only once, right at the end. Since that is not possible with the sorting networks, they had to be called directly when the partitioning resulted in a partition of 16 elements or less. Also we determined the pivot using the 3-element Bose Nelson parameter network instead of using \verb|if-else| and \verb|std::swap|. \\
		The sorters were measured using benchmark \ref{algo:normal} with parameters
		\begin{itemize}
			\item $\mathtt{numberOfIterations} = 50$
			\item $\mathtt{numberOfMeasures} = 200$
			\item $\mathtt{arraySize} = 1024 \times 16 = 16384 = 2^{14}$.
		\end{itemize}
		To have a basis of comparison we also measured sorting with \verb|std::sort|. These times can be taken from figures \ref{plot:quicksort:A}, \ref{plot:quicksort:B} and \ref{plot:quicksort:C}. \\
		The \verb|QSort -Q KR Def| sorter is just a direct copy of the STL sort doing a final insertion sort at the end. That was measured to see that our code copy does as well as \verb|std::sort| before doing the modifications.
	\clearpage
		\begin{table}[!h]
			\begin{center}
			\begin{small}
			\begin{tabular}{ c | c | c | c }
				& A:~~ \verb|N Best -Q KR Cla| & B:~~ \verb|N Best -Q KR Cla| & C:~~ \verb|N Best -Q KR Cla| \\ \hline
				\verb|I -Q KR Def| & 1.76\% & 2.1\% & 8.76\% \\
				\verb|I -Q KR POp| & 3.99\% & 2.58\% & 6.47\%\\
				\verb|StdSort  -Q| & 12.3\% & 10.6\% & 14\% \\
			\end{tabular}
			\end{small}
			\end{center}
			\caption{Average speed-ups of the fastest sorting network over the fastest insertion sort as base case in quicksort and unmodified std::sort} \label{table:completesort:speedups}
		\end{table} 
		Speed-ups of including sorting networks into a sorting algorithm like quicksort can be seen in table \ref{table:completesort:speedups}. \\
		What is notable is that the variants with insertion sort at the base are faster than the one with the final insertion sort, which should come from the fact that they are already specialized for the item they sort and do not require a predicate for the sorting. Also, the base case is called right after the partitioning is at a low enough level, which means that the elements are still present in the first- or second-level cache. That also explains why the \verb|Cla| conditional swap performs the best with quicksort, while we saw in the last section that this is not necessarily the case when we have a cache miss.
		
		Recalling the results from the previous sections, we appeared to be achieving great improvements in reducing the time needed for sorting sets of 2-16 items. By measuring only the sorting of the small sets we have exploited the networks' strength: not containing conditional branches. The results from the measurements with quicksort highlight the networks' weakness: The larger code size. \\
		When integrating the sorting networks into quicksort for sorting the base cases, every time a partition results in one part having 16 elements or less, we switch from the code for quicksort to the code for the sorting network. Thus, the code for quicksort is partly removed from the L1 instruction cache and replaced with the code for the sorting network. Because the network's code is just a flat sequence of conditional swaps, each line of code is accessed exactly once per sort. That means it caused a lot of quicksort's code to be removed from the instruction cache without gaining a speed-up because its code is now in the cache, and will be removed again when quicksort is handed back the flow of control and loads its code back into the instruction cache. \\
		We can see that effect especially for machines A and B which have 32 KiB of L1 instruction cache, where the speed-up is hardly over 2\% for the best network base case over the best insertion sort base case. Where we got a much more improvement is on machine C, which has double the space in its L1 instruction cache. Here we achieved a speed-up of almost 6.5\% when making use of the best networks. \\
		It is no surprise that we do not see improvements similar to those in section \ref{section:experiments:normal} or \ref{section:experiments:inrow} because the partitioning that quicksort performs takes the same amount of time no matter which base case sorter is used, representing a part of the algorithm that is not optimizable through using sorting networks.	
	\begin{table}[!h]
		\begin{center}
			\begin{small}
				\begin{tabular}{ c | c | c | c }
					& A:~~ \verb|N BoNeL -s332 4CS| & B:~~ \verb|N BoNeL -s332 4CS| & C:~~ \verb|N BoNeL -s332 4CS| \\ \hline
					\verb|I -s332 Def| & 17.4\% & 17.5\% & 29.2\% \\
					\verb|StdSort  -s| & 43.6\% & 43.49\% & 51\% \\
				\end{tabular}
			\end{small}
		\end{center}
		\caption{Average speed-ups of the fastest sorting network over the fastest insertion sort as base case in sample sort and unmodified std::sort} \label{table:samplesort:speedups}
	\end{table}
	\subsection{Sorting a Medium-Sized Set of Items with Sample Sort} \label{section:experiments:samplesort}
\begin{figure}[!p]
			\includegraphics[width=1.0\textwidth]{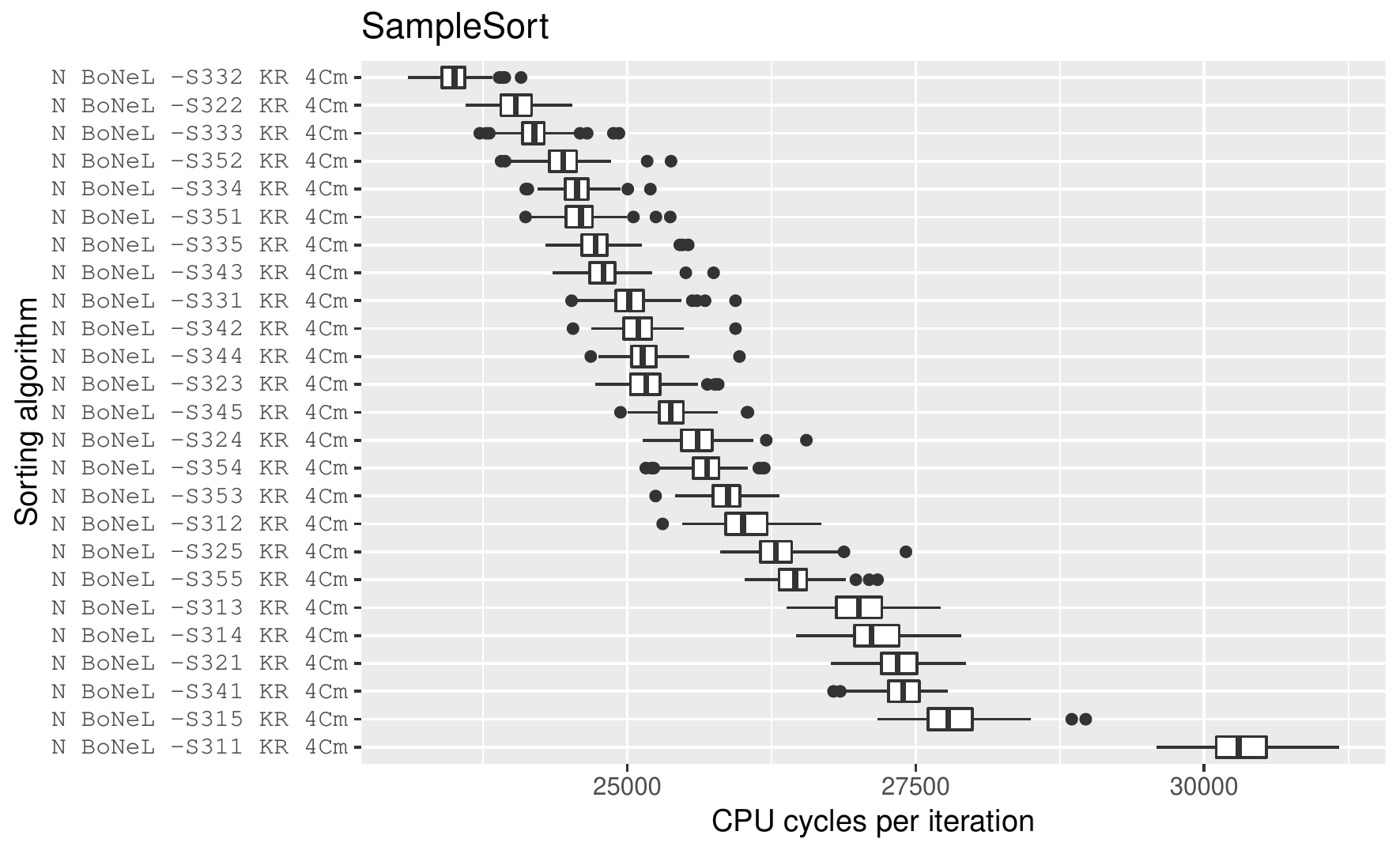}
			\caption[Sample sort on machine A with 256 items and different configurations]{Sample sort on machine A with 256 items. -Sxyz has parameters x = \texttt{numberOfSplitters}, y = \texttt{oversamplingFactor} and z = \texttt{blockSize}} \label{plot:samplesort:bonel:A}
		\end{figure}

\begin{figure}[!p]
			\includegraphics[width=1.0\textwidth]{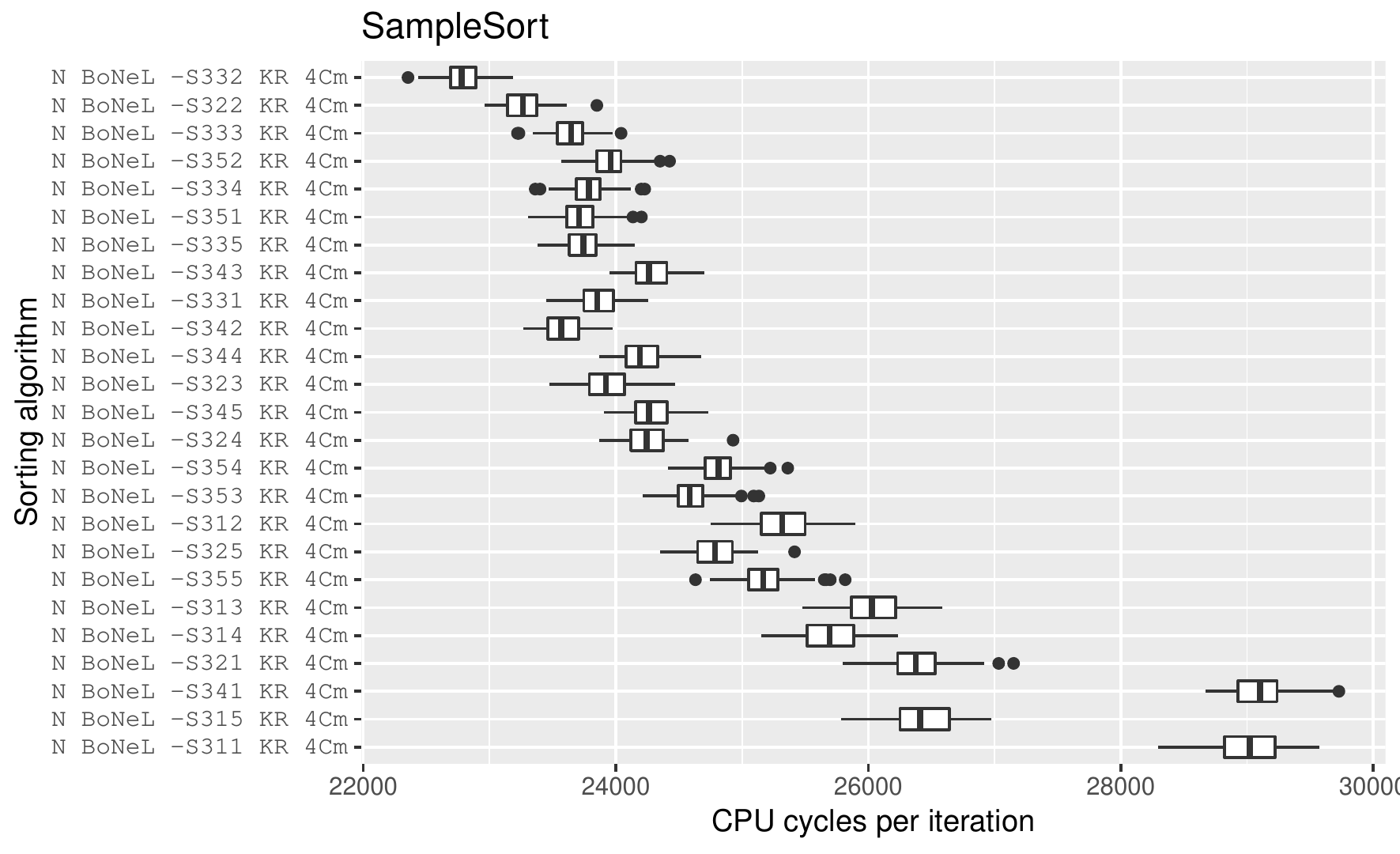}
			\caption[Sample sort on machine B with 256 items and different configurations]{Sample sort on machine B with 256 items. -Sxyz has parameters x = \texttt{numberOfSplitters}, y = \texttt{oversamplingFactor} and z = \texttt{blockSize}} \label{plot:samplesort:bonel:B}
		\end{figure}

\begin{figure}[!p]
			\includegraphics[width=1.0\textwidth]{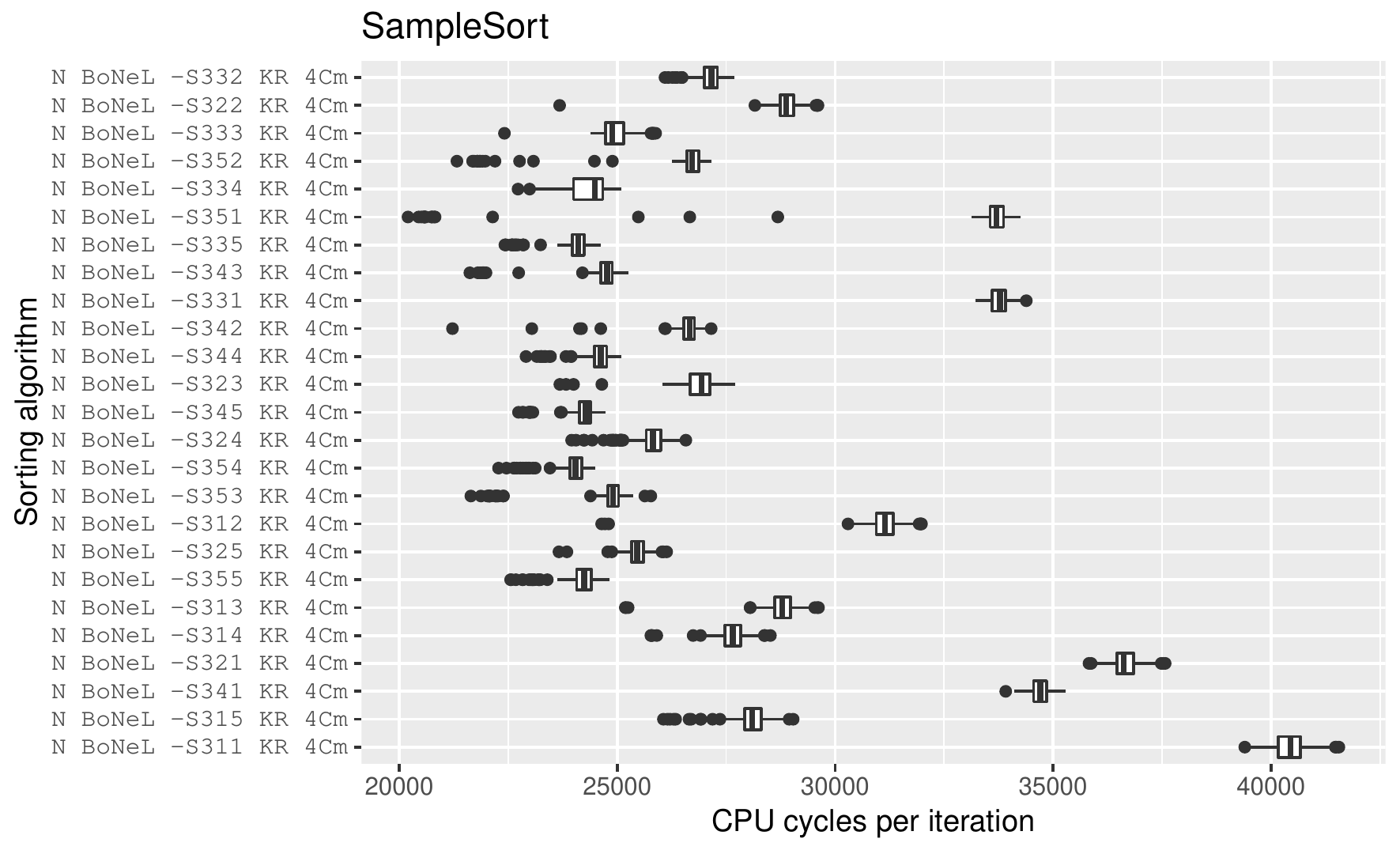}
			\caption[Sample sort on machine C with 256 items and different configurations]{Sample sort on machine C with 256 items. -Sxyz has parameters x = \texttt{numberOfSplitters}, y = \texttt{oversamplingFactor} and z = \texttt{blockSize}} \label{plot:samplesort:bonel:C}
		\end{figure}

		Sample sort was measured using benchmark \ref{algo:normal} with parameters:
		\begin{itemize}
			\item $\mathtt{numberOfIterations} = 50$
			\item $\mathtt{numberOfMeasures} = 200$
			\item $\mathtt{arraySize} = 256$.
		\end{itemize}
		The measurements were done with two different goals in mind:
		The first was to see which parameters work best for the machines used and the array size set. This can be seen in figures \ref{plot:samplesort:bonel:A}, \ref{plot:samplesort:bonel:B} and \ref{plot:samplesort:bonel:C} for the Bose Nelson networks optimizing locality. To be able to compare the results on the different machines, the configurations were ordered based on the times from machine A, and are in the same order in the other two plots. An oversampling factor of 3 and block size of 2 performed best on machine A and B. That configuration also performs best when using the other networks or insertion sort as a base case. \\
		On machine C block sizes larger than 2 performed better (on average) along with an oversampling factors of 3 or greater. We measured larger variances and got a lot more outliers, so here choosing a \enquote{best} configuration was not so easy. When looking at the other networks and insertion sort as base case, consistently well performing parameters are an oversampling factor of 3 and a block size of 4, but with very little lead over other configurations. That is interesting to see because all three machines run x86 assembly instructions and have the same number of general purpose registers available. What comes into play here is the size of the instruction cache: Machine C has double the amount of L1 instruction cache of what machines A and B have. We can only assume that the instructions for classifying three elements need more space than the smaller 32 KiB instruction caches can provide, while the 64 KiB instruction cache that machine C has fits the instructions for classifying four and / or almost five elements at once, considering that block size 5 also performs well. \\
		
\begin{figure}[!p]
			\includegraphics[width=1.0\textwidth]{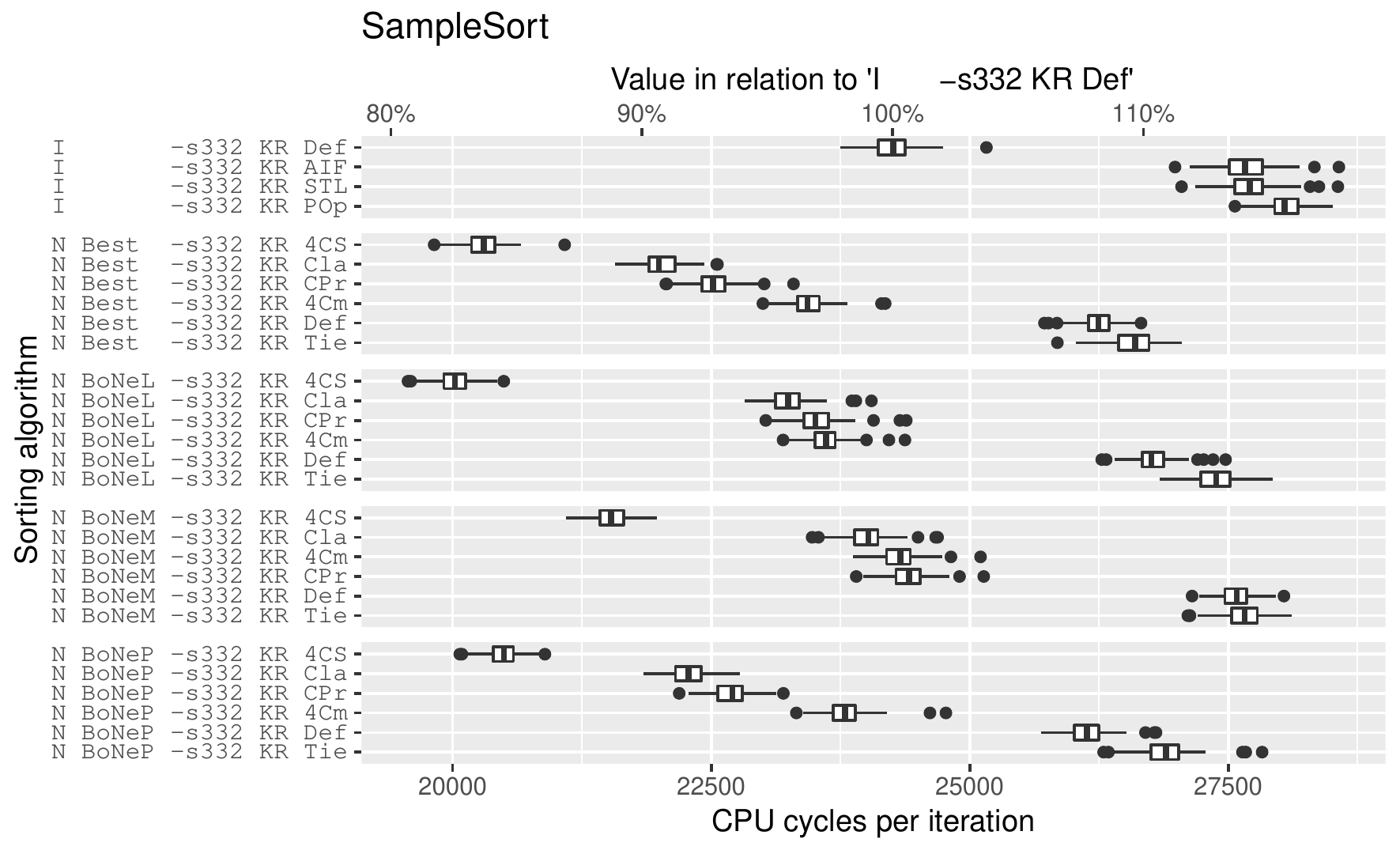}
			\caption{Sample sort 332 with different base cases on machine A} \label{plot:samplesort:s332:A}
		\end{figure}

\begin{figure}[!p]
			\includegraphics[width=1.0\textwidth]{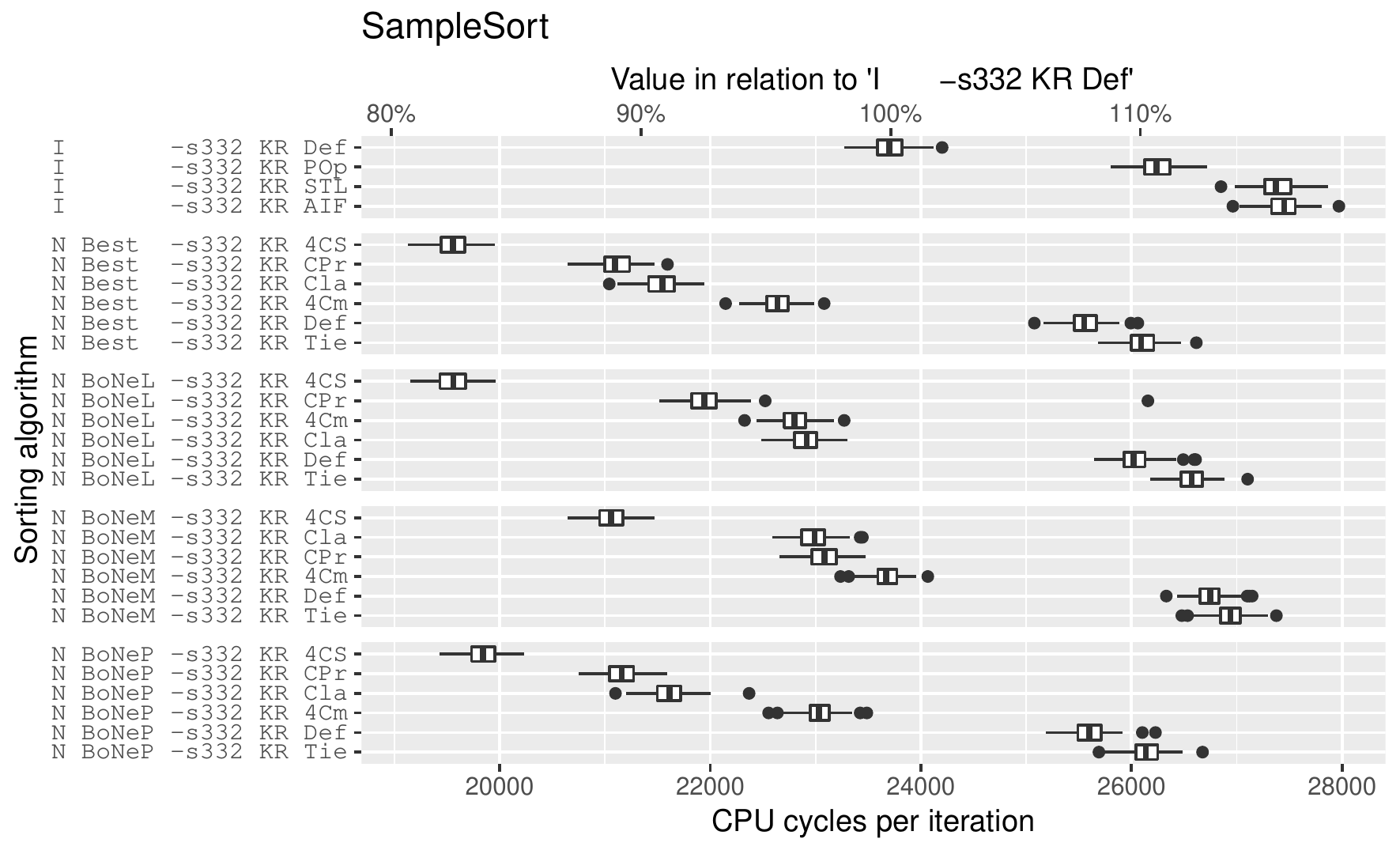}
			\caption{Sample sort 332 with different base cases on machine B} \label{plot:samplesort:s332:B}
		\end{figure}

\begin{figure}[!p]
			\includegraphics[width=1.0\textwidth]{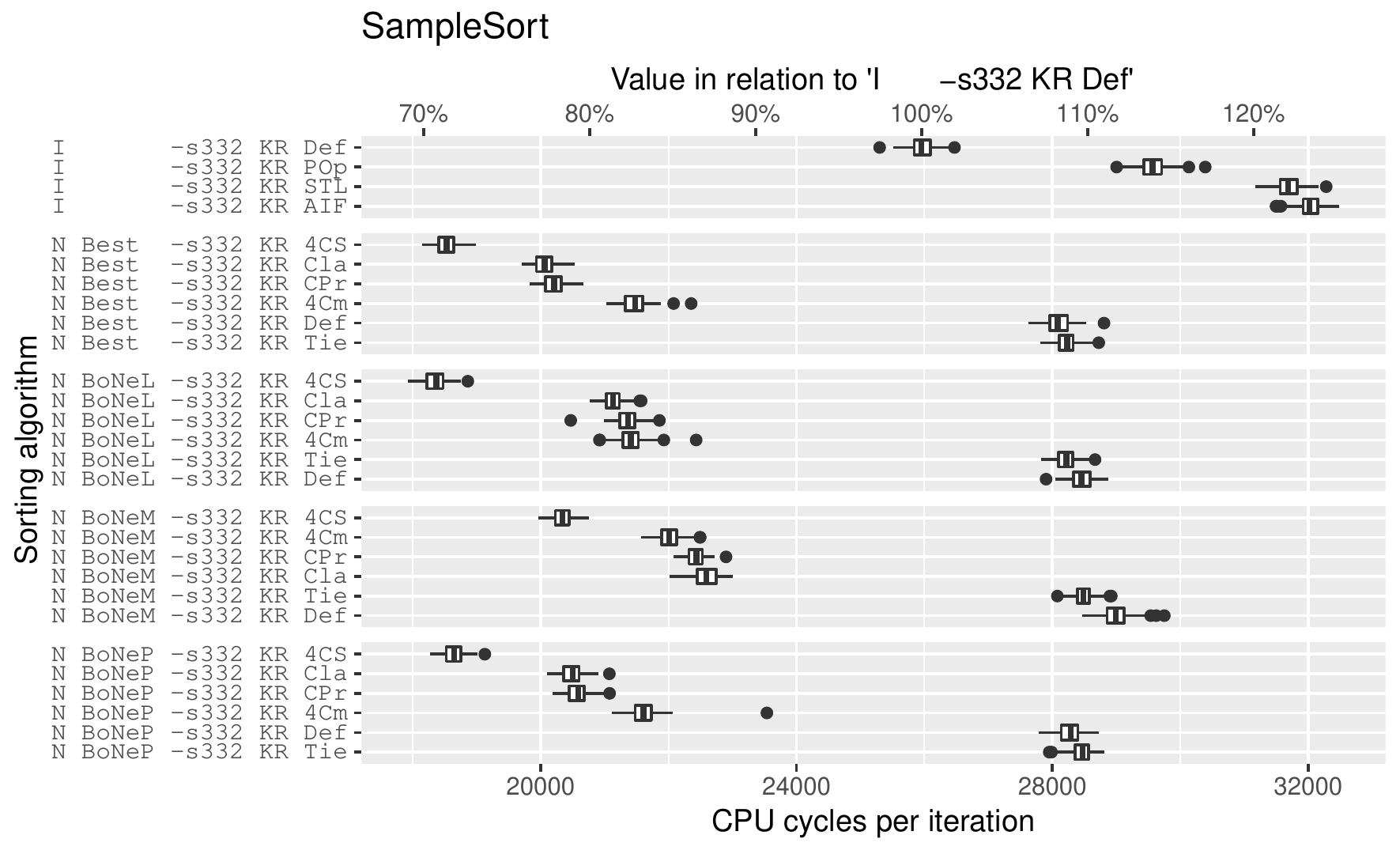}
			\caption{Sample sort 332 with different base cases on machine C} \label{plot:samplesort:s332:C}
		\end{figure}

		The second goal was to see if the results from section \ref{section:experiments:normal}, \ref{section:experiments:inrow} and \ref{section:experiments:quicksort} would relate to the results from using sample sort with the sorting networks as base cases. These results can be seen in figures \ref{plot:samplesort:s332:A}, \ref{plot:samplesort:s332:B} and \ref{plot:samplesort:s332:C} for the 332 configuration. All measurements were made with a base case limit of 16. Here, too, a single outlier was excluded from the dataset for scaling purposes: A value of $40177$ measured on machine B for the \texttt{'N BoNeP -s332 KR 4Cm'} sorter. \\
		The achieved speed-ups of using the sorting networks are given in table \ref{table:samplesort:speedups}. On the left we see sample sort with insertion sort as base case and std::sort that was also measured sorting 256 elements. On the top we see the best performing network \texttt{'N BoNeL -s332 4CS'} as a base case for sample sort on all three machines. The number indicates the speed-up of sample sort with the network over sample sort with insertion sort and over std::sort. \\
		Again we see that due to machine C having a larger L1 instruction cache the performance gain is almost double that for the other machines. Unlike in the previous section though we got much greater speed-ups as a result of using the sorting networks as a base case. That comes from the fact that sample sort has no unpredictable branches classifying the elements, as opposed to quicksort having to deal with conditional branches during the partitioning, while both need to invest the same time to sort all the base cases. So with sample sort, the base case sorting takes up a larger time slot of the whole execution than it does with quicksort. We also see that with very few conditional branches we can get up to 50\% faster than std::sort (for sets of up to 256 items at least).
		
	\clearpage
	\subsection{Sorting a Large Set of Items with \ipso} \label{section:experiments:ipso}
\begin{figure}[!tbp]
			\includegraphics[width=1.0\textwidth]{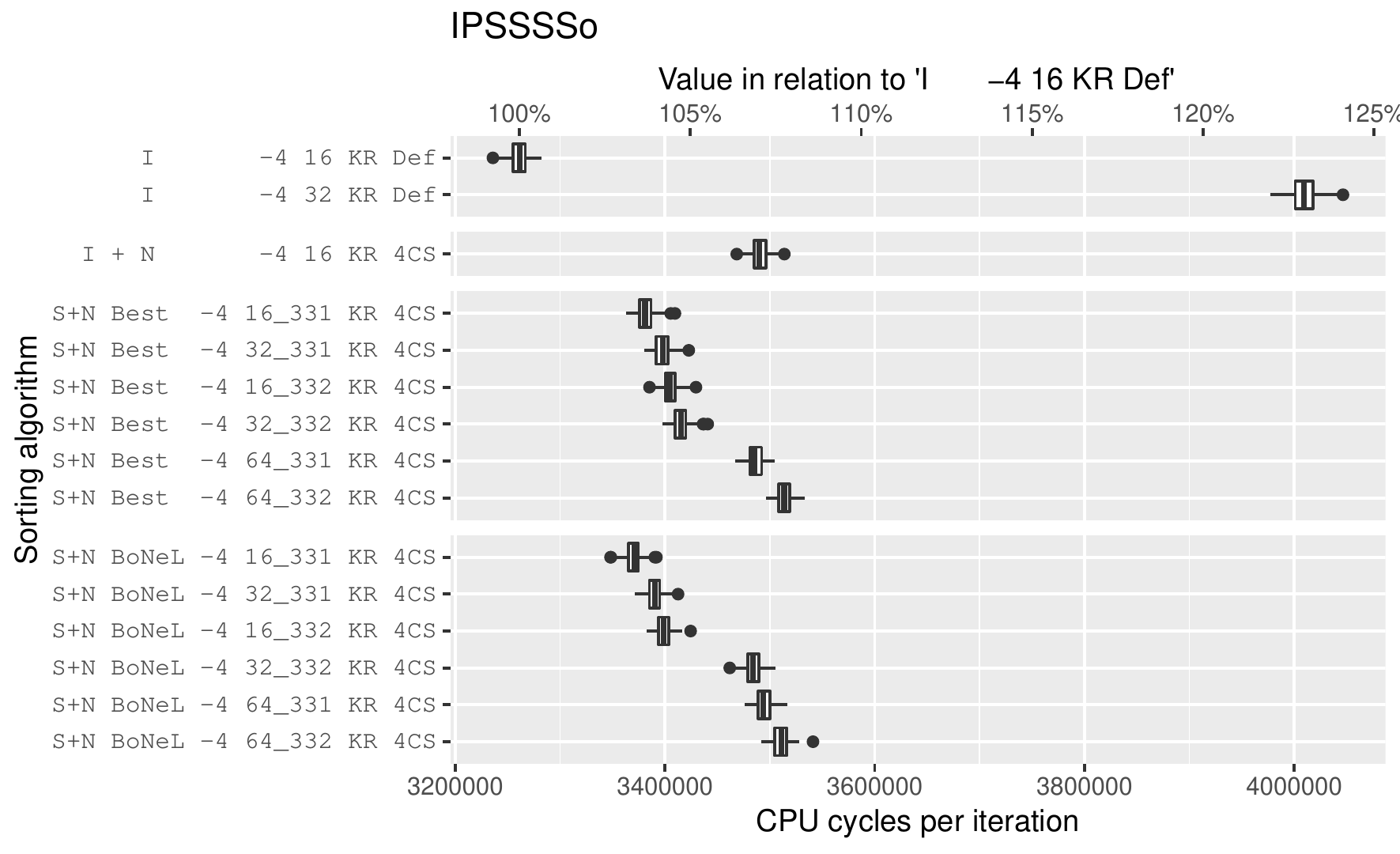}
			\caption{Sorting times for \ipso\, on machine A with different base cases and base case sizes} \label{plot:ipso:A}
		\end{figure}

\begin{figure}[!tbp]
			\includegraphics[width=1.0\textwidth]{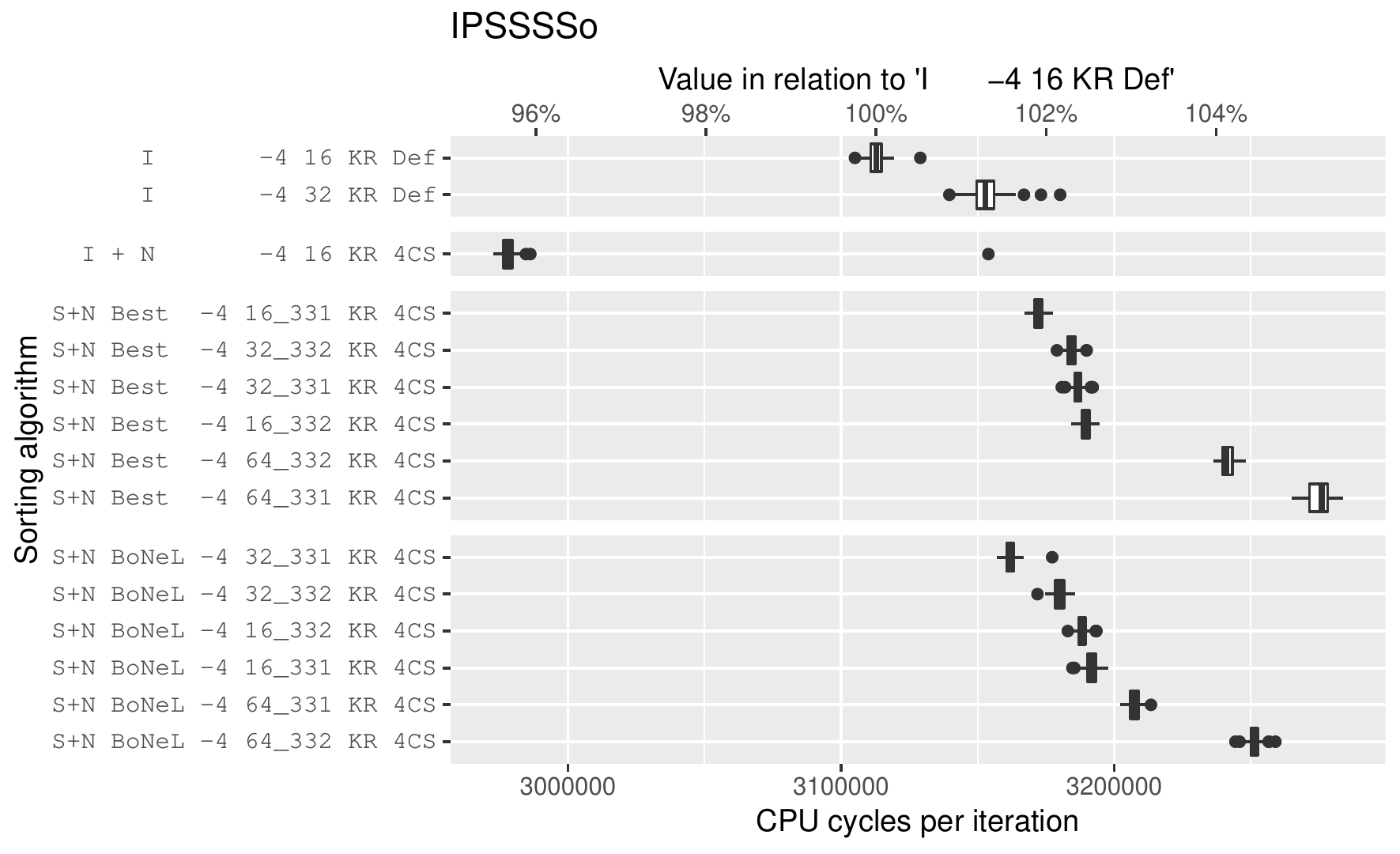}
			\caption{Sorting times for \ipso\, on machine B with different base cases and base case sizes} \label{plot:ipso:B}
		\end{figure}

\begin{figure}[!tbp]
			\includegraphics[width=1.0\textwidth]{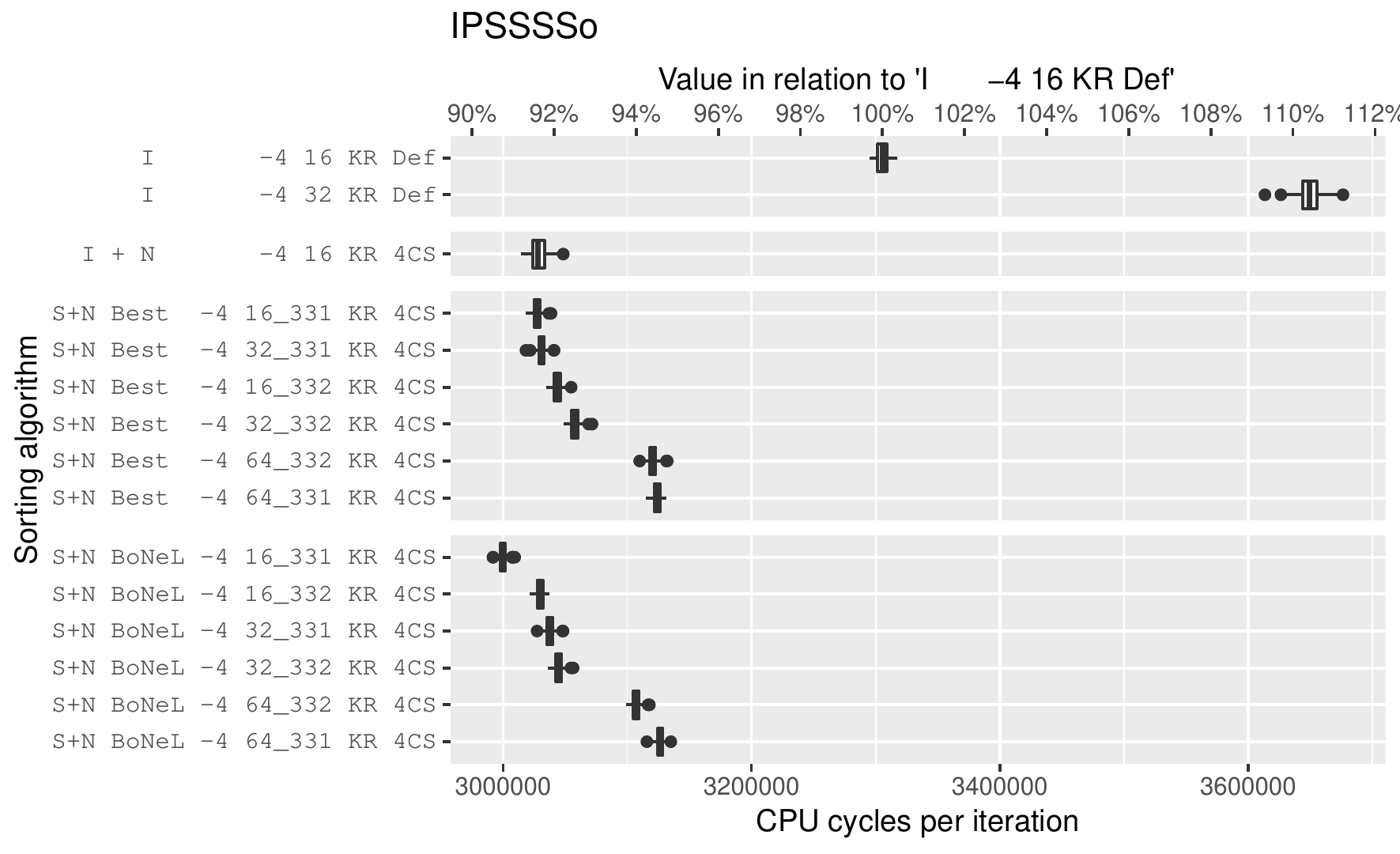}
			\caption{Sorting times for \ipso\, on machine C with different base cases and base case sizes} \label{plot:ipso:C}
		\end{figure}

		\newcommand{\bcsize}{\texttt{BaseCaseSize}${}_4$}
		With the efficient implementation of sample sort for medium-sized sets we can now include the new base case sorters into a complex sorting algorithm. The \textit{In-Place Parallel Super Scalar Samplesort} (\ipso) \cite{DBLP:conf/esa/AxtmannWF017} was executed without introducing parallelism. The algorithm has many parameters that can be adjusted. The important parameter for us was the \bcsize\footnote{we will use the ${}_4$ to distinguish \ipso's \texttt{BaseCaseSize} from Register Sample Sort's base case size}: it tells \ipso~ to aim for base case sizes that are smaller or equal to \bcsize. Even though that is the goal, for a large-scale sorter like \ipso~ it would be far less efficient to partition e.g. 32 elements into many buckets, that might end up not containing many elements each, than just using the base case sorter for these situations, even though the number of items is larger than the specified \bcsize. \\
		That was the reason to develop Register Sample Sort that can break those medium-sized sets down into sizes that can be sorted using the sorting networks. \\
		We started the measuring using the best combination of sample sort from section \ref{section:experiments:samplesort} as a base case for \ipso, together with using the default \bcsize = 16, but that turned out to perform worse than just insertion sort. \\
		The distribution of the base case array sizes can be seen in figure \ref{plot:distr:16} for \bcsize\, = 16 and figure \ref{plot:distr:32} for \bcsize\, = 32. From that it was evident that in most of the instances with parameter \bcsize\, = 16  the base case sorter was being invoked on sets smaller than even 32 elements. That also meant that sample sort had to deal with a larger overhead than insertion sort, not justified by a larger amount of items. \\
		In addition to that the size of the instruction cache that had already had a great influence on the measurements of quicksort seemed to be another factor for the bad performance of Register Sample Sort as a base case. \\
		That is why we decided to measure the following setups:
		\begin{itemize}
			\item Pure insertion sort as base case (\verb|I|) with
			\begin{itemize}
				\item \bcsize = 16 and 32
			\end{itemize}
			\item Register sample sort as base case (\verb|S+N|) with
			\begin{itemize}
				\item \bcsize = 16, 32, and 64,
				\item Configurations 331 and 332, and
				\item Best networks and Bose Nelson networks (optimizing locality) as base case for Register Sample Sort, with the \verb|4CS| conditional swap and base case size 16
			\end{itemize}
			\item A combination of the sorting networks and insertion sort (\verb|I+N|): \\
			Since the base case sizes were often smaller than 16, we wanted to make use of that by using the sorting networks, while not having to rely on Register Sample Sort with its larger overhead for the slightly larger base cases. The solution was to use the Bose Nelson networks (optimizing locality) if the set had 16 elements or less, and insertion sort otherwise.
		\end{itemize}
		\begin{figure}[!p]
			\includegraphics[width=1.0\textwidth]{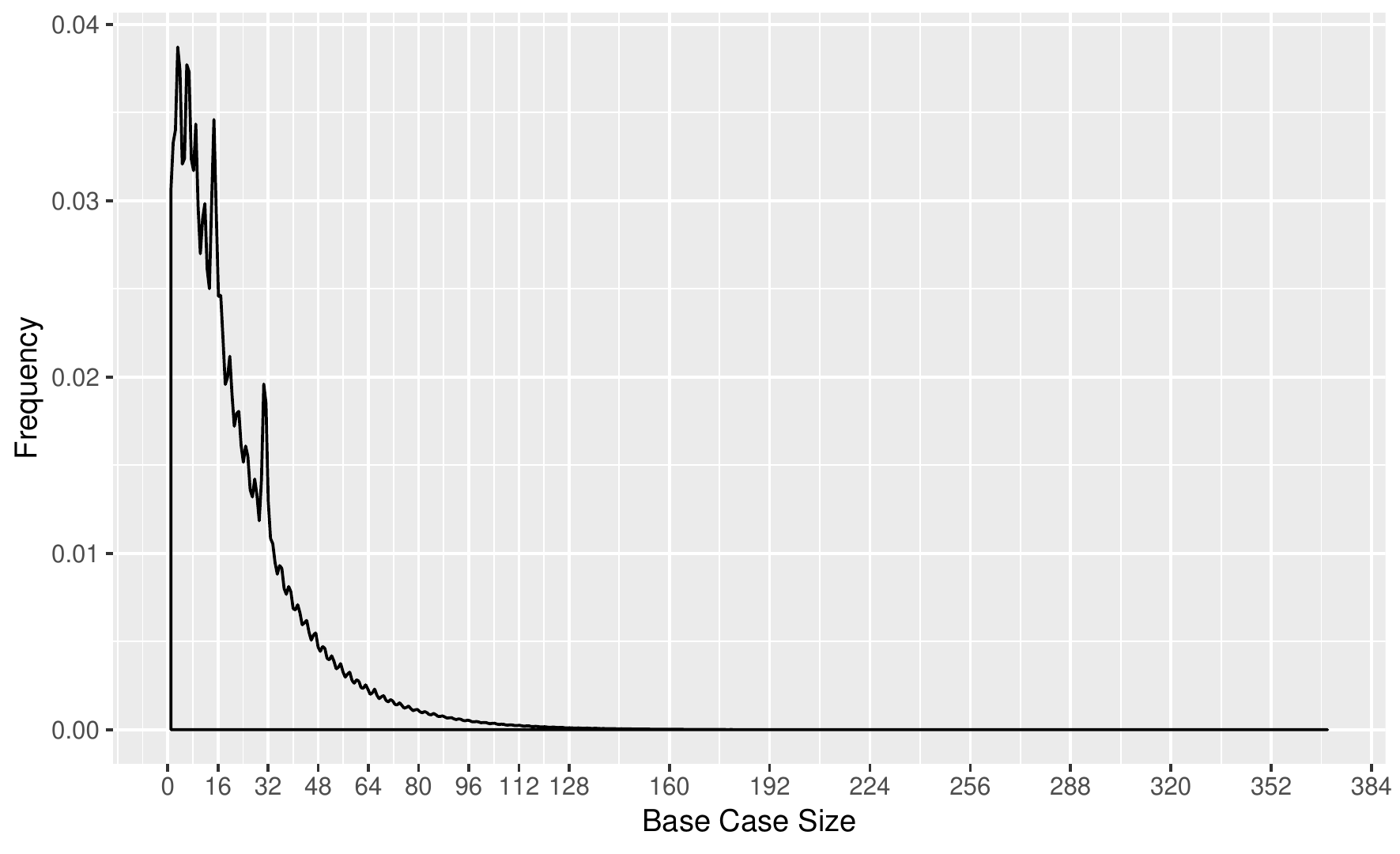}
			\caption{Distribution of the size of the array passed to the base case sorter when executing \ipso\, with parameter \bcsize\, = 16} \label{plot:distr:16}
		\end{figure}
		\begin{figure}[!p]
			\includegraphics[width=1.0\textwidth]{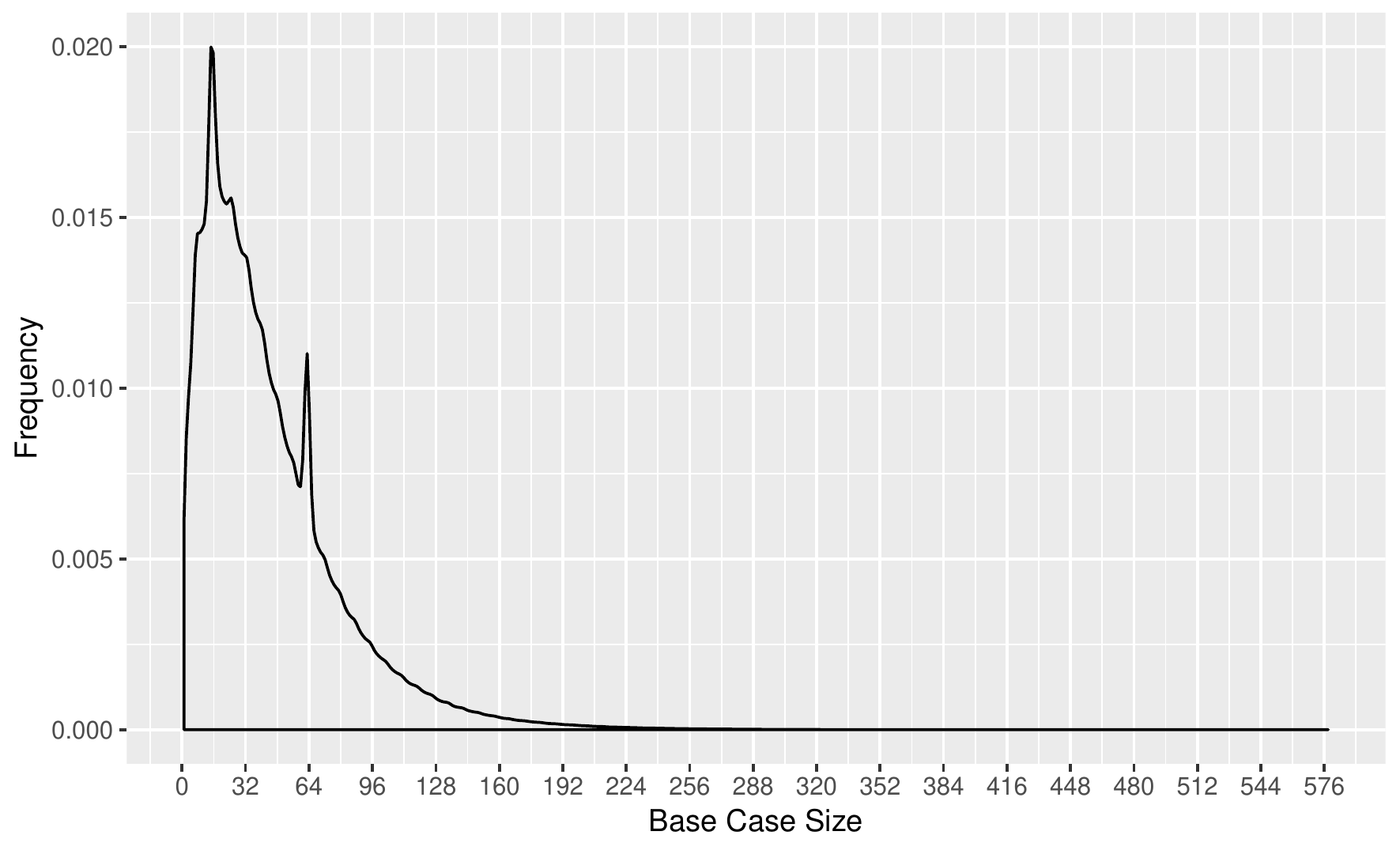}
			\caption{Distribution of the size of the array passed to the base case sorter when executing \ipso\, with parameter \bcsize\, = 32} \label{plot:distr:32}
		\end{figure}
		\clearpage
		Figures \ref{plot:ipso:A}, \ref{plot:ipso:B} and \ref{plot:ipso:C} display the results from the measurements with the above variants. The \bcsize~was appended after the \verb|-4|, along with an underscore followed by the Register Sample Sort configuration. \\
		The the benchmark from algorithm \ref{algo:normal} was used with parameters
		\begin{itemize}
			\item $\mathtt{numberOfIterations} = 50$
			\item $\mathtt{numberOfMeasures} = 200$
			\item $\mathtt{arraySize} = 1024 \times 32 = 32768 = 2^{15}$.
		\end{itemize}
		As already seen in \cite{DBLP:conf/esa/AxtmannWF017}, we get a speed-up of over 59\% over \verb|std::sort| with unchanged \ipso\, on all machines. On machine A, none of the variants we tried led to an improvement in sorting speed over the default use of insertion sort at \bcsize~ 16. For machine B, interestingly, using Register Sample Sort did not lead to an improvement, but the combination of insertion sort and Bose Nelson networks did manage to reduce the sorting time by 4.3\%. For machine C we see the impact of the large L1 instruction cache in the visible improvement of 9.2\% for having Register Sample Sort as a base case instead of insertion sort, though the combinations of insertion sort and the sorting network also performed well. It is notable to see that, while Register Sample Sort by itself did well with blockSizes of 4 or 5, here it is beneficial to use blockSize = 1, having a smaller impact on the instruction cache.
		\begin{table}[!h]
			\begin{center}
			\begin{small}
			\begin{tabular}{ c | c | c | c }
				& A:~~ \verb|S+N BoNeL 16_331 4CS| & B:~~ \verb|I+N 16| & C:~~ \verb|S+N BoNeL 16_331 4CS| \\ \hline
				\verb|I    16 Def| & \red{-3.4\%} & 4.3\% & 9.2\% \\
				\verb|StdSort  -s| & 59.1\% & 61,7\% & 65\% \\
			\end{tabular}
			\end{small}
			\end{center}
			\caption{Average speed-ups of the fastest sorting network over the fastest insertion sort as base case in \ipso\, and unmodified std::sort} \label{table:ipso:speedups}
		\end{table}
		
\clearpage
\section{Conclusion} \label{section:conclusion} 

	\subsection{Results and Assessment} 
		In this thesis we have seen that for sorting sets of up to 16 elements it can be viable to use sorting algorithms other than insertion sort. We looked at sorting networks in particular, paying special attention to the implementation of the conditional swap and giving multiple alternative ways of realizing that implementation. \\
		After seeing that the sorting networks outperform insertion sort each on their own for a specific array size in section \ref{section:experiments:normal} and \ref{section:experiments:inrow}, we saw in section \ref{section:experiments:quicksort} that this improvement does not necessarily transfer to sorting networks being used as base case sorter in quicksort. Because the networks have a larger code size, the code for quicksort is removed from the instruction cache and the advantage of not having conditional branches is impaired by that larger code size. But we also saw that for machines with larger instruction caches using sorting networks with quicksort can lead to visible improvements of about 6.4\%. \\
		After that we integrated the sorting networks into a very advanced sorter like \ipso, which was possible by adding an intermediate sorter into the procedure. For that we created Register Sample Sort, which is an implementation of Super Scalar Sample Sort that holds the splitters in general-purpose registers instead of an array. When measuring \ipso\, with Register Sample Sort as a base case, we found that the instruction cache makes even more of a difference, because we now add the code size for Register Sample Sort on top of the code size for the sorting networks. \\
		We proposed an additional alternative to Register Sample Sort, using a combination of insertion sort and sorting networks: For base cases of 16 elements or less, we used the sorting network, for any size above that insertion sort. \\
		On one of the machines with a smaller instruction cache of 32 KiB we could not achieve a speed-up with any of the variants, on the other the combination of insertion sort and sorting networks led to an improvement in sorting time of 4.3\%. The only substantial improvement we achieved with \ipso\, was on the machine with 64 KiB of L1 instruction cache, where using Register Sample Sort led to an improvement of 9.2\% over plain insertion sort. 
		
		In closing, we want to mention that this particular implementation only compiles when using the gcc C++ compiler due to compiler-dependent inline-assembly statements. This also means that the code is probably not as fast as it could be due to the inline-assembly not being optimized by the compiler. The complete project is available on github at\\ \url{https://github.com/JMarianczuk/SmallSorters}.
	
	\subsection{Experiences and Hurdles}
		The greatest hurdle we encountered during this project was, as mentioned in section \ref{section:measurements}, the fact that the compiler reduces its optimizations with increasing compilation effort, when compiling only a single source file. That can lead to performance variations that happen for no \enquote{apparent} reason, and is especially tricky when dealing with templated methods that can not be moved from header files into source files. The solution was to use code generation and to include all logically coherent method invocations in one wrapper method that is then placed in its own source file, to not have different parts of the program influencing each other over the decision which one gets to be optimized and which one not.
	
	\subsection{Possible Additions} 
		In addition to the work in this thesis, we would like to explore further possibilities to implement the conditional swap for the sorting networks, as well as seeing which of the C++ compilers generate conditional moves when using portable C++ code instead of compiler-dependent inline-assembly. That also includes looking at conditional swaps for elements that differ from the 64-bit key and reference value pair that we looked at in this thesis. \\
		Furthermore we would like to take a look at implementing sorting networks in a way that they take up less code space, and what the trade-off for that decreased code size would be. \\
		
		Apart from the sorting networks we would also like to take another look at Register Sample Sort to find out if using seven splitters instead of three can be more practical when increasing the input size to sizes larger than 256.
	
\clearpage


\bibliographystyle{alphadin}

\end{document}